\documentclass[11pt]{article}

\usepackage{amsthm,amsmath,amssymb,bbm}
\usepackage{natbib}
\usepackage{multirow}
\usepackage{subfigure}
\usepackage{makecell}
\usepackage{booktabs}
\usepackage{array}
\usepackage{tabularx}
\usepackage{tabulary}
\usepackage{caption}
\usepackage{booktabs}
\usepackage{url}
\usepackage{algorithm}
\usepackage{algorithmic}
\usepackage{bm}
\usepackage{wrapfig}
\usepackage{lipsum}
\usepackage{mathrsfs}
\usepackage{dsfont}
\usepackage{titling}
\usepackage{txfonts}

\usepackage{xr}

\usepackage[usenames,dvipsnames,svgnames,table]{xcolor}
\usepackage[colorlinks,
linkcolor=blue,
anchorcolor=blue,
urlcolor  = blue,
citecolor=blue
]{hyperref}

\usepackage{smile}

\newcommand*{\supp}{\mathrm{supp}}
\newcommand*{\var}{\textnormal{var}}

\newcommand{\nn}{\nonumber}

\def\##1\#{\begin{align}#1\end{align}}
\def\$#1\${\begin{align*}#1\end{align*}}

\def\sn{\sum_{i=1}^n}
\def\Sb{\mathbf{S}}

\newcommand{\BB}{\mathbb{B}}

\usepackage{relsize}
\newcommand{\T}{{\mathsmaller {\rm T}}}

\newcommand{\bfsym}[1]{\ensuremath{\boldsymbol{#1}}}
\def \balpha   {\bfsym{\alpha}}       \def \bbeta    {\bfsym{\beta}}
       \def \bdelta   {\bfsym{\delta}}

\newcommand{\Rom}[1]{\text{\uppercase\expandafter{\romannumeral #1\relax}}}

\usepackage{geometry}
\usepackage{enumitem}

\newcommand{\cc}{{\rm c}}

\numberwithin{equation}{section}

\begin{document}

\title{Retire:~Robust Expectile Regression in High Dimensions}

\author{Rebeka Man\thanks{Department of Statistics, University of Michigan, Ann Arbor, Michigan 48109, USA. E-mail:\href{mailto:mrebeka@umich.edu}{\textsf{mrebeka@umich.edu}}.},~~Kean Ming Tan\thanks{Department of Statistics, University of Michigan, Ann Arbor, Michigan 48109, USA. E-mail:\href{mailto:keanming@umich.edu}{\textsf{keanming@umich.edu}}.},~~Zian  Wang\thanks{Department of Mathematics, University of California, San Diego, La Jolla, CA 92093, USA. E-mail:\href{mailto:ziw105@ucsd.edu}{\textsf{ziw105@ucsd.edu}}.}~~and~Wen-Xin Zhou\thanks{Department of Mathematics, University of California, San Diego, La Jolla, CA 92093, USA. E-mail:\href{mailto:wez243@ucsd.edu}{\textsf{wez243@ucsd.edu}}.} }

\date{}
\maketitle

\vspace{-0.5in}

\begin{abstract}

High-dimensional data can often display heterogeneity due to heteroscedastic variance or inhomogeneous covariate effects.  Penalized quantile and expectile regression methods offer useful tools to detect
heteroscedasticity in high-dimensional data.  The former is computationally challenging due to the non-smooth nature of the check loss, and the latter is sensitive to heavy-tailed error distributions.  In this paper, we propose and study (penalized) \underline{r}obust \underline{e}xpec\underline{ti}le \underline{re}gression (\texttt{retire}), with a focus on iteratively reweighted $\ell_1$-penalization which reduces the estimation bias from $\ell_1$-penalization and  leads to oracle properties.   Theoretically,  we establish the statistical properties of the retire estimator under two regimes: (i) low-dimensional regime in which $d \ll n$; (ii) high-dimensional regime in which $s\ll n\ll d$ with $s$ denoting the number of significant predictors.   In the high-dimensional setting, we carefully characterize the  solution path of the iteratively reweighted $\ell_1$-penalized \texttt{retire} estimation,  adapted from the local linear approximation algorithm for folded-concave regularization.  Under a mild minimum signal strength condition,   we show that after as many as $\log(\log d)$ iterations the final iterate enjoys the oracle convergence rate.  At each iteration, the weighted $\ell_1$-penalized convex program can be efficiently solved by a semismooth Newton coordinate descent algorithm.   Numerical studies demonstrate the competitive performance of the proposed procedure compared with either non-robust or quantile regression based alternatives.  \end{abstract}

\noindent
{\bf Keywords}: Concave regularization; Convolution; Minimum signal strength; Oracle property; Quantile regression

\section{Introduction}
\label{sec:1}
 
Penalized least squares has become a baseline approach for fitting sparse linear models in high dimensions.  Its focus is primarily on inferring the conditional mean of the response given the a large number of predictors/covariates.  
In many economic applications, however,  more aspects than the mean of the conditional distribution of the response given the covariates are of interest, and that the covariate effects may be inhomogeneous and/or the noise variables exhibit heavy-tailed and asymmetric tails.   For instance, in the Job Training Partners Act studied in \cite{AAI2002},  one is more interested in the lower tail  than the mean of the conditional distribution of income given predictors such as enrollment in a subsidized training program and demographic variables.
To capture heterogeneity in the set of covariates at different locations of the response distribution, methods such as quantile regression \citep{KB1978} and asymmetric least squares regression (expectile regression) \citep{NP1987} have been widely used. 
We refer the reader to \cite{KB1978}, \citet{K2005}, and \citet{KCHP2017} for a comprehensive overview of quantile regression, and \cite{NP1987} and \cite{GZ2016} for conventional and penalized expectile regressions.



%
Both quantiles and expectiles are useful descriptors of the tail behavior of a distribution in the same way as
the median and mean are related to its central behavior.   As shown by \cite{J1994}, expectiles are exactly quantiles of a transformed version of the original distribution. In fact, the expectile regression can be interpreted as a least squares analogue of regression quantile estimation \citep{NP1987}.
Quantile regression is naturally more dominant in the literature due to the fact that expectiles lack an intuitive interpretation while quantiles are the inverse of the distribution function and directly indicate relative frequency. The key advantage of expectile  regression is its computational expediency and the 
the asymptotic covariance matrix can be estimated without the need of estimating the conditional density function (nonparametrically). Therefore, it offers a convenient and relatively efficient method of summarizing the conditional response distribution.

Expectile regression has found applications in various fields, including risk analysis \citep{T2008,KYH2009,XZW2014,BB2017,DGS2018}, as well as the study of  determinants of inflation \citep{BCM2021} and life expectancy and economic production \citep{SE2009}. In finance applications, the expectile, also known as the expectile-Value at Risk, represents the minimum amount of capital needed to add to a position in order to achieve a specified high gain-loss ratio \citep{BB2017, GZ2019}. 
In contrast to quantile, the expectile is a coherent measure of risk that is desirable  in finance applications \citep{KYH2009}. 
While the expected shortfall is another popular coherent risk measure in finance
\citep{AT2002}, it is not elicitable \citep{G2011}, which poses challenges for its estimation.  In fact, the expectile is the only risk measure that is both coherent and elicitable, making it a valuable tool for financial risk management and decision-making \citep{BB2015,Z2016,BB2017}.


Compared to quantile regression, the expectile regression involves minimizing a quadratic loss and can be sensitive to response that is heavy-tailed. 
To address this limitation,   we introduce a robust expectile regression approach that is capable of handling heavy-tailed response distributions in low- as well as high-dimensional models. In high-dimensional regression models,  the number of covariates/regressors,  $d$, is substantially greater than the number of  observations,  denoted by $n$.   
High-dimensional data analysis greatly benefits from the sparsity assumption---only a small number of significant predictors are associated with the response.  
This motivates the use of various convex and non-convex penalty functions so as to achieve a desirable trade-off between model complexity and statistical accuracy \citep{BG2011, W2019,FLZZ2020}.  The most widely used penalty functions include the $\ell_1$/Lasso penalty \citep{T1996},  the smoothly clipped absolute deviation (SCAD) penalty \citep{FL2001}, and the minimax concave penalty (MCP) \citep{Z2010}.  Albeit being computationally efficient and statistically (near-)optimal under $\ell_2$-errors,  the $\ell_1$ penalty induces non-negligible estimation bias, which may prevent consistent variable selection.   The  selected model  with a relatively small prediction error tends to include many false positives,  unless stringent assumptions are imposed on the design matrix \citep{ZL2008,ZZ2012,  SBC2017, L2021}.     

Folded-concave (non-convex) penalty functions, on the other hand, have been designed to reduce the bias induced by the $\ell_1$ penalty. 
With either the $\ell_2$ or a robust loss function, the resulting folded-concave penalized estimator is proven to achieve the oracle property provided the signals are sufficiently strong, i.e.,  the estimator has the same rate of convergence as that of the oracle estimator obtained by fitting the regression model with true active predictors that are unknown in practice \citep{FL2001,ZL2008,ZZ2012,LW2015,L2017}.
Due to non-convexity, directly minimizing the concave penalized loss raises numerical instabilities. Standard gradient-based algorithms are often guaranteed to find a stationary point, while oracle results are primarily derived for the hypothetical global minimum.  \cite{ZL2008} proposed a unified algorithm for folded-concave penalized estimation based on local linear approximation (LLA). It relaxes the non-convex optimization problem into a sequence of iteratively reweighted $\ell_1$-penalized subproblems. The statistical properties of the final iterate have been studied by \cite{Z2010b}, \cite{FXZ2014},  \cite{FLSZ2018} and \cite{PSZ2021} under different regression models.  We refer to \citet{W2019} and \citet{FLZZ2020}, and the references therein, for a comprehensive introduction of penalized $M$-estimation based on various convex and folded-concave (non-convex) penalty functions.

For sparse quantile regression (QR) in high dimensions, \citet{BC2011} studied $\ell_1$-penalized quantile regression process, and established the uniform (over a range of quantile levels) rate of convergence.
To alleviate the bias induced by the $\ell_1$ penalty, \citet{WWL2012} proposed concave penalized quantile regression,  and showed that the oracle estimator is a local solution to the resulting optimization problem.
Via the one-step LLA algorithm, \citet{FXZ2014} proved that the oracle estimator can be obtained (with high probability) as long as the magnitude of true nonzero regression coefficients is at least of order $\sqrt{s \log(d)/n}$.   We refer to \citet{Wang2019} for a unified analysis of global and local optima of penalized quantile  regressions.   While quantile regression offers the flexibility to model the conditional response distribution and is robust to outliers, together the non-differentiability of the check function and the non-convexity of the penalty pose  substantial technical and computational challenges.
To our knowledge, the theoretical guarantee of the convergence of a computationally efficient algorithm to the oracle QR estimator under the weak minimum signal strength condition---$\min_{j \in \cS } |\beta^*_j| \gtrsim \sqrt{\log(d)/n}$ with $\cS= {\rm supp}(\bbeta^*)$---remains unclear.

In high-dimensional sparse models, \cite{GZ2016} considered the penalized expectile regression using both convex and concave penalty functions. 
Since the expectile loss is convex and twice-differentiable, scalable algorithms, such as the cyclic coordinate decent and proximal gradient descent, can be employed to solve the resulting optimization problem.
Theoretically, the consistency of penalized expectile regression in the high-dimensional regime ``$\log (d) \ll n \ll d$" requires {\em sub-Gaussian} error distributions \citep{GZ2016}. This is in strong contrast to penalized QR, the consistency of which requires no moment condition \citep{BC2011, Wang2019} although certain regularity conditions on the conditional density function are still needed. This lack of robustness to heavy-tailedness is also observed in numerical studies.
Since expectile regression is primarily introduced to explore the tail behavior of the conditional response distribution, its sensitivity to the tails of the error distributions, particularly in the presence of high-dimensional covariates, raises a major concern from a robustness viewpoint.

In this work, we aim to shrink the gap between quantile and expectile regressions, specifically in high dimensions, by proposing a robust expectile regression (\texttt{retire}) method that inherits the computational expediency and statistical efficiency of expectile regression and is nearly as robust as quantile regression against heavy-tailed response distributions. The main idea, which is adapted from \cite{SZF2020}, is to replace the asymmetric squared loss associated with expectile regression with a Lipschitz and locally quadratic robust alternative, parameterized by a data-dependent parameter to achieve a desirable trade-off between bias and robustness.  Under the low-dimensional regime ``$d\ll n$", we provide nonasymptotic high probability error bounds, Bahadur representation, and a Berry-Esseen bound (normal approximation) for the retire estimator when the noise variable has bounded variance.

In the high-dimensional sparse setting ``$\max\{ s, \log(d) \} \ll n \ll d$", we propose an iteratively reweighted $\ell_1$-penalized (IRW-$\ell_1)$ algorithm to obtain the penalized retire estimator, where $s$ denotes the number of significant predictors.
   The problem boils down to iteratively minimizing convex loss functions (proven to be locally strongly convex with high probability), solvable by (but not limited to) a semismooth Newton coordinate descent type algorithm proposed by \citet{YH2017}.  Theoretically, we provide explicit error bounds (in high probability) for the solution path of IRW-$\ell_1$.   More specifically, we first obtain the statistical error of the $\ell_1$-penalized retire estimator, i.e., the first iterate of the IRW-$\ell_1$ algorithm initialized at zero.  
We then show that the statistical error for the subsequent estimators can be improved sequentially by a $\delta$-fraction at each iteration for some constant $\delta \in (0,1)$.  Under a near necessary and sufficient minimum signal strength condition, we show that the IRW-$\ell_1$ algorithm with $\cO\{\log(\log d)\}$ iterations delivers an estimator that achieves the oracle rate of convergence with high probability.

The rest of the paper is organized as follows. In Section~\ref{sec2}, we briefly revisit the connection and distinction between quantile and expectile regressions. We describe the proposed method in Section~\ref{sec3}, where we construct the new loss function and detail the semismooth Newton algorithm to solve the resulting optimization problem. The theoretical properties of the proposed estimator are presented in Section~\ref{sec:theory}. Sections~\ref{sec:numericalstudies} and~\ref{sec:data} consist of extensive numerical studies and two data applications, respectively. The proofs of the theoretical results are given in the online supplementary material.

\section{Background and Problem Setup}
\label{sec2}
Let $y\in\RR$ be a scalar response variable and $\bx=(x_1,\ldots,x_d)^\T \in \RR^d$ be a $d$-dimensional vector of covariates. The training data  $(y_1, \bx_1) , \ldots, (y_n , \bx_n)$ are independent copies of $(y,\bx)$. 
Given a location parameter $\tau \in (0,1)$, we consider the linear model 
\begin{equation}
\label{eq:linearmodel}
y_i = \bx_i^{\T}\bbeta^*(\tau) + \varepsilon_i(\tau),
\end{equation}
where $\bbeta^*(\tau)$ is the unknown $d$-dimensional vector of regression coefficients, and $\varepsilon_i(\tau)$'s are independent random noise.
Model~\eqref{eq:linearmodel} allows the regression coefficients $\bbeta^*(\tau)$ to vary across different values of $\tau$, and thereby offers insights into the entire conditional distribution of $y$ given $\bx$.
Throughout the paper, we let $x_{1} = 1$ so that $\beta_1^*$ denotes the intercept term.  We suppress the dependency of $\bbeta^*(\tau)$ and $\varepsilon(\tau)$ on $\tau$ whenever there is no ambiguity.

The most natural way to relate the conditional distribution of $y$ given $\bx$ and the parameter process $\{ \bbeta^*(\tau), \tau \in (0,1)\}$ is through quantile regression,  under the assumption that $F_{y_i|\bx_i}^{-1}(\tau) = \bx_i^\T \bbeta^*(\tau)$, or equivalently, $\PP\{ \varepsilon_i(\tau) \le 0 \, | \bx_i \} = \tau$.  
Fitting a conditional quantile model involves minimizing a non-smooth piecewise linear loss function, $ \varphi_\tau(u)= u\{\tau- \mathbbm{1}(u<0)\}$, typically recast as a linear program, solvable by the simplex algorithm or interior-point methods. For the latter, \citet{PK1997} showed that the average-case computational complexity grows as a cubic function of the dimension $d$, and thus, is computationally demanding for problems with large dimensions.

Adapted from the concept of quantiles,  \cite{NP1987} and \cite{E1991} separately proposed an alternative class of location measures of a distribution, named the expectile according to the former.
The resulting regression methods are referred to as the expectile regression or the asymmetric least squares regression, which are easy to compute and reasonably efficient under normality conditions. 
We start with some basic notation and facts for expectile regression. 
Let $Z \in \RR$ be a random variable with finite moment, i.e., $\EE(|Z|) < \infty$.  The $\tau$-th expectile or $\tau$-mean of $Z$ is defined as
\begin{equation}
	e_\tau(Z) := \argmin_{u \in \RR}  \EE \bigl\{ \eta_\tau(Z- u) - \eta_\tau(Z) \bigr\}, \qquad \tau \in (0,1) \label{expectile.def},
\end{equation}
where 
\#
 	\eta_\tau( u ) = |  \tau - \mathbbm{1}(u<0) | \cdot \frac{u^2}{2} =  \frac{\tau}{2} \{\max(u,0)\}^2  + \frac{1-\tau }{2}\{\max(-u,0)\}^2 
\label{als.loss}
\#
is the asymmetric squared/$\ell_2$ loss \citep{NP1987}.  
The quantity $e_\tau(Z)$ is well defined as long as $\EE |Z|$ is finite. When $\tau=1/2$, it can be easily seen that $e_{1/2}(Z) = \EE(Z)$. Therefore, expectiles can be viewed as an asymmetric generalization of the mean, and the term ``expectile" stems from a combination of ``expectation'' and ``quantile". Moreover, expectiles are uniquely identified by the first-order condition
$$
	\tau \cdot \EE ( Z - e_\tau(Z) )_+ = (1-\tau)\cdot \EE (Z - e_\tau(Z) )_-,
$$
where $x_+ = \max(x, 0)$ and $x_- = \max(-x, 0)$.
Note also that the $\tau$-expectile of $Z$ defined in \eqref{expectile.def} is equivalent to Efron's $\omega$-mean with $\omega= \tau/(1-\tau)$ \citep{E1991}.

The notion of expectiles is a least squares counterpart of quantiles, and can be viewed as an alternative measure of  ``locations" of the random variable $Z$. Respectively, $1/2$-expectile and $1/2$-quantile correspond to the mean and median, both of which are related to the central behavior.
In general, $\tau$-expectile and $\tau$-quantile with $\tau$ close to zero and one describe the lower and higher regions of the distribution of $Z$, respectively.  The latter is the point below which 100$\tau$\% of the  mass of $Z$ lies, whereas the former specifies the position, say $e_\tau$, such that the average distance from $Z$ below $e_\tau$ to $e_\tau$ itself is 100$\tau$\% of the average distance between $Z$ and $e_\tau$.

Given independent observations $Z_1, \ldots, Z_n$ from $Z$, the expectile location estimator is given by $\hat e_\tau = \argmin_{u \in \RR} \sn \eta_\tau(Z_i - u)$, which is uniquely defined due to the strong convexity of the asymmetric $\ell_2$-loss. The expectile estimator $\hat e_\tau$ can also be interpreted as a maximum likelihood estimator of a normal distributed sample with unequal weights given to disturbances of differing signs, with a larger relative weight given to less variable disturbances \citep{AAP1976}.

Essentially the asymmetric squared loss $\eta_\tau(\cdot)$ is an $\ell_2$-version of the check function $\varphi_\tau(\cdot)$ for quantile regression.  Given train data from the linear model \eqref{eq:linearmodel} subject to $e_\tau(\varepsilon_i|\bx_i)=0$, the  expectile regression estimator \citep{NP1987} is defined as the minimizer of the following convex optimization problem
 \#
 \underset{\bbeta \in \RR^d}{\mathrm{minimize}} ~  \frac{1}{n} \sn \eta_\tau( y_i - \bx_i^{{\T}} \bbeta)   ,  \label{als.reg} 
 \#
 which consistently estimates $\bbeta^*$ when $d=o(n)$ as $n\to \infty$.
In particular,  expectile regression with $\tau=0.5$ reduces to the ordinary least squares regression.

\section{Retire: \underline{R}obust \underline{E}xpec\underline{ti}le \underline{Re}gression}
\label{sec3}
\subsection{A Class of Asymmetric Robust Squared Losses}

Despite its computational advantage over quantile regression,   expectile regression~\eqref{als.reg} is much more sensitive to heavy-tailed distributions due to the squared loss component in~\eqref{als.loss}. 
This lack of robustness is amplified in the presence of high-dimensional covariates, and therefore  necessitates the development of a new class of asymmetric loss functions that preserves the robustness of the check loss to a degree.

To this end, we construct a class of asymmetric robust loss functions  that is more resistant against heavy-tailed error/response distributions.
The main idea is to replace the quadratic component in \eqref{als.loss} with a Lipschitz and locally strongly convex alternative,  typified by the Huber loss \citep{H1964} that is a hybrid $\ell_1$/$\ell_2$ function. The proposed loss function, $\ell_{\gamma}(u)$, contains a tuning parameter $\gamma>0$ that is to be chosen to achieve a balanced trade-off between the robustification bias and the degree of robustness.
At a high level,  we focus on the class of loss functions that satisfies Condition~\ref{def:general.loss} below.
%
\begin{condition}\label{def:general.loss}
Let $\ell_\gamma(u) = \gamma^2 \ell(u/\gamma)$ for $u\in \RR$, where the function $\ell : \RR\mapsto [0,\infty)$ satisfies: (i) $\ell'(0)=0$ and $| \ell'(u) | \leq \min (a_1, |u|) $ for all $u\in \RR$; (ii) $\ell''(0)=1$ and $\ell''(u) \geq a_2$ for all $|u| \leq a_3$; and (iii) $|\ell'(u) - u| \leq  u^2$ for all $u\in \RR$,  
where $a_1$, $a_2$, and $a_3$ are positive constants. 
\end{condition} 

Condition~\ref{def:general.loss} encompasses many commonly used robust loss functions such as the Huber loss $\ell(u) = \min\{ u^2/2, |u| - 1/2 \}$ \citep{H1964}, pseudo-Huber losses $\ell(u) = \sqrt{1+u^2} - 1$ and $\ell(u) = \log( e^u/2 + e^{-u}/2 ) $, smoothed Huber losses $\ell(u) = \min\{ u^2/2 - |u|^3/6, |u|/2 - 1/6\}$ and $\ell(u) = \min\{ u^2/2 - u^4/24, (2\sqrt{2}/3) |u| - 1/2 \}$, among other smooth approximations of the Huber loss \citep{L1990}. 
Consequently,  we consider the following asymmetric robust loss
 \#
 	L_{\tau, \gamma}(u) := | \tau - \mathbbm{1}(u<0) | \cdot \ell_\gamma(u),
	  \label{retire.loss}
 \#
where $\ell_{\gamma}(\cdot)$ is subject to Condition~\ref{def:general.loss}. 

In Section~\ref{subsec:lowd}, we consider the robust expectile regression (retire) estimator based on the robust loss \eqref{retire.loss} in the classical setting that $d<n$. Its statistical properties, both asymptotic and nonasymptotic, will be given in Section~\ref{subsec:lowdim} under the so-called ``many regressors" model \citep{BCCK2015} in which the dimension $d=d_n$ is allowed to grow with $n$ subject to the constraint $d_n = o(n^a)$ for some $0<a \leq 1$. To deal with high-dimensional data for which $d$ can be much larger than $n$, we propose penalized retire estimators in Section~\ref{subsec:highdim} with statistical guarantees (under sparsity) provided in Section~\ref{subsec:theory:highd}.

\subsection{Retire Estimator in Low Dimensions}
\label{subsec:lowd}
Given a location parameter $\tau \in (0,1)$, we define the retire estimator (when $d<n$) as
\begin{align} 
\label{ld estimator}
\hat{\bbeta} = \hat \bbeta_\gamma =  \argmin_{\bbeta \in \RR^d} \frac{1}{n} \sum_{i=1}^n L_{\tau, \gamma}(y_i - \bx_i^{\T} \bbeta),
\end{align}
where $\gamma>0$ is a robustification parameter that will be calibrated adaptively from data as we detail in Section~\ref{sec:simulate}.  Numerically, the optimization problem \eqref{ld estimator} can be efficiently solved by either gradient descent or quasi-Newton methods \citep{NW1999}, such as the Broyden-Fletcher-Goldfarb-Shanno (BFGS) algorithm that can be implemented as on option of the base function \texttt{optim()} in \texttt{R}.

Recall that the population parameter $\bbeta^*$ is uniquely identified as
$$
\bbeta^* = \argmin_{\bbeta \in \RR^d}~ \EE \{L_{\tau,\infty}(y-\bx^{\T}\bbeta)\} ~\mbox{ with }~ L_{\tau,\infty} (u) := | \tau - \mathbbm{1}(u<0) | \cdot u^2/2.
$$
On the other hand,  $\hat{\bbeta}$ can be viewed an $M$-estimator of the following population parameter
$$
\bbeta^*_\gamma  :=  \argmin_{\bbeta \in \RR^d} \EE \{L_{\tau, \gamma}(y-\bx^{\T}\bbeta)\}.
$$
It is worth pointing out that $\bbeta^*_\gamma$ typically differs from $\bbeta^*$ for any given $\gamma>0$.   To see this, note that the convexity of the robust loss $L_{\tau, \gamma}:\RR^d \to \RR$ implies the first-order condition, that is, $\EE \{|\tau - \mathbbm{1}(y < \bx^{\T}\bbeta^*_\gamma )| \cdot \ell'_{\tau, \gamma}(y-\bx^{\T}\bbeta^*_\gamma)\bx \} = \mathbf{0}$. On the other hand, we have $e_\tau(\varepsilon |\bx) = e_\tau(y-\bx^{\T}\bbeta^* |\bx) =0$, implying
$\EE \{ | \tau - \mathbbm{1}(\varepsilon<0) |\cdot \varepsilon  \bx  \}=  \mathbf{0}$. 
Since the random error $\varepsilon$ given $\bx$ is asymmetric around zero, in general we have
$$
 \mathbf{0} \neq \EE \{|\tau - \mathbbm{1}(\varepsilon < 0)| \cdot \ell'_{\tau, \gamma}(\varepsilon)\bx \}   = \EE \{|\tau - \mathbbm{1}(y < \bx^{\T}\bbeta^*  )| \cdot \ell'_{\tau, \gamma}(y-\bx^{\T}\bbeta^*)\bx \} ,
$$
which in turn implies that  $\bbeta^* \neq \bbeta^*_\gamma$. We refer to the difference  $\|\bbeta^*_\gamma - \bbeta^* \|_2$ as the robustification bias.
In Section \ref{subsec:lowdim}, we will  show that  under mild conditions, the  robustification bias is of the order $\cO(\gamma^{-1})$, and a properly chosen $\gamma$ balances bias and robustness.

To perform statistical inference on $\beta^*_j$'s, we construct normal-based confidence intervals based on the asymptotic theory developed in Section~\ref{subsec:lowd}. 
To this end, we first introduce some additional notation. 
Let  $\hat{\varepsilon}_i = y_i-\bx_i^{\T} \hat{\bbeta}$ be the residuals from the fitted model and let $\eb_j \in \RR^d $ be the canonical basis vector, i.e., the $j$-th entry equals one and all other entries equal zero.  
Let $\hat{\mathbf{J}} = n^{-1}\sn |\tau - \mathbbm{1}(\hat{\varepsilon}_i < 0 )|  \cdot \bx_i \bx_i^{\T}$. 
An approximate $95\%$ confidence interval for $\beta_j^*$ can thus be constructed as 
$$
\Bigg[ \hat{\beta}_j - 1.96 {\frac{\hat{\sigma}(\eb_j)}{\sqrt{n}}},~ \hat{\beta}_j - 1.96 {\frac{\hat{\sigma}(\eb_j)}{\sqrt{n}}} \Bigg],
$$
where 
$$
\hat{\sigma}^2(\eb_j) := \eb_j^{\T} \, \hat{\mathbf{J}}^{-1} \Bigg[ \frac{1}{n} \sn \zeta^2(\hat{\varepsilon}_i) \bx_i \bx_i^{\T} \Bigg] \hat{\mathbf{J}}^{-1} \eb_j,
$$
and $\zeta (u) = L'_{\tau,\gamma}(u) =  | \tau - \mathbbm{1}(u<0) | \cdot \ell'_\gamma(u) $ is the first-order derivative of $L_{\tau,\gamma}(\cdot)$ given in \eqref{retire.loss}.

 \subsection{Penalized Retire Estimator in High Dimensions}
\label{subsec:highdim}
In this section, we propose the penalized retire estimator for modeling high-dimensional data 
with $d>n$, obtained by minimizing the robust loss in \eqref{ld estimator} plus a penalty function that induces sparsity on the regression coefficients.
As mentioned in Section~\ref{sec:1},  the non-negligible estimation bias introduced by convex penalties (e.g., the Lasso penalty) can be reduced by folded-concave regularization when the signals are sufficiently strong,  that is, the minimum of magnitudes of all nonzero coefficients are away from zero to some extent.
The latter, however,  is computationally more challenging and unstable due to non-convexity.

Adapted from the local linear approximation algorithm proposed by \cite{ZL2008},  we apply an iteratively reweighted $\ell_1$-penalized algorithm for  fitting sparse robust expectile regression models with the robust loss $L_{\tau, \gamma}(\cdot)$. At each iteration, the penalty weights depend on the previous iterate and the choice of a (folded) concave regularizer satisfying Condition~\ref{def:concavepenalty} \citep{ZZ2012} below.   
Some popular examples  include the smoothly clipped absolute deviation (SCAD) penalty \citep{FL2001},  the minimax concave penalty \citep{Z2010},  and the capped-$\ell_1$ penalty. We refer the reader to \citet{ZZ2012} and Section~4.4 of \cite{FLZZ2020} for more details.  

\begin{condition}
\label{def:concavepenalty}
The penalty function $p_\lambda$ ($\lambda>0$) is of the form $p_\lambda(t) = \lambda^2 p_0(t/\lambda)$ for $t\geq 0$, where the function $p_0: \RR_+ \to \RR_+$ satisfies:
(i) $p_0(\cdot)$ is non-decreasing on $[0,\infty)$ with $p_0(0)=0$; (ii) $p_0(\cdot)$ is differentiable almost everywhere on $(0,\infty)$ and $\lim_{t \downarrow 0} p_0'(t) = 1$; (iv) $p_0'(t_1) \leq p_0'(t_2)$ for all $t_1 \geq t_2> 0$.
\end{condition}

Let $p_{\lambda}(\cdot)$ be a prespecified concave regularizer that satisfies Condition~\ref{def:concavepenalty}, and let  $p'_{\lambda}(\cdot)$ be its first-order derivative.
Starting at iteration 0 an initial estimate $\hat \bbeta^{(0)}$,  we sequentially solve the following weighted $\ell_1$-penalized convex optimization problems:
 \#
\hat{\bbeta}^{(t)} \in \underset{\bbeta\in\RR^d}{\mathrm{minimize}}~\left\{\frac{1}{n} \sum_{i=1}^n L_{\tau,\gamma}(y_i-\bx_i^{\T}\bbeta)+ \sum_{j=2}^d p'_\lambda( | \hat{\beta}_j^{(t-1)} | ) |\beta_j| \right\},
\label{retire.est.convex}
 \# 
 where $\hat{\bbeta}^{(t)}= (\hat{\beta}_1^{(t)}, \ldots, \hat{\beta}_d^{(t)})^\T$.
At each iteration, $\hat{\bbeta}^{(t)}$ is a weighted $\ell_1$-penalized robust expectile regression estimate, where the weight $p'_\lambda( | \hat{\beta}_j^{(t-1)} | ) |\beta_j|$ can be viewed as a local linear approximation of the concave regularizer $p_\lambda(|\beta_j|)$ around $| \hat{\beta}_j^{(t-1)} |$. With the trivial initialization $\hat{\bbeta}^{(0)}=\mathbf{0}$,  the first optimization problem~\eqref{retire.est.convex} (when $t=1$) reduces to the $\ell_1$-penalized robust expectile regression because $p'_{\lambda}(0) =\lambda$.
This iterative procedure outputs a sequence of estimates $\hat{\bbeta}^{(1)},\ldots,\hat{\bbeta}^{(T)}$,  where the number of iterations $T$ can either be set before running the algorithm or depend on a stopping criterion.
Throughout this paper, we refer to the sequence of estimates $\{ \hat{\bbeta}^{(t)} \}_{t=1,\ldots, T}$ given in \eqref{retire.est.convex} as the {\it iteratively reweighted  $\ell_1$-penalized retire} estimators.
We will characterize their statistical properties in Section~\ref{subsec:theory:highd},  including the theoretical choice of $T$ in order to obtain a statistically optimal estimator.

We now outline a coordinate descent type algorithm,    the semismooth Newton coordinate descent (SNCD) algorithm proposed by \citet{YH2017}, to solve the weighted $\ell_1$-penalized convex optimization problem in~\eqref{retire.est.convex}.  Recall that the key component of the asymmetric loss function $L_{\tau,\gamma} (\cdot)$ is the robust loss $\ell_{\gamma}(u) = \gamma^2\ell(u/\gamma)$.   For convenience,   we focus on the Huber loss for which $\ell(u) = u^2/2 \cdot \mathbbm{1}(|u| \leq 1) + (|u| - 1/2)\cdot  \mathbbm{1}(|u| >1)$.  
The main crux of the SNCD algorithm is to combine the semismooth Newton method and the cyclic coordinate descent algorithm to iteratively update the parameter of interest one at a time via a Newton-type step until convergence.  
In the following, we provide a brief derivation of the algorithm, and defer the details to Section~\ref{sec:algorithm:derivation} of the Appendix.

Let $L'_{\tau,\gamma}(u)$ and $L''_{\tau,\gamma}(u)$ be the first and second order derivatives (with respect to $u$) of the loss function in~\eqref{retire.loss}, respectively.  
For notational convenience,  let $\lambda_j^{(t)} = p'_{\lambda}(  |\hat{\beta}_j^{(t-1)} | )$ be the penalty weights at the $t$-th iteration.
Then, the Karush-Kuhn-Tucker conditions for~\eqref{retire.est.convex} read
\begin{equation}
\label{eq:kkt}
\begin{split}
\begin{cases}-\frac{1}{n} \sum_{i=1}^n L'_{\tau,\gamma}(y_i-\bx_i^{\T} \hat{\bbeta}) = 0 \quad \mathrm{for}~ j=1, \\
-\frac{1}{n} \sum_{i=1}^n L'_{\tau,\gamma}(y_i-\bx_i^{\T} \hat{\bbeta})x_{ij} +\lambda_j^{(t)} \hat{z}_j= 0\quad \mathrm{for}~ j=2,\dots,d,\\
\hat{\beta}_j - S(\hat{\beta}_j+\hat{z}_j) = 0 \quad \mathrm{for}~ j=2,\ldots,d,
\end{cases}
\end{split}
\end{equation}
where $\hat{z}_j \in \partial |\hat{\beta}_j|$ is a subgradient of the absolute value function, and $S(u) = \mathrm{sign}(u) \max(|u|-1,0)$.  Finding the optimum to the optimization problem~\eqref{retire.est.convex} is equivalent to solving the system of equations \eqref{eq:kkt}. 
The latter can be done iteratively in a cyclic fashion. That is, at each iteration, we update the pair of parameters $(\beta_j,z_j)$ by solving the corresponding equations in~\eqref{eq:kkt} while keeping the remaining parameters fixed.  
Each pair of parameters is updated by a semismooth Newton step, which we detail in Section~\ref{sec:algorithm:derivation} of the Appendix.
The whole procedure is summarized in Algorithm~\ref{Alg:general}. 

\begin{algorithm}[!htp]
\small
\caption{ Semismooth Newton Coordinate Descent Algorithm for Solving \eqref{retire.est.convex} with a Huber Loss.}
\label{Alg:general}
\textbf{Input:} regularization parameter $\lambda$, Huber loss tuning parameter $\gamma$, and  convergence criterion $\epsilon$.\\
\textbf{Initialization:}  $\hat{\bbeta}^{0}=\boldsymbol{0}$.\\
\textbf{Iterate:} the following until the stopping criterion $\|\hat{\bbeta}^{k}-\hat{\bbeta}^{k-1}\|_2 \le \epsilon$ is met, where $\hat{\bbeta}^{k}$ is the value of $\bbeta$ obtained at the $k$-th iteration.  
\begin{enumerate}
\item $\hat{\beta}_1^{k+1} \leftarrow\hat{\beta}_1^{k} + \{\sum_{i=1}^n L'_{\tau,\gamma}(y_i-\bx_i^{\T} \hat{\bbeta}^{k})\}/\{\sum_{i=1}^n L''_{\tau,\gamma}(y_i-\bx_i^{\T} \hat{\bbeta}^{k})\}$.
\item for $j=2,\ldots,d$, update the pair of parameters $(\beta_j,z_j)$ as follows:
\[
\begin{bmatrix}
\hat{\beta}_j^{k+1}  \\ 
\hat{z}_j^{k+1}
\end{bmatrix}
=
\begin{cases} 
\begin{bmatrix} 
\hat{\beta}_j^{k} + \frac{ \sum_{i=1}^n L'_{\tau,\gamma}(y_i-\bx_i^{\T}\hat{\bbeta}^{k}) x_{ij} - \lambda_j^{(t)} \cdot \mathrm{sign}(\hat{\beta}_j^{k}+\hat{z}_j^{k} )  }{ \sum_{i=1}^n L''_{\tau,\gamma}(y_i-\bx_i^{\T}\hat{\bbeta}^k)) x_{ij}^2 } \\  
\mathrm{sign}(\hat{\beta}_j^{k}+\hat{z}_j^{k} )  
\end{bmatrix}, 
& \mathrm{if~} |\hat{\beta}_j^{k} + \hat{z}_j^{k}|>1, \\ 
\begin{bmatrix} 0  \\  
(n \lambda_j^{(t)})^{-1} \sum_{i=1}^n L'_{\tau,\gamma}(y_i-\bx_i^{\T}\hat{\bbeta}^{k})x_{ij}  + (n \lambda_j^{(t)})^{-1} \hat{\beta}_j^{k} \sum_{i=1}^n L''_{\tau,\gamma}(y_i-\bx_i^{\T}\hat{\bbeta}^{k})x_{ij}^2  
\end{bmatrix}, 
& \mathrm{if~} |\hat{\beta}_j^{k} + \hat{z}_j^{k}|\le 1.
\end{cases}
\]
\end{enumerate}
\textbf{Output:} the final iterate $\hat{\bbeta}^{k}$.
\end{algorithm}

For the Huber loss,  the first- and second-order derivatives of $L_{\tau,\gamma} (u)$ are
\begin{equation*}
\label{eq:lfirstderiv}
L'_{\tau,\gamma}(u) = \begin{cases}
	- (1 - \tau)  \gamma, &{\rm if }~ u < -\gamma,\\
	(1 - \tau)  u, &{\rm if }~ -\gamma \leq u < 0,\\
	\tau  u, &{\rm if }~ 0 \leq u \leq \gamma,\\
	\tau \gamma, &{\rm if }~ u > \gamma,
	\end{cases}
	\quad \mathrm{and}\quad 	
L''_{\tau,\gamma}(u) = \begin{cases}
	1 - \tau, &{\rm if }~ -\gamma \leq u < 0,\\
	\tau, &{\rm if }~ 0 \leq u \leq \gamma,\\
	0, & \text{otherwise},
	\end{cases}
\end{equation*}
respectively.  
In Algorithm~\ref{Alg:general}, the update $\hat{\beta}_j^{k+1}$ involves the second derivative of the loss function, $\sum_{i=1}^nL''_{\tau,\gamma}(y_i - \bx_i^{\T}\hat{\bbeta}^{k})$, in the denominator. 
For extreme values of $\tau$ that are near zero or 1, the denominator may be close to zero, causing instability.
To address this issue, \cite{YH2017} implemented a continuity approximation in their \texttt{R} package \texttt{hqreg}, and we adopt the same technique to implement Algorithm~\ref{Alg:general}.  
In particular,  if the sum of second derivatives is equal to zero or the percentage of residuals with magnitude below $\gamma$ is less than $5\%$ or $n^{-1}$,  we instead substitute the sum of second derivatives by the quantity $\sum_{i=1}^n (| y_i - \bx_i^{\T}\hat{\bbeta}^{k}|)^{-1} \mathbbm{1}(| y_i - \bx_i^{\T}\hat{\bbeta}^{k}| > \gamma)$.
Such a continuity approximation works well across all of the numerical settings that we have considered.

\section{Theoretical Results}
\label{sec:theory} 
We provide an explicit characterization of the estimation error for the retire estimator $\hat{\bbeta}$ defined in \eqref{ld estimator} (low-dimensional setting), and the sequence of penalized retire estimators $\{ \hat{\bbeta}^{(t)} \}_{t=1,\ldots, T}$ defined in \eqref{retire.est.convex} (high-dimensional setting) in Sections~\ref{subsec:lowdim} and~\ref{subsec:theory:highd}, respectively. 
Our proposed estimator relies on the choice of robust loss function  in Condition~\ref{def:general.loss}. 
For simplicity, we focus on the Huber loss $\ell(u) = u^2/2 \cdot \mathbbm{1}(|u| \leq 1) + (|u| - 1/2)\cdot  \mathbbm{1}(|u| >1)$ throughout our analysis, i.e., $a_1=a_2=a_3=1$ in Condition~\ref{def:general.loss}, but note that similar results hold for any robust loss that satisfies Condition~\ref{def:general.loss}. 
Throughout the theoretical analysis, we assume that the location measure $\tau \in (0,1)$ is fixed.

We first defined the empirical loss function and its gradient as 
\begin{equation*}
	\cR_n(\bbeta) = \frac{1}{n} \sn L_{\tau,\gamma}(y_i - \bx_i^\T \bbeta) ~~\mbox{ and}~~ \nabla \cR_n(\bbeta) = -\frac{1}{n} \sn L_{\tau,\gamma}'(y_i - \bx_i^\T  \bbeta) \bx_i ,
\end{equation*}
respectively.
Moreover, we impose some common conditions on the random covariates $\bx$ and the random noise $\varepsilon$ for both low-and high-dimensional settings.
In particular, we assume that the random covariates $\bx \in \RR^d$ are sub-exponential and that the random noise $\varepsilon$ is heavy-tailed with finite second moment. 

\begin{condition}
\label{cond:covariates}    
Let $\bSigma = \EE(\bx\bx^{\T})$ be a positive definite matrix with $\lambda_u \geq \lambda_{\max}(\bSigma) \geq \lambda_{\min}(\bSigma) \ge \lambda_{l} > 0$ and assume that $\lambda_l = 1$ for simplicity.
There exists $\nu_0 \ge 1$ such that $\PP(| \bu^{\T}\bSigma^{-1/2}\bx|\geq \nu_0\|\bu\|_{2}\cdot t)\leq e^{-t}$ for all $t \in \RR$ and $\bu \in \RR^d$.
 For notational convenience, let $\sigma_{\bx}^2 = \max_{1\le j \le d} \sigma_{jj}$, where $\sigma_{jj}$ is the $j$-th diagonal entry of $\bSigma$.
\end{condition}

\begin{condition}
\label{cond:randomnoise}
The random noise $\varepsilon$ has a finite second moment, i.e., 
 $\EE(\varepsilon^2|\bx)\leq \sigma_{\varepsilon}^2<\infty$.  Moreover, the conditional $\tau$-expectile of $\varepsilon$ satisfies $\EE[w_\tau (\varepsilon)\varepsilon|\bx]=0$, where $w_\tau(u) := | \tau - \mathbbm{1}(u<0)|$.
\end{condition}

\subsection{Statistical Theory for the Retire Estimator in~\eqref{ld estimator} }
\label{subsec:lowdim}

In this section, we provide nonasymptotic error bounds for the retire estimator, $\hat{\bbeta}$,  under the regime in which $n>d$ but $d$ is allowed to diverge. 
Moreover, we establish a nonasymptotic Bahadur representation for $\hat{\bbeta}-\bbeta^*$, based on which we construct a Berry-Esseen bound for a normal approximation. 
As mentioned in Section~\ref{subsec:lowd}, the robustification bias $||\bbeta^*_\gamma - \bbeta^* ||_2$ is inevitable due to the asymmetry nature of error term $\varepsilon$. Let $\underbar{$\tau$}  =\min(\tau, 1-\tau)$, $\bar{\tau} = \max(\tau, 1-\tau)$, and $A_1 \geq 1$ be a constant satisfying $\EE  (\bu^{\T} \bSigma^{-1/2} \bx) ^4 \leq A_1^4 \|\bu\|^4_2$ for all $\bu \in \RR^d$. The following proposition reveals the fact that the robustification bias scales at the rate $\gamma^{-1}$, which decays as $\gamma$ grows.

\begin{proposition}
\label{ld prop approx err}
Assume Conditions \ref{def:general.loss}, \ref{cond:covariates}, and \ref{cond:randomnoise} hold. Provided that $\gamma \geq 2 \sigma_\varepsilon A_1^2 \bar{\tau} / \underline{\tau}$, we have 
\[
\| \bSigma^{1/2} (\bbeta^*_\gamma - \bbeta^*  ) \|_{2}   \leq 2 \gamma^{-1} (\bar{\tau}/\underline{\tau}) \sigma^2_\varepsilon. 
\]
\end{proposition}

The key to our subsequent analysis for the retire estimator $\hat{\bbeta}$ is the strong convexity property of the empirical loss function $\cR_n(\cdot)$ uniformly over a local ellipsoid centered at $\bbeta^*$ with high probability. Let $\kappa_1 = \min_{|u| \leq 1} \ell''(u)$, $\BB_{\bSigma} (r) = \{ \bdelta \in \RR^d: \| \bSigma^{1/2} \bdelta   \|_{2} \leq r \}$ be an ellipsoid. We characterize the strong convexity of $\cR_n(\cdot)$ in Lemma~\ref{ld lem:RSC}. With the aid of Lemma~\ref{ld lem:RSC}, we establish a non-asymptotic error bound for the retire estimator $\hat{\bbeta}$ in Theorem~\ref{ld thm 1 l2errorbound}.

\begin{lemma}
\label{ld lem:RSC}
Let $(\gamma, n)$ satisfy $\gamma \geq 4\sqrt{2}  \max \{\sigma_{\varepsilon}, 2A_1^2r \}$ and $n \gtrsim  ({\gamma}/{r})^2(d+t)$. Under Conditions \ref{def:general.loss}, \ref{cond:covariates}, and \ref{cond:randomnoise}, with probability at least $1 - e^{-t}$, we have
\begin{align*}
  \langle \nabla  \cR_n(\bbeta) -  \nabla \cR_n (\bbeta^*) , \bbeta - \bbeta^* \rangle 
  \geq \frac{1}{2} \kappa_1 \underbar{$\tau$}\| \bSigma^{1/2}(\bbeta-\bbeta^*) \|_{2}^2 \mbox{~~uniformly over~~} \bbeta \in  \bbeta^* + \BB_{\bSigma}(r) .
\end{align*}
\end{lemma}

\begin{theorem}
\label{ld thm 1 l2errorbound}
Assume  Conditions \ref{def:general.loss}, \ref{cond:covariates}, and \ref{cond:randomnoise} hold. For any $t>0$, the retire estimator $\hat{\bbeta}$ in \eqref{ld estimator} with $\gamma = \sigma_\varepsilon \sqrt{n/(d+t)}$ satisfies the bound
$$
\| \bSigma^{1/2} (\hat{\bbeta} - \bbeta^*) \|_{2} \leq C(\bar{\tau}/\underline{\tau})\kappa_1^{-1} \sigma_\varepsilon v_0 \sqrt{\frac{d+t}{n}},
$$
with probability at least $1-2e^{-t}$ as long as $n \gtrsim d+t$, where $C>0$ is an absolute constant.
\end{theorem}

Theorem \ref{ld thm 1 l2errorbound} shows that under the sub-exponential design with heavy-tailed random errors with bounded second moment, the retire estimator $\hat{\bbeta}$  exhibits a sub-Gaussian type deviation bound, provided that the robustification parameter is properly chosen, i.e., $\gamma=  \sigma_\varepsilon \sqrt{n/(d+t)}$. In other words, the proposed retire estimator gains robustness to heavy-tailed random noise without compromising statistical accuracy.

\begin{remark}
The choice of $\gamma = \sigma_\varepsilon \sqrt{n/(d+t)}$ in Theorem~\ref{ld thm 1 l2errorbound} is a reflection of the bias and robustness trade-off for the retire estimator $\hat{\bbeta}$. Intuitively, a large $\gamma$ creates less robustification bias but sacrifices robustness. More specifically, we shall see from the proof of Theorem \ref{ld thm 1 l2errorbound} that conditioning on the event $\{\hat{\bbeta} \in \bbeta^* + \BB_{\bSigma}(r_{\rm loc}) \}$,
$$
\| \bSigma^{1/2} (\hat{\bbeta} - \bbeta^*) \|_{2} \lesssim
\underbrace{\frac{\sigma^2_\varepsilon}{\gamma}}_{{\rm robustification~bias}} + \, \underbrace{   \sigma_\varepsilon \sqrt{\frac{  d}{n}} + \gamma \frac{ d}{n} }_{{\rm statistical~error}}
$$ 
with high probability. Therefore, we choose $\gamma = \sigma_\varepsilon \sqrt{n/(d+t)}$ to minimize the right-hand side as a function of $\gamma$.
\end{remark}


Next, we establish nonasymptotic Bahadur representation for the difference $\hat{\bbeta} - \bbeta^*$. 
To this end, we need slightly stronger conditions on both the random covariates $\bx$ and the random noise $\varepsilon$.  
In particular, we require that the random covariates $\bx$ to be sub-Gaussian and that the conditional density of the random noise $\varepsilon$ is upper bounded. We formalize the above into the following conditions.


\begin{condition}
\label{cond:covariates ld}    
There exists $\nu_1 \ge 1$ such that $\PP(| \bu^{\T}  \bSigma^{-1/2}\bx| \geq v_1 ||\bu||_2 t) \leq 2e^{-t^2/2}$ for all $t \in \RR$ and $\bu \in \RR^d$. 
\end{condition}


\begin{condition}
\label{cond:randomnoise ld 1}
Let $f_{\varepsilon|\bx}(\cdot)$ be the conditional density function of the random noise $\varepsilon$. There exists $\bar{f}_{\varepsilon|\bx} > 0$ such that $\sup_{u \in \RR}f_{\varepsilon|\bx}(u) \leq \bar{f}_{\varepsilon|\bx}$ almost surely (for all $\bx$).
\end{condition}

Recall that $w_\tau(u) = | \tau - \mathbbm{1}(u<0)|$ and that $\zeta(u) = L_{\tau, \gamma}'(u) =  w_\tau(u) \ell_\gamma'(u) $. Moreover, let ${\mathbf{J}} = \EE  \{ w_\tau(\varepsilon) \bx \bx^{\T} \}$ be the Hessian matrix. Theorem~\ref{ld thm 2 bahadur representation} establishes the Bahadur representation of the retire estimator $\hat{\bbeta}$. 
Specifically, we show that the remainder of the Bahadur representation for $\hat{\bbeta}$ exhibits sub-exponential tails, which we will use to establish the Berry-Esseen bound for linear projections of $\hat{\bbeta}- \bbeta^*$ in Theorem~\ref{ld thm 3 asym normality}.

\begin{theorem}
\label{ld thm 2 bahadur representation}
Assume Conditions \ref{def:general.loss}, \ref{cond:randomnoise}, \ref{cond:covariates ld}, and \ref{cond:randomnoise ld 1} hold. 
For any $t>0$, the retire estimator $\hat{\bbeta}$ given in \eqref{ld estimator} with $\gamma = \sigma_\varepsilon \sqrt{n/(d+t)}$ satisfies the following nonasymptotic Bahadur representation
\begin{align}
\label{ld thm 2a}
\bigg\|    \bSigma^{-1/2} {\mathbf{J}} (\hat{\bbeta} - \bbeta^*) -  \frac{1}{n} \sum_{i=1}^n  \zeta(\varepsilon_i) \bSigma^{-1/2}\bx_i    \bigg\|_2  \leq C \sigma_\varepsilon \cdot {\frac{d+t}{n}}
\end{align}
with probability at least $1-3e^{-t}$ as long as $n \gtrsim d+t$ , where $C>0$ is a constant independent of $(n,d)$ and $t$.
\end{theorem}

\begin{theorem}
\label{ld thm 3 asym normality}
Under the same set of conditions as in Theorem \ref{ld thm 2 bahadur representation}, assume further that $\EE ( |\varepsilon|^3 | \bx ) \leq  v_3 < \infty$ (almost surely). 
Then, the retire estimator $\hat{\bbeta}$ in \eqref{ld estimator} with $\gamma = \sigma_\varepsilon \sqrt{n/(d+\log n)}$ satisfies
\begin{align*}
\sup_{\bu \in \RR^d, z \in \RR} \big| \PP(n^{1/2} \langle \bu , \hat{\bbeta} - \bbeta^* \rangle \leq \sigma z) - \Phi(z) \big| \lesssim \frac{d + \log n}{\sqrt{n}},
\end{align*}
where $\sigma^2 = \sigma^2(\bu) := \bu^{\T} {\mathbf{J}}^{-1} \EE \{ \zeta^2(\varepsilon) \bx \bx^{\T} \} {\mathbf{J}}^{-1} \bu$ and $\Phi(\cdot)$ is the standard normal cumulative distribution function.
\end{theorem}

Theorem \ref{ld thm 3 asym normality} shows that with a diverging parameter $\gamma = \sigma_\varepsilon \sqrt{n/(d+\log n)}$, for any $\bu \in \RR^d$, the linear projection of $\hat{\bbeta}-\bbeta^*$ is asymptotically normal after some standardization as long as $(n,d)$ satisfies the scaling condition $d = o(\sqrt{n})$.

\subsection{Statistical Theory for the Iteratively Reweighted $\ell_1$-Penalized Retire Estimator }
\label{subsec:theory:highd}
In this section, we analyze the sequence of estimators  $\{\hat{\bbeta}^{(t)}\}_{t=1}^T$ obtained in~\eqref{retire.est.convex} under the high-dimensional regime in which $d >n$. 
Throughout the theoretical analysis, we assume that the regression parameter $\bbeta^*\in \RR^d$ in model~\eqref{eq:linearmodel} is exactly sparse,  i.e.,  $\bbeta^*$ has $s$ non-zero coordinates.
Let $\cS = \{1\le j \le d: \beta_j^*\ne 0\}$ be the active set of $\bbeta^*$ with cardinality $|\cS|=s$.
Recall that $\underbar{$\tau$}  =\min(\tau, 1-\tau)$, $\kappa_1 = \min_{|u| \leq 1} \ell''(u)$ and $A_1>0$ is a constant that satisfies $\EE  (\bu^{\T} \bSigma^{-1/2} \bx) ^4 \leq A_1^4 \|\bu\|^4_2$ for all $\bu \in \RR^d$, where $\bx$ satisfies Condition \ref{cond:covariates}. Similar to the low-dimensional setting, the key to our high-dimensional analysis is an event $\cE_{\rm{rsc}}$ that characterizes the local restricted strong convexity property of the empirical loss function $\cR_n(\cdot)$ over the intersection of an $\ell_1$-cone and a local $\ell_2$-ball centered at $\bbeta^*$  \citep{LW2015}. 
Lemma~\ref{lem:RSC} below shows that the event $\cE_{\rm{rsc}}$ occurs with high probability for suitably chosen parameters.

\begin{definition}
\label{def:rsc}
Given radii parameters $r , L >0$ and a curvature parameter $\kappa>0$, define  the event
\begin{align*}
\cE_{\rm{rsc}}(r,L,\kappa)  = \Bigg\{ \inf_{\bbeta \in \bbeta^* + \BB(r) \cap  \CC(L)}  \frac{\langle  \nabla \cR_n(\bbeta) -  \nabla \cR_n(\bbeta^*) , \bbeta - \bbeta^*  \rangle  }{\| \bbeta - \bbeta^*  \|_2^2 } \geq \kappa   \Bigg\},
\end{align*}
where $\BB (r) = \{ \bdelta \in \RR^d: \| \bdelta   \|_2 \leq r \}$ is an $\ell_2$-ball with radius $r$, and  $\CC(L)=\{ \bdelta: \| \bdelta  \|_1 \leq L\| \bdelta   \|_2 \}$ is an $\ell_1$-cone.
\end{definition}

\begin{lemma} 
\label{lem:RSC}
 Let the radii parameters $(r,L)$ and the robustification parameter $\gamma$ satisfy
\[
\gamma \geq 4\sqrt{2} \lambda_u  \max \{\sigma_{\varepsilon}, 2A_1^2r \} \qquad \mathrm{and} \qquad n \gtrsim  ({\sigma_{\bx} \nu_0 \gamma}/{r})^2 (L^2\log d + t).
\]
Then, under Conditions~\ref{def:general.loss}, \ref{cond:covariates}, and \ref{cond:randomnoise},  event $\cE_{\rm{rsc}}(r,L,\kappa)$ with $\kappa= \kappa_1  \underbar{$\tau$}/ 2$ occurs with probability at least    $1-e^{-t}$. 
\end{lemma}

Under the local restricted strong convexity,  in Theorem~\ref{thm:l1.retire},
we  provide an upper bound on the estimation error of $\hat{\bbeta}^{(1)}$, i.e., the $\ell_1$-penalized retire estimator.

\begin{theorem} 
\label{thm:l1.retire}
Assume Conditions~\ref{def:general.loss}, \ref{cond:covariates}, and \ref{cond:randomnoise} hold.  Then, the $\ell_1$-penalized retire estimator  $\hat{\bbeta}^{(1)}$ 
with 
$\gamma = \sigma_{\varepsilon}\sqrt{{n}/{(\log d + t)}}$ and $\lambda \asymp \sqrt{{(\log d + t)}/{n}}$ satisfies the bounds
\begin{equation*}
\| \hat{\bbeta}^{(1)} - \bbeta^* \|_2 \leq 3(\kappa_1 \underbar{$\tau$} )^{-1} s^{1/2} \lambda\qquad  \mathrm{and}\qquad 
\| \hat{\bbeta}^{(1)} - \bbeta^* \|_1\leq 12(\kappa_1 \underbar{$\tau$} )^{-1} s \lambda,
\end{equation*}
with probability as least $1-3e^{-t}$.
\end{theorem} 

Theorem~\ref{thm:l1.retire} shows that with an appropriate choice of the tuning parameters $\gamma$ and $\lambda$, the $\ell_1$-penalized robust expectile regression satisfies exponential deviation bounds with near-optimal convergence rate as if sub-Gaussian random noise were assumed \citep{GZ2016}.

\begin{remark}
Condition~\ref{cond:randomnoise} can be further relaxed to accommodate heavy-tailed random error with finite $(1+\phi)$ moment with $0<\phi<1$. Specifically, it can be shown that under the $\ell_2$ norm, the estimation error of the $\ell_1$-penalized Huber regression estimator takes the form $s^{1/2} \{\log (d)/n\}^{\min\{\phi/(1+\phi),1/2\}}$ \citep{SZF2020,TSW2022}. Similar results can be obtained for the proposed $\ell_1$-penalized retire estimator and we leave it for future work. 
\end{remark}

\begin{remark}
Throughout this section, we assume that the underlying regression parameter $\bbeta^* \in \RR^d$ is exactly sparse. In this case,  iteratively reweighted $\ell_1$-penalization  helps reduce the estimation bias from $\ell_1$-penalization as signal strengthens. For weakly sparse vectors $\bbeta^*$ satisfying
$\sum_{j=1}^d |\beta_j^*|^q  \leq R_q$ for some $0< q\leq 1$ and $R_q >0$, \cite{FLW2017} showed that the convergence rate (under $\ell_2$-norm) of the $\ell_1$-penalized adaptive Huber estimator with a suitably chosen robustification parameter is of order $\cO(\sigma \sqrt{R_q} \, \{ \log(d) / n \}^{1/2 - q/4 })$.
Using the same argument, the results in Theorem~\ref{thm:l1.retire} can be directly extended to the weakly sparse case where $\bbeta^*$ belongs to an $L_q$-ball for some $0<q \leq 1$. For recovering weakly sparse signals, folded-concave penalization no longer improves upon $\ell_1$-penalization, and therefore we will not provide details on such an extension.
\end{remark}

Next, we establish the statistical properties for the entire sequence of estimators $\hat{\bbeta}^{(1)}, \hat{\bbeta}^{(2)},\ldots, \hat{\bbeta}^{(T)}$ obtained from solving the convex optimization problem~\eqref{retire.est.convex} iteratively.
Let $\|\bbeta_{\cS}^*\|_{\min}=  \min_{j\in\cS} |\beta_j^*|$ be the smallest (in absolute value) non-zero regression coefficient.  
Under a beta-min condition, we show that the estimation error of $\hat{\bbeta}^{(1)}$ stated in Theorem~\ref{thm:l1.retire} can be refined.  More generally,  given the previous iterate $\hat{\bbeta}^{(T-1)}$,  the estimation error of the subsequent estimator, $\hat{\bbeta}^{(T)}$, can be improved by a $\delta$-fraction for some constant $\delta\in (0,1)$.

\begin{theorem} \label{thm:random}
 Let $p_0(\cdot)$ be a penalty function satisfying Condition~\ref{def:concavepenalty}.
 Under Conditions \ref{def:general.loss}, \ref{cond:covariates} and \ref{cond:randomnoise}, assume there exist  some constants $a_1 > a_0 >0$  such that
\begin{align*}
a_0 > {\sqrt{5}}/(\kappa_1 \underbar{$\tau$}),~~~~ p'_0(a_0 ) >0, ~~~~ p'_0(a_1) =0.
\end{align*}
Assume further the minimum signal strength condition $\| \bbeta^*_{\cS} \|_{\min } \geq (a_0+a_1) \lambda$  and the sample size requirement $n\gtrsim s \log d + t$.  Picking $\gamma \asymp  \sigma_{\varepsilon} \sqrt{n/(s + \log d + t)}$ and $\lambda \asymp    \sigma_{\varepsilon} \sqrt{(\log d  +t) / n}$, we have 
\begin{align*}
\| \hat{ \bbeta}^{(T)} - \bbeta^* \|_2
\lesssim
\delta^{T-1} \sigma_{\varepsilon} \sqrt{\frac{s\, (\log d + t)}{n}} + \frac{\sigma_{\varepsilon}}{1-\delta}    \sqrt{\frac{s+\log d +t}{n}}, 
\end{align*}
with probability at least $1-4e^{-t}$.
Furthermore, setting $T \gtrsim \frac{\log\{ \log(d)  + t \}}{\log (1/\delta)}$, we have 
\begin{align} 
\| \hat{ \bbeta}^{(T)} - \bbeta^* \|_2 
 \lesssim  
\sigma_{\varepsilon}    \sqrt{\frac{s+\log d +t}{n}}  
  \label{test}  \\
\mbox{ and }~ \| \hat{ \bbeta}^{(T)} - \bbeta^* \|_1 
 \lesssim    
 \sigma_{\varepsilon}   s^{1/2}  \sqrt{\frac{s+\log d +t}{n}}   
 \label{oracle1}
\end{align}
with probability at least $1-4e^{-t}$, where $\delta = \sqrt{5}/( a_0 \kappa_1 \underbar{$\tau$}) < 1$. 
\end{theorem}

Theorem~\ref{thm:random} shows that under the beta-min condition $\| \bbeta^*_{\cS} \|_{\min } \gtrsim \sqrt{\log(d)/n}$,  the iteratively reweighted $\ell_1$-penalized retire estimator $\hat{\bbeta}^{(T)}$  with $T \asymp  \log \{\log  (d)\} $ achieves the near-oracle convergence rate, i.e.,  the convergence rate of the oracle estimator that has access to the true support of $\bbeta^*$.
This is also known as the weak oracle property.
Picking $t= \log d$, we see that iteratively reweighted $\ell_1$-penalization refines the statistical rate from $\sqrt{s\log(d)/n}$ for $\hat{\bbeta}^{(1)}$ to $\sqrt{(s+ \log d) /n}$ for $\hat{\bbeta}^{(T)}$.

\begin{remark}
Theorem~\ref{thm:random} reveals the so-called {\it weak oracle property} in the sense that the  regularized estimator $\hat{ \bbeta}^{(T)}$ enjoys the same convergence rate as the oracle estimator defined by regressing only on the significant predictors. To obtain such a result, the required minimum signal strength $\| \bbeta^*_{\cS} \|_{\min } \gtrsim \sqrt{\log(d)/n}$ is almost necessary and sufficient. To see this, consider the linear model $y_i = \bx_i^\T \bbeta^*+ \varepsilon_i$ with $\varepsilon_i \sim N(0, \sigma^2)$ independent of $\bx_i$, and define the parameter space $\Omega_{s, a} = \{ \bbeta \in \RR^d: \| \bbeta \|_0 \leq s, \min_{j : \beta_j \neq 0} |\beta_j| \geq a\}$ for $a>0$. 
Under the assumption that the design matrix $\mathbb{X} = (\bx_1, \ldots, \bx_n)^\T \in \RR^{n\times d}$ satisfies a restricted isometry property and has normalized columns, \cite{N2019} derived the following sharp lower bounds for the minimax risk $\psi(s,a) := \inf_{\hat \bbeta} \sup_{\bbeta^* \in \Omega_{s,a}} \EE \| \hat \bbeta - \bbeta^* \|_2^2$: for any $\epsilon\in (0,1)$,
\$
\psi(s,a)  \geq \{ 1 + o(1) \} \frac{2\sigma^2 s \log(e d/s)}{n} ~\mbox{ for any } a \leq (1-\epsilon) \sigma \sqrt{\frac{2\log(ed/s)}{n}} \\
\mbox{ and }~ \psi(s,a) \geq  \{ 1 + o(1) \} \frac{ \sigma^2 s }{n} ~\mbox{ for any } a \geq (1 + \epsilon) \sigma \sqrt{\frac{2\log(ed/s)}{n}} ,
\$
where the limit corresponds to $s/d \to 0$ and $s\log(ed/s) /n \to 0$. The minimax rate $2\sigma^2 s \log(ed/s)/n$ is attainable by both Lasso and Slope \citep{BLT2018},  while the oracle rate $\sigma^2 s/n$ can only be achieved when the magnitude of the minimum signal is of order $\sigma \sqrt{\log(d/s)/n}$. 
The beta-min condition imposed in Theorem~\ref{thm:random} is thus (nearly) necessary and sufficient, and is the weakest possible within constant factors.

Under a stronger beta-min condition $\| \bbeta^*_{\cS} \|_{\min } \gtrsim \sqrt{s\log(d)/n}$, \cite{GZ2016} showed that with high probability, the IRW-$\ell_1$ expectile regression estimator (initialized by zero) coincides with the oracle estimator after three iterations. This is known as the {\it strong oracle property}. Based on the more refined analysis by \cite{PSZ2021}, we conjecture that
the IRW-$\ell_1$ retire estimator $\hat \bbeta^{(T)}$ with $T\asymp \log(s \vee \log d)$ achieves the strong oracle property provided $\| \bbeta^*_{\cS} \|_{\min } \gtrsim \sqrt{\log(d)/n}$ without the $\sqrt{s}$-factor.
\end{remark}

\section{Numerical Studies}
\label{sec:numericalstudies}


\subsection{Simulated Data}
\label{sec:simulate}
We evaluate the performance of the proposed IRW-$\ell_1$-penalized \texttt{retire} estimator \emph{via} extensive numerical studies.  
We implement the $\ell_1$-penalized \texttt{retire} and the IRW-$\ell_1$-penalized   \texttt{retire} using SCAD-based weights with $T=3$, which we compare to three other competitive methods: (i) $\ell_1$-penalized Huber regression (\texttt{huber}); (ii)  $\ell_1$-penalized asymmetric least squares regression (\texttt{sales}) proposed by \citet{GZ2016}, and (iii) $\ell_1$-penalized quantile regression (\texttt{qr}) implemented via the \texttt{R} package \texttt{rqPen} \citep{SM2020}. 
To assess the performance across different methods,  we report the estimation error under the $\ell_2$-norm, i.e., $\|\hat{\bbeta}-\bbeta^*\|_2$, the true positive rate (TPR), and the false positive rate (FPR).  Here, TPR is defined as the proportion of the number of correctly identified non-zeros and the false positive rate is calculated as the proportion of the number of incorrectly identified nonzeros. 

Note that \texttt{huber} and \texttt{sales} are special cases of \texttt{retire} by taking $\tau =  0.5$    and $\gamma \rightarrow \infty$, respectively.  
Thus, both \texttt{huber} and \texttt{sales} can be implemented via Algorithm~\ref{Alg:general}. 
For all methods, the sparsity inducing tuning parameter $\lambda$ is selected via ten-fold cross-validation. Specifically, for methods \texttt{retire}, \texttt{huber}, and \texttt{sales}, we select the largest tuning parameter that yields a value of the asymmetric least squares loss that is less than the minimum of the asymmetric least squares loss plus one standard error. 
For \texttt{qr}, we use the default cross validation function in \texttt{R} package \texttt{rqPen} to select the largest tuning parameter that yields a value of its corresponding loss function that is the minimum of the quantile loss function. 

Both \texttt{huber} and $\ell_1$-penalized \texttt{retire} require tuning an additional robustness parameter $\gamma$. We propose to select $\gamma$  using a heuristic tuning method that involves updating $\gamma$ at the beginning of each iteration in Algorithm~\ref{Alg:general}.   
Let $r_i^k = y_i - \bx_i^{\T} \hat{\bbeta}^{k-1}, i = 1,\ldots,n$ be the residuals, where $\hat{\bbeta}^{k-1}$ is obtained from the $(k-1)$-th iteration of Algorithm~\ref{Alg:general}.  
Let  $\tilde{r}_i^k = (1-\tau)r_i^k \mathbbm{1}_{r_i^k \leq 0} + \tau r_i^k \mathbbm{1}_{r_i^k > 0}$ be the asymmetric residuals, and let 
$\tilde{\rb}^k = (\tilde{r}_1^k,\ldots,\tilde{r}_n^k)^{\T}$.  We define $\mathrm{mad}(\tilde{\rb}^k) = \{\Phi^{-1}(0.75)\}^{-1}\text{median}(|\tilde{\rb}^k - \text{median}(\tilde{\rb}^k)|)$ as the median absolute deviation of the asymmetric residuals, adjusted by a factor $\Phi^{-1}(0.75)$.
We start with setting $\gamma=\sqrt{ n / \log(n p)}$.  
At the $k$-th iteration of Algorithm~\ref{Alg:general}, we update the robustification parameter by  
\begin{equation}\label{heurgamma}
\gamma^{k} = \text{mad}(\tilde{\rb}^{k}) \cdot \sqrt{\frac{n}{\log{(n p)}}}.
\end{equation}
Throughout our numerical studies, we have found that $\gamma$ chosen using the above heuristic approach works well across different scenarios.

For all of the numerical studies, we generate the covariates $\bx_i$ from a multivariate normal distribution $N(\mathbf{0}, \bSigma = (\sigma_{jk})_{1\leq j, k \leq d})$ with $\sigma_{jk} = 0.5^{| j-k|}$.  We then generate the response variable $y_i$ from one of the following three models:
\begin{enumerate}\label{models}
\item Homoscedastic model:  
\begin{equation}\label{hommodel}
y_i = \bx_i^{\T} \bbeta^*+\epsilon_i,
\end{equation}

\item Quantile heteroscedastic model:  
\begin{equation}\label{hetmodel} 
	y_i = \bx_i^{\T} \bbeta^*+{ (0.5 |x_{id}| +0.5) }\{ \epsilon_{i} - F_{\epsilon_{i}}^{-1}(\tau) \},
\end{equation}

\item Expectile heteroscedastic model: 

\begin{equation}\label{exphetmodel}
	y_i = \bx_i^{\T} \bbeta^*+{ (0.5 |x_{id}| +0.5) }\{ \epsilon_{i} - e_{\tau}(\epsilon_{i})\},
\end{equation}
\end{enumerate}
where $\epsilon_i$ is the random noise, $F_{\epsilon_{i}}^{-1}(\cdot)$ denotes the inverse cumulative distribution function of $\epsilon_i$, and $e_{\tau}(\epsilon_{i})$ denotes the inverse of the expectile function of $\epsilon_i$. Note that under Gaussian and t-distributed noises, the two models~\eqref{exphetmodel} and~\eqref{hetmodel} are the same for $\tau=0.5$.
We set the regression coefficient vector $\bbeta^*= (\beta_1^*, \beta_2^*,\ldots, \beta_d^*)^{\T}$ as $\beta_1^* = 2$ (intercept), $\beta^*_j =\{1.8,1.6,1.4,1.2,1,-1,-1.2,-1.4,-1.6,-1.8\}$ for $j = 2,4,\ldots,20$, and 0 otherwise.
The random noise is generated from either a Gaussian distribution, $N(0,2)$,  or a $t$ distribution with 2.1 degrees of freedom.  For the heteroscedastic models,  we consider two quantile/expectile levels $\tau=\{0.5,0.8\}$. 
The results, averaged over 100 repetitions, are reported in Tables~\ref{tab:method4CV}--\ref{tab:method5} for the moderate- ($n=400, ~d=200$) and high-dimensional ($n=400, ~d=500$) settings.

Table~\ref{tab:method4CV} contains results ($\tau=0.5$) under the homoscedastic model with normally and $t$-distributed noise.  For Gaussian noise,   the four $\ell_1$-penalized estimators have similar performance,  and both the estimation error and FPR of IRW \texttt{retire} (with SCAD) are notably reduced.
Under the $t_{2.1}$ noise, we see that \texttt{retire} gains considerable advantage over \texttt{sales} in both estimation and model selection accuracy, suggesting that the proposed estimator gains robustness without compromising statistical accuracy.

Tables~\ref{tab:method1CV} and~\ref{tab:method3CV} show results under the quantile heteroscedastic model with the Gaussian and $t_{2.1}$ noise, respectively.   Two quantile levels $\tau=\{0.5, 0.8\}$ are considered.
We see that \texttt{huber} and $\ell_1$-penalized \texttt{retire} have the same performance when $\tau=0.5$ since they are equivalent for the case when $\tau=0.5$.
Moreover, IRW \texttt{retire} has the lowest estimation error among all methods under the Gaussian noise.  When $\tau=0.8$, the performance of \texttt{huber} deteriorates since \texttt{huber} implicitly assumes $\tau=0.5$ and there is a non-negligible bias when $\tau=0.8$. 
Finally, from Table~\ref{tab:method5} under the expectile heteroscedastic model, we see that the proposed estimator has an even lower estimation error than that of the \texttt{qr}.

We want to point out that in general, under the $t_{2.1}$ noise, the quantile regression method \texttt{qr} has an advantage because the quantile loss is more robust to outliers than all of the other methods. 
While \texttt{qr} exhibits an advantage in terms of estimation error, it is not as computationally efficient as \texttt{retire}, which we will show in Section~\ref{timingcomp}.
In summary,  the numerical studies confirm IRW \texttt{retire} as a robust alternative to its least squares counterpart \texttt{sales} and as a computationally efficient surrogate for the penalized quantile regression approach.

\begin{table}[!htp]
	\fontsize{9}{9.5}\selectfont
	\centering
	\caption{Homoscedastic model \eqref{hommodel} with Gaussian noise ($\epsilon \sim N(0, 2)$) and $t_{2.1}$ noise ($ \epsilon \sim t_{2.1})$.  Estimation error under $\ell_2$-norm (and its standard deviation),  true positive rate (TPR) and false positive rate (FPR),   averaged over 100 repetitions, are reported.}
	\label{tab:method4CV}
	\begin{tabular}{c | c | c c c | c c c}
		\hline
		& & \multicolumn{3}{c|}{$n = 400, d = 200$} & \multicolumn{3}{c}{$n = 400, d = 500$} \\
		Noise & Method & $\ell_2$ error & TPR & FPR & $\ell_2$ error & TPR & FPR \\
		\hline
		Gaussian&$\ell_1$ retire&0.577 (0.009)&1.000 (0.000)&0.026 (0.002)&0.615 (0.009)&1.000 (0.000)&0.013 (0.001)\\
		&IRW retire (SCAD)&\textbf{0.258} (0.006)&1.000 (0.000)&0.011 (0.001)&\textbf{0.251} (0.005)&1.000 (0.000)&0.005 (0.001)\\         
		&$\ell_1$ huber&0.577 (0.009)&1.000 (0.000)&0.026 (0.002)&0.615 (0.009)&1.000 (0.000)&0.013 (0.001)\\
		&$\ell_1$ sales&0.577 (0.009)&1.000 (0.000)&0.026 (0.002)&0.614 (0.009)&1.000 (0.000)&0.013 (0.001)\\
		&$\ell_1$ qr&0.604 (0.010)&1.000 (0.000)&0.159 (0.008)&0.681 (0.010)&1.000 (0.000)&0.085 (0.005)\\
		\hline 
		$t_{2.1}$&$\ell_1$ retire&1.307 (0.039)&0.994 (0.003)&0.006 (0.001)&1.328 (0.040)&0.995 (0.002)&0.003 (0.000)\\
		&IRW  retire (SCAD)&\textbf{0.780} (0.052)&0.982 (0.005)&0.000 (0.000)&\textbf{0.788} (0.052)&0.983 (0.005)&0.000 (0.000)\\       
		&$\ell_1$ huber&1.307 (0.039)&0.994 (0.003)&0.006 (0.001)&1.328 (0.040)&0.995 (0.002)&0.003 (0.000)\\
		&$\ell_1$ sales&1.424 (0.046)&0.990 (0.003)&0.012 (0.002)&1.460 (0.046)&0.987 (0.004)&0.005 (0.001)\\
		&$\ell_1$ qr&0.505 (0.010)&1.000 (0.000)&0.142 (0.009)&0.563 (0.010)&1.000 (0.000)&0.078 (0.004)\\
		\hline
	\end{tabular}
\end{table}

\begin{table}[!htp]
	\fontsize{9}{9.5}\selectfont
	\centering
	\caption{Heteroscedastic model \eqref{hetmodel} with Gaussian noise ($\epsilon \sim N(0, 2)$) and quantile levels $\tau = \{ 0.5, 0.8\}$. }
	\label{tab:method1CV}
	\begin{tabular}{c | c | c c c | c c c}
		\hline
		& & \multicolumn{3}{c|}{$n = 400, d = 200$} & \multicolumn{3}{c}{$n = 400, d = 500$} \\
		$\tau$ & {Method} & $\ell_2$ error & {TPR} & {FPR} & $\ell_2$ error & {TPR} & {FPR} \\
		\hline
		0.5&$\ell_1$ retire&0.570 (0.009)&1.000 (0.000)&0.021 (0.002)&0.597 (0.009)&1.000 (0.000)&0.011 (0.001)\\
		&IRW  retire (SCAD)&\textbf{0.235} (0.006)&1.000 (0.000)&0.006 (0.001)&\textbf{0.230} (0.006)&1.000 (0.000)&0.005 (0.001)\\         
		&$\ell_1$ huber&0.570 (0.009)&1.000 (0.000)&0.021 (0.002)&0.597 (0.009)&1.000 (0.000)&0.011 (0.001)\\
		&$\ell_1$ sales&0.575 (0.009)&1.000 (0.000)&0.021 (0.002)&0.599 (0.009)&1.000 (0.000)&0.011 (0.001)\\
		&$\ell_1$ qr&0.498 (0.008)&1.000 (0.000)&0.146 (0.007)&0.562 (0.008)&1.000 (0.000)&0.086 (0.004)\\
		\hline
		0.8&$\ell_1$ retire&0.581 (0.008)&1.000 (0.000)&0.061 (0.005)&0.624 (0.011)&1.000 (0.000)&0.064 (0.005)\\
		& IRW retire (SCAD)&\textbf{0.448} (0.012)&1.000 (0.000)&0.040 (0.004)&\textbf{0.588} (0.025)&1.000 (0.000)&0.047 (0.004)\\         
		&$\ell_1$ huber&1.210 (0.008)&1.000 (0.000)&0.019 (0.002)&1.233 (0.008)&1.000 (0.000)&0.009 (0.001)\\
		&$\ell_1$ sales&0.636 (0.008)&1.000 (0.000)&0.051 (0.004)&0.661 (0.010)&1.000 (0.000)&0.047 (0.004)\\
		&$\ell_1$ qr&0.574 (0.009)&1.000 (0.000)&0.138 (0.007)&0.639 (0.011)&1.000 (0.000)&0.073 (0.004)\\
		\hline
	\end{tabular}
\end{table}

\begin{table}[!htp]
	\fontsize{9}{9.5}\selectfont
	\centering
	\caption{Heteroscedastic model \eqref{hetmodel} with $t_{2.1}$ noise ($\epsilon \sim t_{2.1}$) and quantile levels $\tau = \{ 0.5, 0.8\}$.  }
	\label{tab:method3CV}
	\begin{tabular}{c | c | c c c | c c c}
		\hline
		& & \multicolumn{3}{c|}{$n = 400, d = 200$} & \multicolumn{3}{c}{$n = 400, d = 500$} \\
		$\tau$ & {Method} & $\ell_2$ error & {TPR} & {FPR} & $\ell_2$ error & {TPR} & {FPR} \\
		\hline
		0.5&$\ell_1$ retire&1.222 (0.039)&0.995 (0.003)&0.006 (0.001)&1.275 (0.042)&0.995 (0.003)&0.003 (0.000)\\
		&IRW  retire (SCAD)&\textbf{0.663} (0.051)&0.988 (0.004)&0.000 (0.000)&\textbf{0.728} (0.055)&0.979 (0.005)&0.000 (0.000)\\
		&$\ell_1$ huber&1.222 (0.039)&0.995 (0.003)&0.006 (0.001)&1.275 (0.042)&0.995 (0.003)&0.003 (0.000)\\
		&$\ell_1$ sales&1.351 (0.051)&0.995 (0.003)&0.011 (0.003)&1.399 (0.047)&0.989 (0.003)&0.004 (0.000)\\
		&$\ell_1$ qr&0.420 (0.008)&1.000 (0.000)&0.150 (0.008)&0.473 (0.008)&1.000 (0.000)&0.075 (0.004)\\
		\hline
		0.8&$\ell_1$ retire&1.052 (0.053)&0.991 (0.004)&0.015 (0.002)&1.065 (0.060)&0.987 (0.006)&0.009 (0.001)\\
		&IRW  retire (SCAD)&\textbf{0.498} (0.045)&0.983 (0.006)&0.003 (0.001)&\textbf{0.487} (0.056)&0.985 (0.006)&0.002 (0.000)\\         
		&$\ell_1$ huber&1.556 (0.032)&0.996 (0.002)&0.007 (0.001)&1.593 (0.034)&0.995 (0.003)&0.003 (0.000)\\
		&$\ell_1$ sales&1.464 (0.102)&0.979 (0.006)&0.039 (0.004)&1.415 (0.060)&0.983 (0.005)&0.026 (0.002)\\
		&$\ell_1$ qr&0.630 (0.013)&1.000 (0.000)&0.132 (0.007)&0.683 (0.014)&1.000 (0.000)&0.070 (0.003)\\
		\hline
	\end{tabular}
\end{table}

\begin{table}[!htp]
	\fontsize{9}{9.5}\selectfont
	\centering
	\caption{Heteroscedastic model \eqref{exphetmodel} with Gaussian noise ($\epsilon \sim N(0, 2)$) and $t_{2.1}$ noise ($\epsilon \sim t_{2.1}$), under the $\tau$-expectile $=0.8$.  }
	\label{tab:method5}
	\begin{tabular}{c | c | c c c | c c c}
		\hline
		& & \multicolumn{3}{c|}{$n = 400, d = 200$} & \multicolumn{3}{c}{$n = 400, d = 500$} \\
		Noise & {Method} & $\ell_2$ error & {TPR} & {FPR} & $\ell_2$ error & {TPR} & {FPR} \\
		\hline
		Gaussian&$\ell_1$ retire&0.534 (0.008)&1.000 (0.000)&0.063 (0.004)&0.557 (0.011)&1.000 (0.000)&0.066 (0.005)\\
		&IRW  retire (SCAD)&\textbf{0.353} (0.012)&1.000 (0.000)&0.042 (0.004)&\textbf{0.501} (0.025)&1.000 (0.000)&0.050 (0.005)\\
		&$\ell_1$ huber&0.898 (0.008)&1.000 (0.000)&0.020 (0.002)&0.924 (0.008)&1.000 (0.000)&0.010 (0.001)\\
		&$\ell_1$ sales&0.538 (0.009)&1.000 (0.000)&0.058 (0.004)&0.548 (0.010)&1.000 (0.000)&0.052 (0.004)\\
		&$\ell_1$ qr&0.671 (0.009)&1.000 (0.000)&0.147 (0.007)&0.716 (0.012)&1.000 (0.000)&0.074 (0.004)\\
		\hline
		$t_{2.1}$&$\ell_1$ retire&1.055 (0.053)&0.991 (0.004)&0.015 (0.002)&1.068 (0.060)&0.987 (0.006)&0.009 (0.001)\\
		&IRW  retire (SCAD)&\textbf{0.487} (0.045)&0.983 (0.006)&0.004 (0.001)&\textbf{0.472} (0.057)&0.985 (0.006)&0.002 (0.000)\\         
		&$\ell_1$ huber&1.535 (0.032)&0.996 (0.002)&0.007 (0.001)&1.573 (0.035)&0.995 (0.003)&0.003 (0.000)\\
		&$\ell_1$ sales&1.470 (0.102)&0.979 (0.006)&0.039 (0.004)&1.420 (0.060)&0.985 (0.005)&0.026 (0.002)\\
		&$\ell_1$ qr&0.638 (0.013)&1.000 (0.000)&0.135 (0.007)&0.688 (0.014)&1.000 (0.000)&0.069 (0.003)\\
		\hline
	\end{tabular}
\end{table}

\subsection{Timing Comparison}
\label{timingcomp}
In this section, we show using additional numerical studies that the proposed $\ell_1$-penalized \texttt{retire} estimator has a significant computational advantage over the $\ell_1$-penalized \texttt{qr}.
We implement \texttt{retire} and \texttt{qr} using the \texttt{R} packages \texttt{adaHuber} and \texttt{rqPen}, respectively.  
For both methods, their corresponding sparsity regularization parameter is selected from a sequence of 50 $\lambda$-values via ten-fold cross-validation.   
The robustification parameter $\gamma$ for \texttt{retire} is selected using the data adaptive procedure described in Section~\ref{sec:simulate}.

\begin{figure}[!t]
	\centering
	\subfigure[Estimation error for model~\eqref{hommodel} with $N(0,2)$ error.]{\label{fig:esterrorgaussian}\includegraphics[width=.47\linewidth]{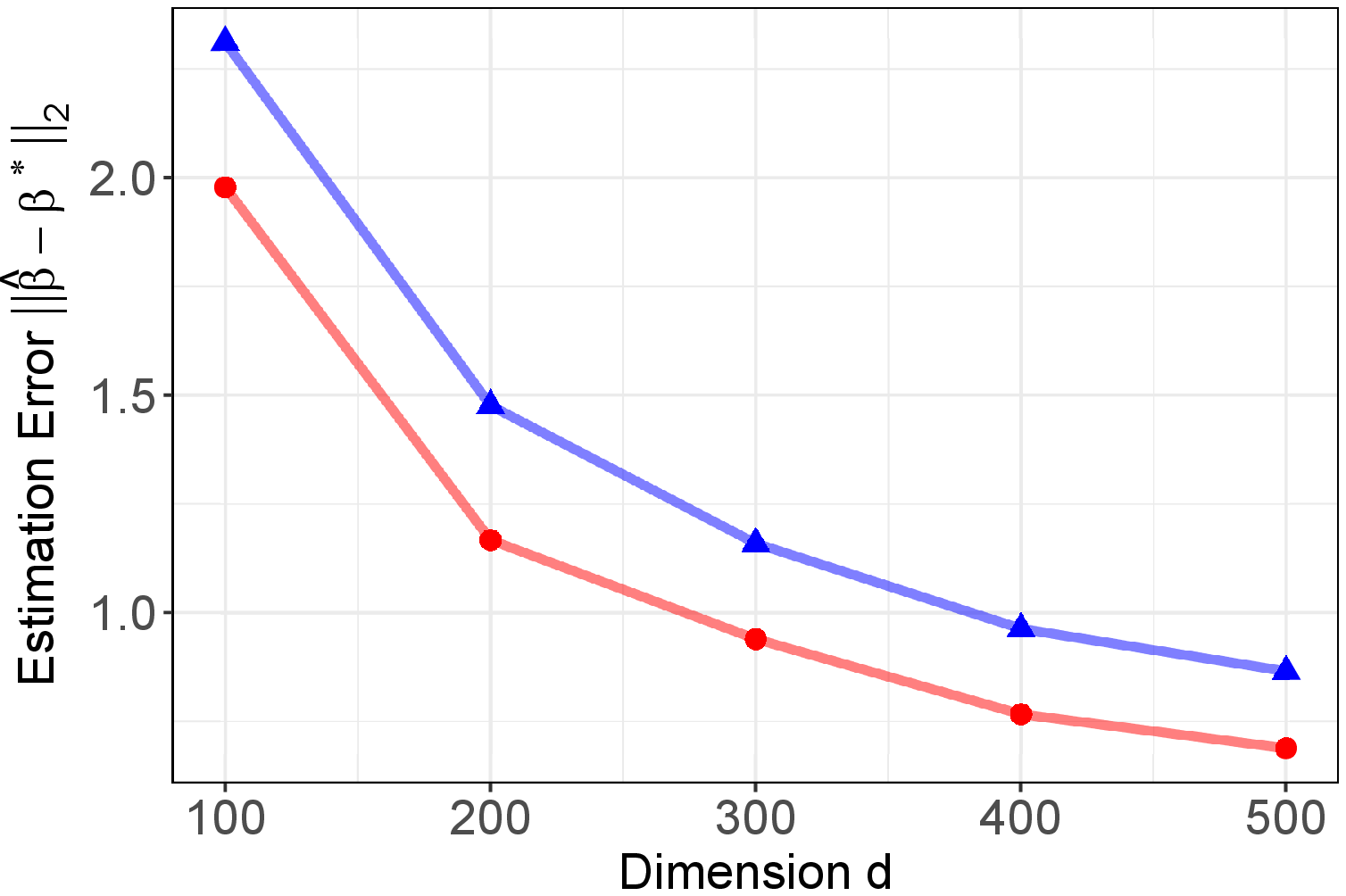}}\qquad
	\subfigure[Elapsed time for model~\eqref{hommodel} with $N(0,2)$ error.]{\label{fig:timecompgaussian}\includegraphics[width=.47\linewidth]{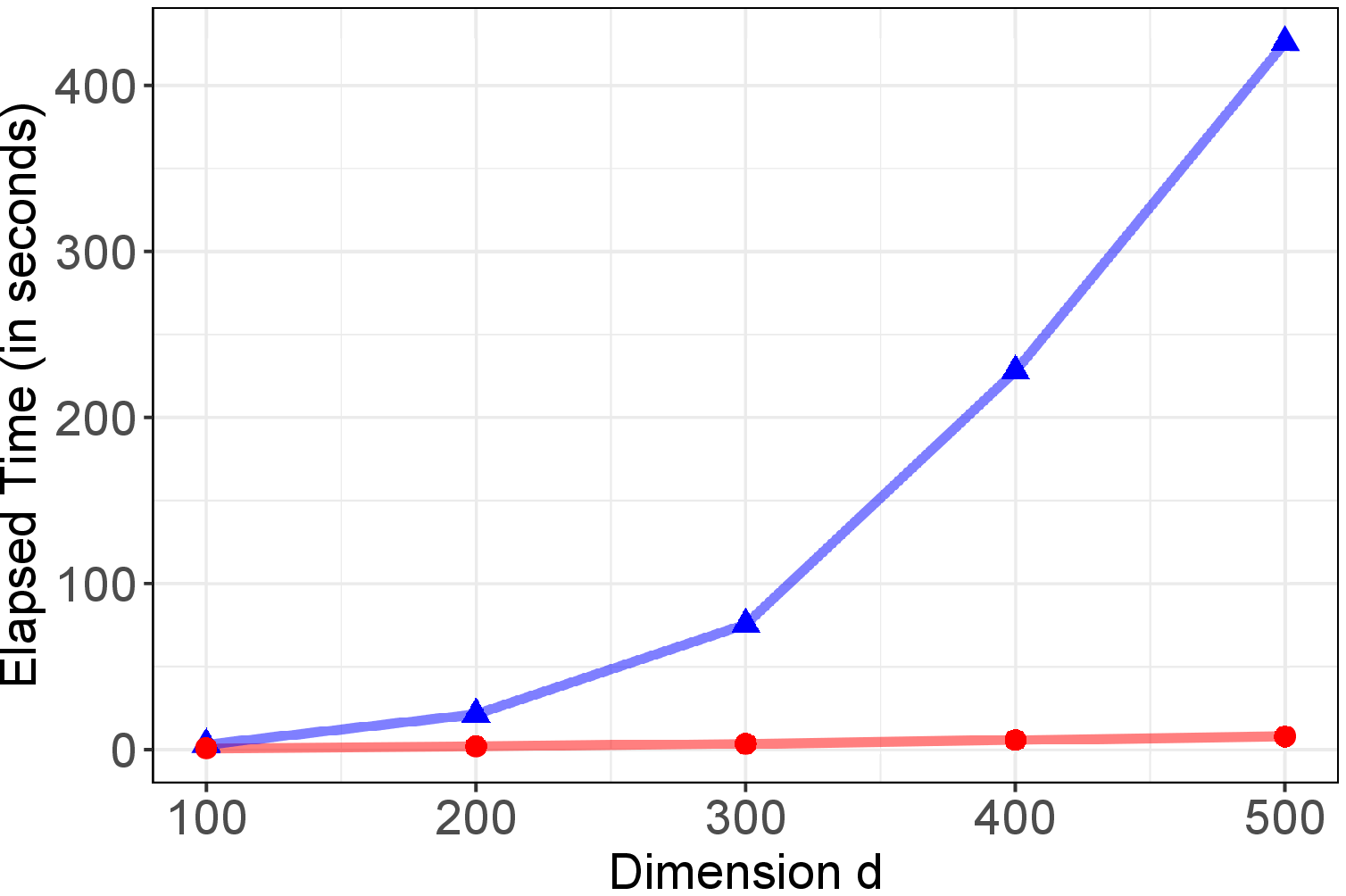}}\\
	\subfigure[Estimation error for model~\eqref{hommodel} with $t_{2.1}$ error.]{\label{fig:esterrort}\includegraphics[width=.47\linewidth]{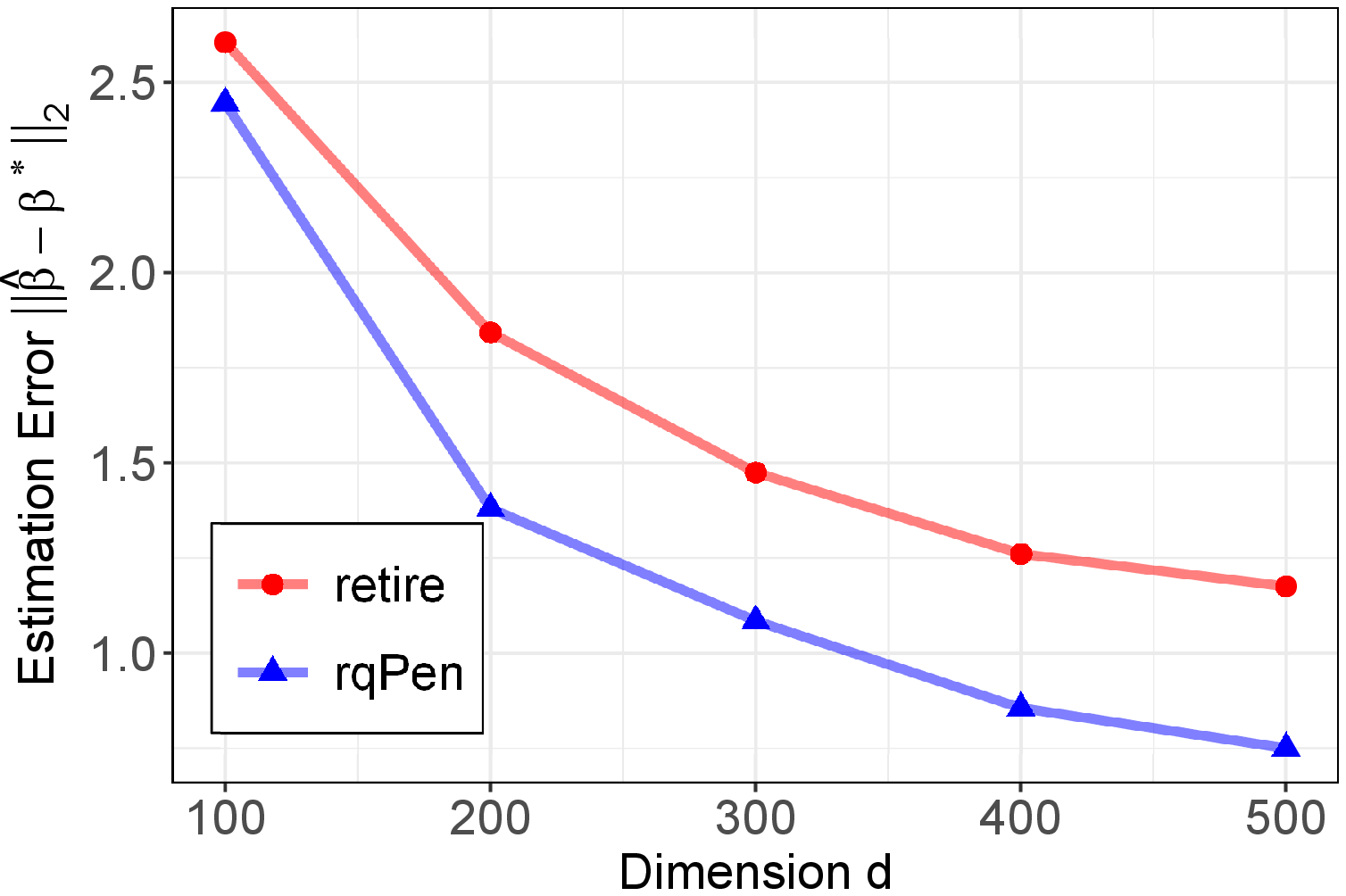}}\qquad
	\subfigure[Elapsed time for model~\eqref{hommodel} with $t_{2.1}$ error.]{\label{fig:timecompt}\includegraphics[width=.47\linewidth]{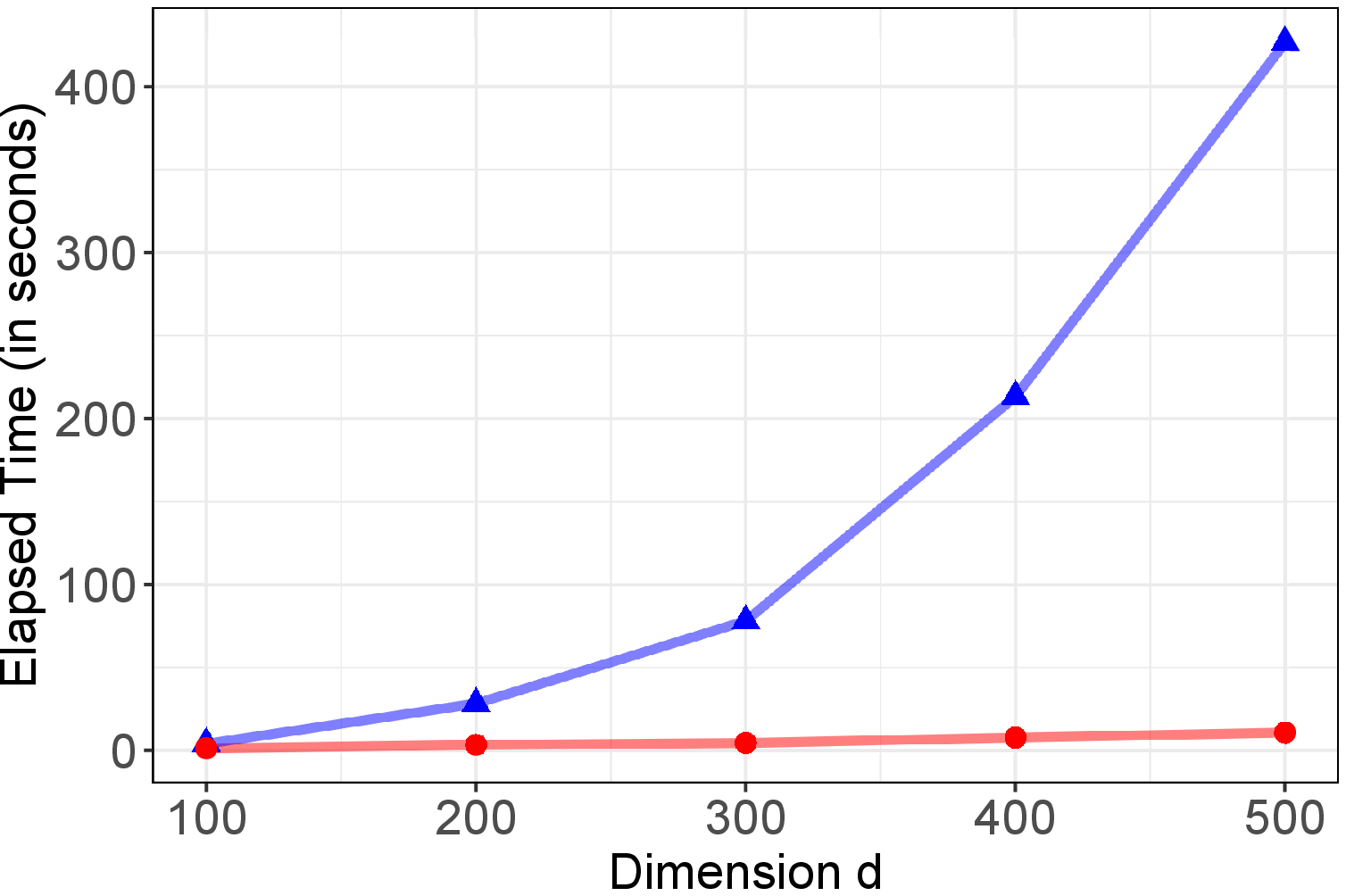}}%
	\caption{Estimation error and elapsed time (in seconds) under model~\eqref{hommodel} with $N(0,2)$ and $t_{2.1}$ random noise and $\tau=0.5$, averaged over 100 data sets for two different methods: (i) the $\ell_1$-penalized \texttt{retire} implemented using the \texttt{R} package \texttt{adaHuber}; (ii) the $\ell_1$-penalized \texttt{qr} implemented using the \texttt{R} package \texttt{rqPen}. The sample size $n$ is set to equal $n = d/2$.}
	\label{fig:comparisons}
\end{figure}

We generate the data from the homoscedastic model~\eqref{hommodel} with the same setup as in Section~\ref{sec:simulate}. Results, averaged over 100 independent data sets, for $n=d/2$ and $d=\{100,200,300,400,500\}$ are summarized in Figure~\ref{fig:comparisons}.
The curves in panels (a) and (c) of Figure~\ref{fig:comparisons} represent the estimation error (under $\ell_2$ norm) as a function of the dimension $d$, and the curves in panels (b) and (d) of Figure~\ref{fig:comparisons} represent the computational time (in seconds) as a function of the dimension $d$.

 Under the Gaussian random noise, $\epsilon \sim N(0,2)$, the $\ell_1$-penalized \texttt{retire} has slightly lower estimation error than $\ell_1$-penalized \texttt{qr}, and  both estimation errors decrease as $n$ and $d$ grow. On the other hand, the $\ell_1$-penalized \texttt{qr} performs better under the $t_{2.1}$ noise since the quantile loss is more robust to outliers than that of the Huber-type loss. 
 Computationally,  the $\ell_1$-penalized \texttt{retire}, implemented via the \texttt{adaHuber} package, exhibits a significant improvement over the $\ell_1$-penalized \texttt{qr}, implemented via the \texttt{rqPen} package, especially when $d$ is large.

\section{Data Application}
\label{sec:data}
\subsection{Job Training Partners Act Data}
We analyze the Job Training Partners Act (JTPA) data, previously studied in \cite{AAI2002}, using the retire estimator proposed in Section~\ref{subsec:lowd}. The JTPA began funding federal training programs in 1983, and its largest component Title II supports training for the economically disadvantaged. 
Specifically, applicants who faced "barriers to employment," the most common of which were high-school dropout status and long periods of unemployment, were typically considered eligible for JTPA training. The services offered as a part of training included classroom training, basic education, on-the-job training, job search assistance, and probationary employment.

In this data set, applicants who applied for training evaluation between November 1987 and September 1989 were randomly selected to enroll for the JTPA training program.  Of the 6,102 adult women in the study, 4,088 were offered training and 2,722 enrolled in the JTPA services, and of the 5,102 adult men in the study, 3,399 were offered training and 2,136 enrolled in the services. 
The goal is to assess the effect of subsidized training program on earnings.  
Motivated by \cite{AAI2002}, we use the 30-month earnings data collected from the Title II JTPA training evaluation study as the response variable.
Moreover, we adjust for the following covariates: individual's sex (male=1, female=0), whether or not the individual graduated high school or obtained a GED (yes=1, no=0), whether or not the individual worked less than 13 weeks in the 12 months preceding random assignment (yes=1, no=0), whether or not the individual is black (yes=1, no=0), whether or not the individual is Hispanic (yes=1, no=0), and marriage status (married=1, not married=0).
We study the conditional distribution of 30-month earnings at different expectile levels $\tau = \{0.1,0.5,0.9\}$.  Our proposed method involves robustification parameter $\gamma$, which we select using the tuning method described in Section~\ref{sec:simulate}. 

The regression coefficients and their associated 95\% confidence intervals are shown in Table~\ref{tab:jtpa}. We find that covariates with positive regression coefficients for all quantile levels are enrollment for JTPA services, individual's sex, high school graduation or GED status, and marriage status. Black, hispanic, and worked less than 13 weeks in the past year had negative regression coefficients. The regression coefficients varied across the three different expectile levels we considered. The positive regression coefficients increase as the $\tau$ level increases and the negative regression coefficients decrease as the $\tau$ level increases. That is, for the lower expectile level of 30-month earnings, the covariates have a smaller in magnitude effect on the individual's earnings compared to the higher expectile level. The regression coefficient for enrollment in JTPA services was 1685.34, 2637.57, and 2714.57 at $\tau=\{0.1,0.5,0.9\}$, respectively. The $\tau$-expectile of 30-month earnings for $\tau=\{0.1,0.5,0.9\}$ is 5068.02, 15815.29, and 32754.89 dollars, respectively. Compared to the expectile at the given $\tau$, the effect of subsidized training was larger for lower expectile levels. Notably, if an individual is a male, conditional on other covariates, their 30-month earnings increase by 5,005 dollars for $\tau=0.5$ and increase by 10,311 dollars for $\tau=0.9$. From the confidence intervals, we see that all variables are statistically significant except Hispanic. 
 
\begin{table}[!htp]
	\fontsize{9}{9.5}\selectfont
	\centering
	\caption{Regression coefficients (and their associated 95\% confidence intervals) for the retire estimator.   }
	\label{tab:jtpa}
	\begin{tabular}{c | c c c }
		\hline
		Variable & $\tau=0.1$ & $\tau=0.5$ & $\tau=0.9$\\
		\hline
		enrolled in services&1685.34 (1401.03, 1969.65)&2637.57 (2079.74, 3195.40)&2714.57 (1766.01, 3663.13)\\
		male&1706.87 (1435.04, 1978.69)&5005.12 (4449.07, 5561.17)&10310.62 (9338.91, 11282.34)\\
		high school or GED&1477.19 (1218.33, 1736.06)&3656.13 (3140.12, 4172.14)&5718.62 (4803.60, 6633.63)\\
		black&-580.04 (-917.86, -242.21)&-1567.03 (-2265.51, -868.56)&-2459.81 (-3686.14, -1233.48)\\
		hispanic&-130.72 (-588.11, 326.66)&-669.76 (-1626.83, 287.32)&-1495.33 (-3306.12, 315.46)\\
		married&1268.30 (933.66, 1602.94)&3343.63 (2668.95, 4018.30)&4518.43 (3376.92, 5659.93)\\
		worked less than 13 wks&-3677.98 (-3957.24, -3398.72)&-6879.14 (-7438.20, -6320.08)&-8206.16 (-9151.81, -7260.50)\\
		\hline
	\end{tabular}
\end{table}

\subsection{Childhood Malnutrition Data}
We apply the IRW $\ell_1$-penalized retire estimator  with SCAD-based weights to the childhood malnutrition data,.  This data set is previously studied in \cite{BCK2019} and \cite{K2011}. The data are collected from the Demographic and Health Surveys (DHS) conducted regularly in more than 75 countries.
Similar to \cite{BCK2019}, in this analysis, we will focus on data collected from India, with a total sample size of 37,623. The children studied are between the ages of zero and five. 

The goal is to study the conditional distribution of children's height in India given the following covariates: the child's age, months of breastfeeding, mother's body mass index, mother's age, mother's education in years, partner's (father's) education in years, number of dead children in family, and multiple categorical variables including but are not limited to child's sex, child's birth order, mother's employment status, family's wealth (whether they are in poorest, poorer, middle, or richer bracket), electricity, television, refrigerator, bicycle, motorcycle, and car. Additionally, interactions between the following variables were considered: child's age, months of breastfeeding, child's sex, whether or not the child was a twin, mother's BMI, mother's age, mother's years of education, father's years of education, mother's employment status, and mother's residence. There are a total of 75 covariates: 30 individual variables and 45 two-way interactions.

We aim to study the conditional distribution of children's height at different expectile levels $\tau = \{0.1,0.5,0.9\}$. Our proposed method involves two tuning parameters $\gamma$ and $\lambda$.  The choice of robustification parameter $\gamma$ was determined by theoretic guidance via a tuning method described in Section~\ref{sec:simulate}.  The choice of sparsity tuning parameter $\lambda$ is selected using a ten-fold cross validation where we select the largest tuning parameter that yields a value of the asymmetric least squares loss that is less than the minimum of the asymmetric least squares loss plus one standard error. For fair comparison, we apply the same sparsity tuning parameter across the three expectile levels. This is achieved by taking the maximum of the sparsity tuning parameters selected using a ten-fold cross-validation for the three different expectile levels. The selected tuning parameter takes value $\lambda = 0.035$.

The regression coefficients that are non-zero for at least one value of $\tau$ are shown in Table~\ref{tab:malnutrition}. There are a total of 38 non-zero coefficients. The regression coefficients for months of breastfeeding vary across the three different expectile levels we consider. At $\tau = 0.1$, the coefficient is 0.445, while at $\tau = \{ 0.5,0.9\}$, the coefficients are 0.397 and 0.378 respectively. 
That is, for lower expectile level of child's height, months of breastfeeding plays a more important role to ensure that the child is not malnourished compared to higher expectile levels. 

Other variables of interest are electricity, television, and motorcycle. For $\tau=0.1$ and $\tau=0.9$, the regression coefficients are 0, suggesting that access to these resources plays less of a role in a child's height at extreme expectile levels since access becomes a given for $\tau=0.9$ and vice versa. For $\tau=0.5$, the coefficients for electricity, television, and motorcycle are 0.647, 0.367, and 0.587 respectively, suggesting that these resources are important.

\begin{table}[!t]
    \fontsize{9}{9.5}\selectfont
    \centering
    \caption{Non-zero regression coefficients for IRW-$\ell_1$-penalized \texttt{retire} with SCAD-based weights across three expectile levels $\tau=\{0.1,0.5,0.9\}$ for the childhood malnutrition data. }
    \label{tab:malnutrition}
    \begin{tabular}{c | c c c | c | c c c}
        \hline
         Variable & $\tau=0.1$ & $\tau=0.5$ & $\tau=0.9$ & Variable & $\tau=0.1$ & $\tau=0.5$ & $\tau=0.9$\\
         \hline
         cage (child age)&0.650&0.686&0.728&breastfeeding*medu&-0.002&-0.001&-0.001\\
         breastfeeding&0.445&0.397&0.378&breastfeeding*edupartner&0.001&0.000&0.000\\
         cbirthorder1&0.012&0.703&0.000&breastfeeding*munemployed&0.000&0.000&-0.002\\
         deadchildren&0.000&-0.345&0.000&breastfeeding*mresidence&0.000&0.000&-0.011\\
         electricity&0.000&0.647&0.000&csex*mbmi&-0.011&-0.015&0.000\\
         television&0.000&0.367&0.000&csex*mage&-0.039&-0.034&-0.040\\
         motorcycle&0.000&0.587&0.000&ctwin*mage&-0.035&-0.032&0.000\\
         cage*breastfeeding&-0.010&-0.008&-0.008&mbmi*mage&0.000&0.002&0.002\\
         cage*csex&0.000&0.001&0.000&mbmi*medu&0.002&0.001&0.004\\
         cage*ctwin&0.000&0.000&-0.028&mbmi*edupartner&0.000&-0.002&0.000\\
         cage*mbmi&0.002&0.001&0.001&mbmi*munemployed&0.000&0.013&0.019\\
         cage*mage&0.001&0.002&0.002&mage*medu&0.002&0.000&-0.001\\
         cage*medu&0.005&0.003&0.002&mage*edupartner&0.002&0.003&0.001\\
         cage*edupartner&0.001&0.001&0.001&mage*munemployed&0.006&0.006&0.004\\
         cage*munemployed&-0.006&-0.007&-0.007&mage*mresidence&-0.010&0.000&0.000\\
         cage*mresidence&0.000&0.003&0.000&medu*edupartner&0.002&0.003&0.005\\
         breastfeeding*csex&-0.005&-0.007&-0.009&medu*munemployed&-0.009&-0.031&-0.050\\
         breastfeeding*mbmi&0.002&0.002&0.002&edupartner*munemployed&-0.017&-0.030&-0.028\\
         breastfeeding*mage&-0.002&-0.002&-0.003&edupartner*mresidence&0.000&-0.019&-0.015\\

         \hline
    \end{tabular}
\end{table}

\section{Discussion}

In this study, we focused on robust estimation and inference for expectile regression  in two scenarios: the low-dimensional setting where  $d \ll n$, and the high-dimensional sparse setting where $s \ll n \ll d$.  For the latter, we developed a robust penalized expectile regression method through iterative reweighted $\ell_1$-penalization and established non-asymptotic high probability bounds.  Performing statistical inference in high dimensions is much more challenging than in low dimensions due to the lack of a tractable limiting distribution of the penalized estimator when $d\gg n$.  In recent years, there has been a rich development of debiased and de-sparsified procedures for penalized regression. These methods lead to estimators with asymptotically normal distributions, as demonstrated by \cite{ZZ2014},  \cite{vdGetal2014},   \cite{JM2014}  and \cite{NL2017}, among others.  However, a complete overview of these methods is beyond the scope of this study.

To conduct statistical inference for $\beta^*_j$ $(2\leq j\leq d)$,  the $j$-th coordinate of $\beta^*$,  we consider the score function
$$
	S_n(\beta_j,  \bbeta_{-j} ,  \bv) = \frac{1}{n} \sn L'_{\tau, \gamma} (y_i - x_{i, j } \beta_j - \bx_{i, -j}^\T \bbeta_{-j} ) (x_{i, j} - \bx_{i, -j}^\T \bv) , 
$$
where $\bbeta_{-j}$ and $\bx_{i, -j}$ are, respectively, the subvectors of $\bbeta \in \RR^d$ and $\bx_i \in \RR^d$ with the $j$-th element removed.   Let $\hat \bbeta^{{\rm init}}$ be an initial penalized estimator of $\bbeta^*$ as described in Section~\ref{subsec:highdim}, and denote by $\hat \bv \in \argmin_{\bv \in \RR^{p-1} } \{  (2n)^{-1} \sn (x_{i, j} - \bx_{i, -j}^\T \bv)^2 + \lambda_v \| \bv \|_1 \}$ a Lasso-type estimator of $\bv^*: = \argmin_{\bv\in \RR^{d-1}} \EE(x_{i,j} - \bx_{i, -j}^\T \bv)^2$,  the linear projection vector of the regressor of interest $x_{i,j}$ on the remaining covariates $\bx_{i, -j}$.   Assuming that $\bv^*$ is $s_v$-sparse,  we conjecture that with a properly chosen robustification parameter $\gamma$,  the ``oracle" score $\sqrt{n} \,S_n(\beta^*_j, \hat \bbeta^{{\rm init}}_{-j} , \hat \bv)$ converges in distribution to a centered normal distribution as $n, d \to \infty$ provided that $\max\{s , s_v \}  \log(d) = o(\sqrt{n})$. Motivated by the classical one-step construction \citep{B1975},  which aims to improve an initial estimator that is consistent but not efficient,  we  propose a debiased estimator 
$$
	\hat \beta_j = \hat \beta^{{\rm init}}_j - S_n(\hat \beta^{{\rm init}}_j , \hat \bbeta^{{\rm init}}_{-j} , \hat \bv ) \Big/  \partial_b S_n( b  , \hat \bbeta^{{\rm init}}_{-j} , \hat \bv ) |_{b=\hat \beta^{{\rm init}}_j }   .
$$
This estimator is conjectured  to follow an asymptotically normal distribution.  With a consistent estimate of its asymptotic variance,  a Wald-type confidence interval can be constructed.  However, a rigorous  theoretical investigation of the debiased estimator and the  accompanying asymptotic variance estimation problem requires a significant amount of future work, which we leave for future studies.


\appendix

%
\section{Derivation of Algorithm~\ref{Alg:general}} 
\label{sec:algorithm:derivation}
In this section, we provide a derivation of Algorithm~\ref{Alg:general} for solving~\eqref{retire.est.convex} under the Huber loss $\ell(u) = u^2/2 \cdot \mathbbm{1}(|u| \leq 1) + (|u| - 1/2) \cdot \mathbbm{1}(|u| > 1)$. 
Under the Huber loss, the loss function $L_{\tau,\gamma}(u)$ takes the form
\begin{equation*}
\label{eq:rethub}
	\begin{split}
	L_{\tau, \gamma}(u) &= 
		\begin{cases}
		(1- \tau) \cdot (\gamma \cdot |u| - \gamma^2/2), & u/\gamma < -1, \\
		(1-\tau) \cdot u^2/2, & -1 \leq u/\gamma < 0, \\
		\tau \cdot u^2/2, & 0 \leq u/\gamma \leq 1, \\
		\tau \cdot (\gamma \cdot |u| - \gamma^2/2), & u/\gamma > 1.
		\end{cases}
	\end{split}
\end{equation*}
For notational simplicity, let $\lambda_j^{(t)} = p_{\lambda}'(\hat{\beta}_j^{(t-1)})$. Then,  
optimization problem~\eqref{retire.est.convex} reduces to 
 \#
\underset{\bbeta\in\RR^d}{\mathrm{minimize}}~\left\{\frac{1}{n} \sum_{i=1}^n L_{\tau,\gamma}(y_i-\bx_i^{\T}\bbeta)+ \sum_{j=2}^d \lambda_j^{(t-1)} |\beta_j| \right\},  
\label{eq:penr}
 \# 

We now derive the SNCD algorithm proposed by \cite{YH2017} to solve~\eqref{eq:penr}. The main idea is to solve the system of equations, i.e., the KKT conditions, in~\eqref{eq:kkt} iteratively in a cyclic fashion using a semismooth Newton algorithm.   
In particular, we update the parameters $\beta_1$, $(\beta_2,z_2),\ldots,(\beta_d,z_d)$ iteratively one at a time while keeping the others fixed. 
In the following, we will derive the explicit updates for the parameters at each iteration.

We start with the intercept term $\beta_1$. 
Let $r_i^{k} = y_i -\bx_i^{\T} \bbeta^{k}$ and 
$F_1^{k}(u) = -\frac{1}{n} \sum_{i=1}^n L'_{\tau,\gamma} (r_i^{k} + \beta_1^{k} -u)$. 
At  the $k$th iteration, we update  the parameter $\beta_1$ as
\begin{align*}
	\beta_1^{k+1} &= \beta_1^{k} - \{H_1^{k}(\beta_1^{k})\}^{-1} F_1^{k}(\beta^{k}_1) = \beta_1^{k} + \frac{\sum_{i=1}^n L'_{\tau,\gamma}( y_i -\bx_i^{\T} \bbeta^{k})}{\sum_{i=1}^n L''_{\tau,\gamma}( y_i -\bx_i^{\T} \bbeta^{k})}, 
	\end{align*}
where 	$H_1^{k}(u)=\frac{1}{n} \sum_{i=1}^n L_{\tau,\gamma}'' (r_i^{k}+\beta_1^{k}-u)$ is the derivative of $F_1^k(u)$.

We now derive the update for $(\beta_j,z_j)$.  Let  
\[
    F_{j}^{k}(v_1,v_2) = \begin{bmatrix} -\frac{1}{n}\sum_{i=1}^nL'_{\tau,\gamma}(r_{i}^{k} + x_{ij}\beta_{j}^{k} - x_{ij}v_{1})x_{ij} + \lambda_j^{(t-1)} v_{2} \\ v_{1} - S(v_{1} + v_{2})
\end{bmatrix}, \qquad\mathrm{for}~j=2,\ldots,d, 
 \]
where $S(u) = \mathrm{sign}(u) \max(|u|-1,0)$ is the soft-thresholding operator.
The derivative of $F_j^k(v_1,v_2)$ then takes the form
\begin{equation}
\label{eq:Hjupdate}
H_j^k (v_1,v_2) = \begin{cases}
 \begin{bmatrix} \frac{1}{n}\sum_{i=1}^nL''_{\tau,\gamma}(r_{i}^{k} + x_{ij}\beta_{j}^{k} - x_{ij}v_{1})x_{ij}^{2} & \lambda_j^{(t-1)} \\ 0 & -1
    \end{bmatrix} &  \mathrm{if} ~|\beta_j^k + z_j^k | > 1\\ 
    \begin{bmatrix} \frac{1}{n}\sum_{i=1}^nL''_{\tau,\gamma}(r_{i}^{k} + x_{ij}\beta_{j}^{k} - x_{ij}v_{1})x_{ij}^{2} & \lambda_j^{(t-1)} \\ 1 & 0
    \end{bmatrix}
     &\mathrm{otherwise}. \\
\end{cases} 
\end{equation}
The update for  $(\beta_j,z_j)$ then takes the form 
\begin{align*}
	\begin{bmatrix} \beta_j^{k+1} \\ z_j^{k+1} \end{bmatrix} &= \begin{bmatrix} \beta_j^{k} \\ z_j^{k} \end{bmatrix} - \{H_j^k(\beta_j^{k},z_j^{k})\}^{-1}F_j^k(\beta_j^{k},z_j^{k}).
\end{align*}
Substituting~\eqref{eq:Hjupdate} into the above equation yields the updates in Algorithm~\ref{Alg:general}.

\section{Proofs} 
\subsection{Preliminary Results}
\label{Sec:Prelim}
Given $\tau \in (0,1)$, let $\{ (y_i, \bx_i)\}_{i=1}^n$ be a sample of independent data vectors from the linear regression model in \eqref{eq:linearmodel},
$y_i = \bx_i^{\T}\bbeta^*(\tau) + \varepsilon_i(\tau)$, where $\varepsilon_i(\tau)$ satisfies $e_\tau(\varepsilon_i | \bx_i )=0$.  
In other words, the conditional $\tau$-mean of $y_i$ give $\bx_i$ is a linear combination of $\bx_i$.
We suppress the dependency of $\bbeta^*(\tau)$ and $\varepsilon (\tau)$ on $\tau$ throughout the Appendix.
Let $w_\tau(u) := | \tau - \mathbbm{1}(u<0)|$ and let $\ell_\gamma(u) = \gamma^2 \ell(u/\gamma)$.  
Recall from~\eqref{retire.loss} that $L (u) :=  L_{\tau,\gamma}(u) = w_\tau (u) \ell_{\gamma}(u)$ and let     
\begin{equation*}
	\cR_n(\bbeta) = \frac{1}{n} \sn L(y_i - \bx_i^\T \bbeta) ~~\mbox{ and}~~ \nabla \cR_n(\bbeta) = -\frac{1}{n} \sn L'(y_i - \bx_i^\T  \bbeta) \bx_i , 
\end{equation*}
where $L'(u) = \gamma w_\tau(u) \ell'(u/\gamma)$ is the first-order derivative of $L(u)$. 

For $\bbeta \in \RR^d$,  let $
\bw(\bbeta) = \nabla \cR_n(\bbeta) - \nabla  \cR (\bbeta)$, 
where $\cR (\bbeta) = \EE \{\cR_n(\bbeta)\}$  is the population loss. Moreover, we define the quantity $\bw^* = \nabla \cR_n(\bbeta^*) - \nabla \cR(\bbeta^*)$
as the centered score function.
Recall from Definition~\ref{def:rsc} that $\CC(L)=\{ \bdelta: \| \bdelta \|_1 \leq L\| \bdelta  \|_2 \}$. Let $\CC_1 :=\{ \bdelta: \| \bdelta_{\cS^{\cc}} \|_1 \leq 3\| \bdelta _{\cS}  \|_1 \}$.
Moreover, define the symmetrized Bregman divergence $\cB: \RR^p\times \RR^p \to [0,\infty)$ associated with the convex function $\cR_n(\cdot)$ evaluated at $\bbeta_1, \bbeta_2$ as 
\#
\label{eq:symmetricdivergence}
	\cB(\bbeta_1, \bbeta_2 ) = \langle \nabla  \cR_n(\bbeta_1) -  \nabla \cR_n (\bbeta_2) , \bbeta_1 - \bbeta_2 \rangle. 
\#
Recall from Condition~\ref{cond:covariates} that $\lambda_u \ge \lambda_{\max} (\bSigma)$, where $\bSigma = \EE(\bx\bx^{\T})$. Also recall from Condition~\ref{cond:randomnoise} that $\EE (\varepsilon^2 | \bx) \le \sigma_{\varepsilon}^2$.

We first present some technical lemmas that are useful for proving theoretical results in the low-dimensional setting for the non-penalized retire estimator in Section~\ref{subsec:lowd}, i.e.,
\begin{equation}
\label{lowdretire}
\hat{\bbeta} = \hat \bbeta_\gamma =  \argmin_{\bbeta \in \RR^d} \frac{1}{n} \sum_{i=1}^n L_{\tau, \gamma}(y_i - \bx_i^{\T} \bbeta).
\end{equation}

\begin{lemma}
\label{ld lem:B1B2}
Under Conditions~\ref{def:general.loss}, \ref{cond:covariates}, and \ref{cond:randomnoise}, we have $ \| \bSigma^{-1/2}\nabla \cR(\bbeta^*) \|_2 \leq \gamma^{-1} \bar{\tau} \sigma_\varepsilon^2$. Moreover, for any $t>0$,
\begin{align*}
\big|\big| \bSigma^{-1/2} \big\{ \nabla \cR_n(\bbeta^*) - \nabla \cR(\bbeta^*) \big\}\big|\big|_2 \leq 3\bar{\tau}v_0 \Bigg(\sigma_\varepsilon \sqrt{\frac{2d+t}{n}} + \gamma \frac{2d+t}{2n} \Bigg)
\end{align*}
with probability at least $1 - e^{-t}$.
\end{lemma}
\begin{lemma}
\label{ld lem:huber}
Let $\varepsilon$ be a real-valued random variable with $\EE (\varepsilon) = \sigma_\varepsilon^2$, $\EE |\varepsilon|^3 = v_3 < \infty$, and $\EE \{w_\tau(\varepsilon)\varepsilon\} = 0$ with $w_\tau(u) = |\tau - \mathbbm{1}(u<0)|$. Let $\ell_\gamma (\cdot)$ be the Huber loss with parameter $\gamma$. We have
$$
\big| \EE \{w_\tau(\varepsilon)\ell_\gamma'(\varepsilon) \}\big| \leq {\bar{\tau}v_3}/{\gamma^2}
\mbox{~~and~~}
\underline{\tau}^2 \bigg( \sigma_\varepsilon^2 - {v_3}/{\gamma} \bigg) \leq \EE \Big\{ w_\tau(\varepsilon)\ell_\gamma'(\varepsilon) \Big\}^2 \leq \bar{\tau}^2 \sigma_\varepsilon^2.
$$
\end{lemma}

Next, we present some technical lemmas that are useful for analyzing the high-dimensional penalized retire estimator.  Recall that the penalized retire estimator is obtain by solving optimization problem~\eqref{retire.est.convex}.  
For notational convenience, throughout the Appendix, we define the minimizer of ~\eqref{retire.est.convex} as 
\begin{equation} 
 \hat \bbeta^{(t)}  \in \argmin_{\bbeta \in \RR^{d}}  \big\{  \cR_n(\bbeta) +  \| \blambda^{(t)}   \circ  \bbeta \|_1  \big\}   ,  \label{general.lasso2}
\end{equation}
where $\blambda^{(t)} = (\lambda_1^{(t)}, \ldots, \lambda_d^{(t)} )^\T$ is a $d$-dimensional vector of tuning parameters with  $\lambda_j^{(t)} = p'_{\lambda}(| \hat{\beta}_j^{(t-1)}|)$, and $\circ$ is the Hadamard product.  Throughout the proof, we drop the superscript from $\hat{\bbeta}^{(t)}$ and $\blambda^{(t)}$ when the context is clear.

The proofs of all of the technical lemmas are deferred to Section~\ref{appendix:technicallemmaproof}.
\begin{lemma}
\label{lem:grad}
Under Conditions~\ref{def:general.loss}, \ref{cond:covariates}, and \ref{cond:randomnoise}, we have $ \| \nabla \cR (\bbeta ^*) \|_2 \leq \gamma^{-1} \bar{\tau}  \lambda_{u}^{1/2}\sigma_{\varepsilon}^2$ and $\| \nabla \cR (\bbeta ^*) \|_\infty \leq \gamma^{-1}  \bar{\tau}  \sigma_{\bx}  \sigma^2_{\varepsilon}$.
Moreover, for any $t\geq 0$,
\begin{align*}
\| \bw^* \|_\infty= \|  \nabla \cR_n (\bbeta ^*) -  \nabla \cR (\bbeta ^*)\|_\infty 
 \leq 
\nu_0 \sigma_{\bx} \bar{\tau} \Bigg(2\sigma_{\varepsilon} \sqrt{\frac{\log d + t }{n}}+ \gamma \frac{\log d + t }{n}\Bigg) 
\end{align*}
holds with probability at least $1-2e^{-t}$.
\end{lemma}

Lemma \ref{lem:grad} reveals the proper range for the penalty level $\lambda$ so that event $ \{\lambda \geq 2\|\nabla \cR_n(\bbeta^*) \|_\infty \}$ occurs with high probability.
   Let $\cS$ be the active set of the true regression parameter $\bbeta^*$, and 
 $\Sb = \EE(\bx_{\cS} \bx_{\cS}^{\T})$ be the $s \times s$ principal submatrix of $\bSigma$.  Denote by $\lambda_{\max}(\Sb)$ the maximal eigenvalue of $\Sb$.  Write $\bw^* = \nabla \cR_n (\bbeta ^*) -  \nabla \cR (\bbeta ^*)$.  The next lemma provides an upper bound for the centered score $\bw^*$,  projected on the true support $\cS$.
 
\begin{lemma}\label{lem:grad_l2}
Under Conditions~\ref{cond:covariates}--\ref{cond:randomnoise}, for any $t > 0$, we have
\begin{align*}
\| \bw^*_{\cS} \|_2 
\leq 3\bar{\tau} \nu_0 \lambda^{1/2}_{\max}(\Sb) \Bigg( \sigma_{\varepsilon} \sqrt{\frac{2s+t}{n}} + \gamma \frac{2s+t}{2n} \Bigg),
\end{align*}
 with probability at least $1-e^{-t}$.
\end{lemma}

The following two lemmas contain some results for the solution of~\eqref{general.lasso2}. Both lemmas are essential for the proof of Proposition~\ref{prop:further steps}.

\begin{lemma} 
\label{lem:l1cone}
Let $\cA$ be a set such that $\cS \subseteq \cA \subseteq [d]$.  For any $\bbeta\in \RR^d$, let $\bbeta_{\cA^{\cc}}=\bf{0}$. Assume that $\| \blambda_{\cA^{\cc}} \|_{\min} > \| \bw(\bbeta) \|_{\infty}$.  
Then, any solution $\hat{\bbeta}$ to the optimization problem~\eqref{general.lasso2} satisfies
\begin{align}
\| (\hat{\bbeta}-\bbeta)_{\cA^{\cc}} \|_1 \leq
\frac{ \big\{ \| \blambda \|_{\infty} + \| \bw(\bbeta) \|_{\infty} \big\} \| (\hat{\bbeta}-\bbeta)_{\cA} \|_1 +  \|  \nabla  \cR  (\bbeta )\|_2 \|\hat{\bbeta}-\bbeta  \|_2}{\| \blambda_{\cA^{\cc}} \|_{\min}-\| \bw(\bbeta) \|_{\infty}}. 
\label{eq:l1cone}
\end{align}

\end{lemma}

\begin{lemma} 
\label{lem:deterministic.error.bound}
Let $\cA$ be a set such that $\cS \subseteq \cA \subseteq [d]$ and $|\cA|=k$.  
Let $\blambda = (\lambda_1,\ldots,\lambda_d)^{\T}$ be a vector of tuning parameters that satisfies $\| \blambda \|_{\infty} \leq \lambda$ and $\| \blambda_{\cA^{\cc}}  \|_{\min} \geq a\lambda$ for some constant $a \in (0,1]$ and $\lambda \geq s^{-1/2} \| \nabla \cR (\bbeta ^*) \|_2$.  Then, under the event $\{a \lambda \geq 2\|\bw^* \|_\infty \}$, any solution $\hat{\bbeta}$ to~\eqref{general.lasso2} satisfies $\hat{\bbeta} \in \bbeta^* + \CC(L)$ with $L=(2+2/a)k^{1/2} + 2 s^{1/2}/a$. 
In addition, let $\kappa, r >0$ satisfy $ r > \kappa^{-1}(2s^{1/2}+ {k^{1/2}a}/{2})\lambda$.   
Then, under the event $\cE_{\rm{rsc}}(r,L,\kappa)$, we have 
\begin{align*}
\| \hat{\bbeta} - \bbeta^* \|_2
&\leq \kappa^{-1} \big\{ (2s^{1/2}+ {k^{1/2}a}/{2})\lambda  \big\} < r.
\end{align*}
\end{lemma}

\subsection{Proof of Theorems}
\subsubsection{Proof of Theorem \ref{ld thm 1 l2errorbound}}
\begin{proof}
Recall from~\eqref{lowdretire} that $\hat{\bbeta} = \argmin~\cR_n(\bbeta)$ and from~\eqref{eq:symmetricdivergence} that $\cB(\bbeta, \bbeta^*) = \langle \nabla \cR_n(\bbeta) - \nabla \cR_n(\bbeta^*) , \bbeta - \bbeta^* \rangle$ is the symmetric Bregman divergence. 
The main idea is to establish lower and upper bounds for $\cB(\hat{\bbeta}, \bbeta^*)$.

We start with obtaining a lower bound for $\cB(\hat{\bbeta}, \bbeta^*)$.   
Let $r_{\rm loc} = \gamma / (8\sqrt{2}A_1^2)$ and define an intermediate quantity $\hat{\bbeta}_\eta = \eta \hat{\bbeta} + (1-\eta)\bbeta^* $, where $\eta = \sup \{ \eta \in [0,1]: \hat{\bbeta}_\eta \in \bbeta^* + \BB_{\bSigma}(r_{\rm loc}) \}$. Then $\hat{\bbeta}_\eta \in \bbeta^* + \partial\BB_{\bSigma}(r_{\rm loc}) $ whenever $\hat{\bbeta} \notin \bbeta^* + \BB_{\bSigma}(r_{\rm loc})$, where $\partial\BB_{\bSigma}(r_{\rm loc})$ is the boundary of $\BB_{\bSigma}(r_{\rm loc})$. On the other hand, $\hat{\bbeta}_\eta = \hat{\bbeta}$ whenever $\hat{\bbeta} \in \bbeta^* + \BB_{\bSigma}(r_{\rm loc})$. By an application of Lemma~\ref{ld lem:RSC}, provided that $\gamma \geq 4\sqrt{2}\sigma_\varepsilon$ and $n \gtrsim d+t$, we obtain
\begin{align}
\label{ld thm 1a}
 \cB(\hat{\bbeta}_\eta, \bbeta^*)\ge \frac{1}{2} \kappa_1 \underline{\tau} \|\hat{\bbeta}_\eta-\bbeta^* \|_{\bSigma}^2,
\end{align}
with probability at least $1 - e^{-t}$.

Next, we proceed to obtain an upper bound of
$\cB(\hat{\bbeta}, \bbeta^*)$.  By an application of 
 Lemma C.1  in \cite{SZF2020} and the first order condition $\nabla \cR_n(\hat{\bbeta}) = \bm{0}$, we have 
\begin{align}
\label{ld thm 1b}
\cB(\hat{\bbeta}_\eta, \bbeta^*) &\leq \eta \cB(\hat{\bbeta}, \bbeta^*)\\
&=
\eta \langle -\nabla \cR_n(\bbeta^*) , \hat{\bbeta} - \bbeta^* \rangle \nn \\
&\leq
||\bSigma^{-1/2} \nabla \cR_n(\bbeta^*) ||_2 \cdot \|\hat{\bbeta}_\eta-\bbeta^* \|_{\bSigma} \nn \\
&\leq
\bigg[ \big|\big| \bSigma^{-1/2} \nabla \cR(\bbeta^*) \big|\big|_2 + \big|\big|\bSigma^{-1/2} \big\{ \nabla \cR_n(\bbeta^*) - \nabla \cR(\bbeta^*) \big\} \big|\big|_2 \bigg] \cdot \|\hat{\bbeta}_\eta-\bbeta^* \|_{\bSigma}.
\end{align}
Combining the above upper and lower bounds in \eqref{ld thm 1a} and \eqref{ld thm 1b},  applying Lemma \ref{ld lem:B1B2}, and picking $\gamma = \sigma_\varepsilon \sqrt{n/(d+t)}$, we have 
\begin{align}
\label{ld thm 1c}
\|\hat{\bbeta}_\eta-\bbeta^* \|_{\bSigma} \leq C(\bar{\tau}/\underline{\tau})\kappa_1^{-1} \sigma_\varepsilon v_0 \sqrt{\frac{d+t}{n}},
\end{align}
 with probability at least $1 - 2e^{-t}$ as long as $n \gtrsim d+t$, where $C$ is an absolute constant.

Lastly, it can be checked that with our proper choice of $\gamma$ and $r_{\rm loc}$, we have $\|\hat{\bbeta}_\eta-\bbeta^* \|_{\bSigma} \lesssim \sigma_\varepsilon \sqrt{(d+t)/n} < \sigma_\varepsilon \sqrt{n/ (d+t)} \asymp r_{\rm loc}$. It immediately implies $\hat{\bbeta}_\eta \in \bbeta^* + \BB_{\bSigma}(r_{\rm loc})$ and $\hat{\bbeta}_\eta = \hat{\bbeta}$ by construction. Thus \eqref{ld thm 1c} also holds when replacing $\hat{\bbeta}_\eta$ by $\hat{\bbeta}$.

\end{proof}


\subsubsection{Proof of Theorem \ref{ld thm 2 bahadur representation}}
\begin{proof}

We consider the following vector-valued random process
$$
\bB(\bbeta) = \bSigma^{-1/2} \{ \nabla \cR_n(\bbeta) - \nabla \cR_n(\bbeta^*) \} - \frac{1}{n} \sum_{i=1}^n \bSigma^{-1/2} \EE w_\tau (\varepsilon_i) \bx_i \bx_i^{\T}  (\bbeta - \bbeta^*).
$$
By the first order condition $\nabla \cR_n(\hat{\bbeta}) = \bm{0}$, it can be shown that the nonasymptotic Bahadur representation in~\eqref{ld thm 2a} takes the form $\|\bB(\hat{\bbeta})\|_2$. By the triangle inequality, we have 
\[
\|\bB(\hat{\bbeta})\|_2 \le \sup_{\bbeta \in \bbeta^* + \BB_{\bSigma}(r)} \|\EE \{\bB(\bbeta)\} \|_2+ \sup_{\bbeta \in \bbeta^* + \BB_{\bSigma}(r)} \|\bB(\bbeta) - \EE \{\bB(\bbeta)\} \|_2
\]
for radius $r$ that satisfies $\hat{\bbeta} \in \bbeta^* + \BB_{\bSigma}(r)$ with high probability.
It suffices to obtain upper bounds for the two terms separately. 

We start with an upper bound on 
$\sup_{\bbeta \in \bbeta^* + \BB_{\bSigma}(r)} \|\EE \{\bB(\bbeta)\} \|_2 $. By the mean value theorem for vector-valued functions (Theorem 12 in Section 2 of \cite{Pugh2015}), we obtain
\begin{align*}
\EE \{\bB(\bbeta) \}
&=
\bSigma^{-1/2} \EE \int_0^1 \nabla^2 \cR_n(\bbeta^*_t)dt(\bbeta - \bbeta^*) - \frac{1}{n} \sum_{i=1}^n \bSigma^{-1/2} \EE w_\tau (\varepsilon_i) \bx_i \bx_i^{\T}  (\bbeta - \bbeta^*) \\
&=
\Big\langle \int_0^1 \Big\{ \bSigma^{-1/2} \EE \nabla^2 \cR_n(\bbeta^*_t) \bSigma^{-1/2} - \frac{1}{n} \sum_{i=1}^n \EE w_\tau (\varepsilon_i) \bz_i \bz_i^{\T} \Big\} dt , \bSigma^{1/2} (\bbeta - \bbeta^*)  \Big \rangle,
\end{align*}
where $\bbeta^*_t = (1-t)\bbeta^* + t\bbeta$ for $(0 \leq t \leq 1)$ and $\bz_i = \bSigma^{-1/2} \bx_i$. 
Let $\bdelta_t = \bSigma^{1/2}(\bbeta^*_t - \bbeta^*)$. 
Since $\bbeta \in \bbeta^* + \BB_{\bSigma}(r)$, we have $||\bdelta_t||_2 \leq r$ and $y_i - \bx_i^{\T} \bbeta^*_t = \varepsilon_i - \bdelta_t^{\T} \bz_i$. For all $\bu \in \SSS^{d-1}$, we obtain
\begin{align*}
&\bigg|\bu^{\T} \Big\{ \bSigma^{-1/2} \EE \nabla^2 \cR_n(\bbeta^*_t) \bSigma^{-1/2} - \frac{1}{n} \sum_{i=1}^n \EE w_\tau (\varepsilon_i) \bz_i \bz_i^{\T} \bigg\}  \bu  \bigg| \\
&=
\bigg| \frac{1}{n} \sum_{i=1}^n \bu^{\T} \bigg[ \EE w_\tau(\varepsilon_i - \bdelta_t^{\T} \bz_i)\Big\{1 - \mathbbm{1}(|\varepsilon_i - \bdelta_t^{\T} \bz_i|>\gamma)\Big\}\bz_i \bz_i^{\T} -  \EE w_\tau (\varepsilon_i) \bz_i \bz_i^{\T} \bigg]  \bu  \bigg| \\
&\leq
\bigg|  \frac{1}{n} \sum_{i=1}^n \EE \bigg[ (\bu^{\T} \bz_i)^2 \EE \Big \{ w_\tau(\varepsilon_i - \bdelta_t^{\T} \bz_i) - w_\tau(\varepsilon_i)|\bz_i \Big\} \bigg]  \bigg| 
+
\bigg|  \frac{1}{n} \sum_{i=1}^n \EE  (\bu^{\T} \bz_i)^2  w_\tau(\varepsilon_i - \bdelta_t^{\T} \bz_i)    \mathbbm{1}(|\varepsilon_i - \bdelta_t^{\T} \bz_i|>\gamma) \bigg| \\
&:= \Pi_1 + \Pi_2.
\end{align*}

For $\Pi_1$, let $f_{\varepsilon|\bx}$ be the conditional density function of $\varepsilon$ given $\bx$, and recall that it is upper bounded by $\bar{f}_{\varepsilon|\bx}$. Moreover, let $m_3>0$ be a constant that satisfies $\sup_{\bu \in \SSS^{d-1}} \EE  |\bu^{\T} \bSigma^{-1/2} \bx| ^3 \leq m_3$. 
We have $\EE \big \{ w_\tau(\varepsilon_i - \bdelta_t^{\T} \bz_i) - w_\tau(\varepsilon_i)|\bz_i \big\}
=
\int_{-\infty}^{\infty} \big \{ w_\tau(u - \bdelta_t^{\T} \bz_i) - w_\tau(u) \big \} f_{\varepsilon | \bx}(u)du 
\leq
(\bar{\tau} - \underline{\tau}) \bar{f}_{\varepsilon|\bx} |\bdelta_t^{\T} \bz_i|$. Consequently,
\begin{align}
\label{ld thm 2b}
\Pi_1 \leq \frac{1}{n} \sum_{i=1}^n (\bar{\tau} - \underline{\tau}) \bar{f}_{\varepsilon|\bx} \EE \Big \{ (\bu^{\T} \bz_i)^2 |\bdelta_t^{\T} \bz_i| \Big \} \leq  (\bar{\tau} - \underline{\tau}) \bar{f}_{\varepsilon|\bx} m_3 rt.
\end{align}

For $\Pi_2$, we first note that $\mathbbm{1}(|\varepsilon_i - \bdelta_t^{\T} \bz_i|>\gamma) \leq \mathbbm{1}(|\varepsilon_i |>\gamma/2) + \mathbbm{1}(| \bdelta_t^{\T} \bz_i|>\gamma/2)$.  By an application of the Markov's inequality, we obtain
\begin{align}
\label{ld thm 2c}
\Pi_2 &\leq \bigg|  \bar{\tau} \EE  (\bu^{\T} \bz)^2 \Big \{ \mathbbm{1}(|\varepsilon |>\gamma/2) + \mathbbm{1}(| \bdelta_t^{\T} \bz|>\gamma/2)\Big \} \bigg| \nn \\
&\leq
\bar{\tau} \bigg| \EE \bigg(\frac{|\varepsilon|}{\gamma/2} \bigg)^2 (\bu^{\T} \bz)^2  \bigg|
+ 
\bar{\tau} \bigg| \EE \frac{| \bdelta_t^{\T} \bz|}{\gamma/2}  (\bu^{\T} \bz)^2  \bigg| \nn \\
&\leq
\frac{4\bar{\tau}\sigma_\varepsilon^2}{\gamma^2}
+ 
\frac{2 \bar{\tau} m_3 r}{\gamma}.  
\end{align}

Combining \eqref{ld thm 2b} and \eqref{ld thm 2c}, we have 
\begin{align}
\label{ld thm 2e}
\sup_{\bbeta \in \bbeta^* + \BB_{\bSigma}(r)} \|\EE \{\bB(\bbeta)\} \|_2
\leq 
\delta(r)r :=
\bigg \{ (\bar{\tau} - \underline{\tau}) \bar{f}_{\varepsilon|\bx} m_3 rt 
+
\frac{4\bar{\tau}\sigma_\varepsilon^2}{\gamma^2} + \frac{2\bar{\tau} m_3 r}{\gamma} \bigg\}r.
\end{align}

Next, we obtain an upper bound for $  \sup_{\bbeta \in \bbeta^* + \BB_{\bSigma}(r)} \|\bB(\bbeta) - \EE \{\bB(\bbeta)\} \|_2$. 
With some abuse of notation, let $
\bar{\bB}(\bdelta) = \bB(\bbeta) - \EE \{\bB(\bbeta)\}$ where $\bdelta = \bSigma^{1/2}(\bbeta - \bbeta^*) \in \BB(r).
$
It can be checked that $\bar{\bB}(\bm{0}) = \bm{0}$, $\EE \{\bar{\bB}(\bdelta)\}=\bm{0}$, and
\begin{align*}
\nabla_{\bdelta} \bar{\bB}(\bdelta)
&=
\frac{1}{n}\sum_{i=1}^n \bigg[ w_\tau(\varepsilon_i - \bdelta^{\T} \bz_i)\ell_\gamma''(\varepsilon_i - \bdelta^{\T} \bz_i) \bz_i \bz_i^{\T} -\EE \Big\{ w_\tau(\varepsilon_i - \bdelta^{\T} \bz_i)\ell_\gamma''(\varepsilon_i - \bdelta^{\T} \bz_i) \bz_i \bz_i^{\T} \Big\} \bigg] \\
&:= \frac{1}{n}\sum_{i=1}^n \bA_i.
\end{align*}

For all $\bu, \bv \in \SSS^{d-1}$ and $\lambda \in \RR$, we see that $\EE \bu^{\T} \bA_i \bv = 0$, $|\bu^{\T} \bA_i \bv| \leq \bar{\tau}|\bu^{\T}\bz_i \bv^{\T}\bz_i| + \bar{\tau}\EE|\bu^{\T}\bz_i \bv^{\T}\bz_i|$ and $|\bu^{\T} \bA_i \bv|^2 \leq 2\bar{\tau}^2 \big(|\bu^{\T}\bz_i \bv^{\T}\bz_i|^2 + \EE^2|\bu^{\T}\bz_i \bv^{\T}\bz_i| \big)$. It then follows from the elementary inequality $|e^z -1-z|\leq z^2e^{|z|}/2$ and bound $\EE|\bu^{\T}\bz_i \bv^{\T}\bz_i| \leq \big\{ \EE(\bu^{\T}\bz_i)^2 \big\}^{1/2} \big\{ \EE(\bv^{\T}\bz_i)^2\big\}^{1/2} \leq 1 $ that
\begin{align}
\label{ld thm 2f}
\EE \exp \Big\{ \lambda \sqrt{n} \bu^{\T} \nabla_{\bdelta} \bar{\bB}(\bdelta) \bv \Big\}
&=
\prod_{i=1}^n \EE \exp \bigg\{\frac{\lambda}{\sqrt{n}} \bu^{\T} \bA_i \bv  \bigg\} \nn \\
&\leq
\prod_{i=1}^n \EE  \Bigg\{ 1 + \frac{\lambda}{\sqrt{n}} \bu^{\T} \bA_i \bv + \bigg( \frac{\lambda}{\sqrt{n}} \bu^{\T} \bA_i \bv \bigg)^2 e^{\big| \frac{\lambda}{\sqrt{n}} \bu^{\T} \bA_i \bv \big|}/2  \Bigg\}  \nn \\
&\leq
\prod_{i=1}^n \EE  \Bigg\{ 1 + \frac{\lambda^2 \bar{\tau}^2}{n} e^{\frac{|\lambda|\bar{\tau}}{\sqrt{n}}} \Big(  e^{ \frac{|\lambda|\bar{\tau}}{\sqrt{n}}  |\bu^{\T}\bz_i \bv^{\T}\bz_i| } + |\bu^{\T}\bz_i \bv^{\T}\bz_i|^2 e^{ \frac{|\lambda|\bar{\tau}}{\sqrt{n}}  |\bu^{\T}\bz_i \bv^{\T}\bz_i| }  \Big)  \Bigg\}.
\end{align}

Here we upper-bound the components appeared in the right-hand side of \eqref{ld thm 2f}. For all $t>0$, it follows from Cauchy-Schwarz inequality and the elementary inequality $ab\leq a^2/2 + b^2/2$ that
\begin{align*}
\EE |\bu^{\T}\bz_i \bv^{\T}\bz_i|^2 e^{ t |\bu^{\T}\bz_i \bv^{\T}\bz_i| } 
&\leq
\EE (\bu^{\T}\bz_i)^2 (\bv^{\T}\bz_i)^2 e^{ t(\bu^{\T}\bz_i)^2/2 + t(\bv^{\T}\bz_i)^2/2 } \\
&\leq
\Big\{  \EE (\bu^{\T}\bz_i)^4 e^{ t(\bu^{\T}\bz_i)^2} \Big\}^{1/2} \Big\{  \EE (\bv^{\T}\bz_i)^4 e^{ t(\bv^{\T}\bz_i)^2} \Big\}^{1/2}. 
\end{align*} 
Consequently $\EE |\bu^{\T}\bz_i \bv^{\T}\bz_i|^2 e^{ t |\bu^{\T}\bz_i \bv^{\T}\bz_i| } \leq \sup_{\bu \in \SSS^{d-1}} \EE (\bu^{\T} \bz_i)^4 e^{t (\bu^{\T} \bz_i)^2}$, and similarly $\EE e^{ t |\bu^{\T}\bz_i \bv^{\T}\bz_i| } \leq \sup_{\bu \in \SSS^{d-1}} \EE  e^{t (\bu^{\T} \bz_i)^2}$. To further upper-bound these supremums, let $\chi := (\bu^{\T}\bz)^2/(2v_1)^2$. Recall the sub-Gaussian condition $\PP(|\langle \bu , \bSigma^{-1/2}\bx \rangle| \geq v_1 ||\bu||_2 t) \leq 2e^{-t^2/2}$, we have $\PP(\chi \geq t) \leq 2e^{-2t}$ (i.e. $\chi$ is sub-Exponential). It follows that $\EE e^{\chi} = 1 + \int_0^{\infty} e^t \PP(\chi \geq t)dt \leq 1 + 2\int_0^{\infty} e^{-t} dt =3 $, and
\begin{align*}
\EE (\chi^2 e^\chi) = \int_0^{\infty} (t^2 + 2t)e^t \PP(\chi \geq t)dt \leq 2 \int_0^{\infty} (t^2 + 2t)e^{-t} dt = 8.
\end{align*}
Along with the monotonicity of exponential function, we conclude that both $  \EE |\bu^{\T}\bz_i \bv^{\T}\bz_i|^2 e^{ \frac{|\lambda|\bar{\tau}}{\sqrt{n}}  |\bu^{\T}\bz_i \bv^{\T}\bz_i| }$ and $ \EE e^{ \frac{|\lambda|\bar{\tau}}{\sqrt{n}}  |\bu^{\T}\bz_i \bv^{\T}\bz_i| }$ can be upper-bounded by some constants $C_1, C_2$ respectively, uniformly over $\bu, \bv \in \SSS^{d-1}$ , as long as $|\lambda| \leq \sqrt{n}/(4v_1^2 \bar{\tau})$. Substituting the above bounds into \eqref{ld thm 2f} yields,
\begin{align*}
\EE \exp \Big\{ \lambda \sqrt{n} \bu^{\T} \nabla_{\bdelta} \bar{\bB}(\bdelta) \bv \Big\}
&\leq
\prod_{i=1}^n   \Bigg[ 1 + \frac{\lambda^2 \bar{\tau}^2}{n} e^{\frac{|\lambda|\bar{\tau}}{\sqrt{n}}} \bigg\{ \sup_{\bu \in \SSS^{d-1}} \EE e^{\frac{|\lambda|\bar{\tau}}{\sqrt{n}} (\bu^{\T} \bz_i)^2}   + \sup_{\bu \in \SSS^{d-1}} \EE (\bu^{\T} \bz_i)^4 e^{\frac{|\lambda|\bar{\tau}}{\sqrt{n}} (\bu^{\T} \bz_i)^2} \bigg \}  \Bigg] \\
&\leq
\exp  \Big\{ {\lambda^2 \bar{\tau}^2} e^{\frac{|\lambda|\bar{\tau}}{\sqrt{n}}} (C_1 + C_2) \Big\} \\
&\leq
\exp \bigg\{ 2(C_1+C_2)\bar{\tau}^2 e^{-4 v_1^2} \cdot \frac{\lambda^2}{2}  \bigg\} \mbox{~~~valid for all~~~} \lambda^2 \leq 2\cdot \frac{n}{32 \bar{\tau}^2  v_1^4}.
\end{align*}

With the above preparations, we apply Theorem A.3 in \cite{Spokoiny2013} with $v_0^2 = 2(C_1 + C_2)\bar{\tau}^2e^{-4v_1^2}$ and $g^2 = n/(32\bar{\tau}^2 v_1^4) $ to yield
\begin{align}
\label{ld thm 2g}
\sup_{\bbeta \in \bbeta^* + \BB_{\bSigma}(r)} ||\bB(\bbeta) - \EE \bB(\bbeta) ||_2 
\leq
12\sqrt{C_1 + C_2} \bar{\tau} e^{-2v_1^2}\sqrt{\frac{2d+t}{n}} \cdot r
\end{align}
with probability at least $1 - e^{-t}$, as long as $n \geq 64\bar{\tau}^2 v_1^4 (2d+t)$.

Lastly, combining \eqref{ld thm 2e} and \eqref{ld thm 2g}, we have
\begin{align}
\label{ld thm 2h}
\sup_{\bbeta \in \bbeta^* + \BB_{\bSigma}(r)} || \bB(\bbeta) ||_2 
\leq \Bigg\{ \delta(r) + 12\sqrt{C_1 + C_2} \bar{\tau} e^{-2v_1^2}\sqrt{\frac{2d+t}{n}} \Bigg\} r
\end{align}
with probability at least $1 - e^{-t}$, as long as $n \geq 64\bar{\tau}^2 v_1^4 (2d+t)$. Recall from the proof of Theorem \ref{ld thm 1 l2errorbound} that we have $\hat{\bbeta} \in \bbeta^* + \BB_{\bSigma}(r_0)$ with probability at least $1 - 2e^{-t}$ for some $r_0 \asymp \sigma_\varepsilon \sqrt{(d+t)/n}$. Taking $r = r_0$ in \eqref{ld thm 2h} and $\gamma = \sigma_\varepsilon \sqrt{n/(d+t)}$ finishes the proof.

\end{proof}


\subsubsection{Proof of Theorem \ref{ld thm 3 asym normality}}
\begin{proof}
Let $\bu \in \RR^d$ be an arbitrary vector, and ${\mathbf{J}} = \EE \{w_\tau(\varepsilon) \bx \bx^{\T}\}$ be the Hessian matrix. Define $S_n = n^{-1/2} \sum_{i=1}^n a_i b_i$ and its centered version $S_n^0 = S_n - \EE (S_n)$, where $a_i = w_\tau(\varepsilon_i)\ell_\gamma'(\varepsilon_i)$ and $b_i = \langle  {\mathbf{J}}^{-1} \bu, \bx_i  \rangle$. 
We first show that the centered partial sum $S_n^0$ is close to the quantity of interest $n^{1/2} \langle \bu , \hat{\bbeta} - \bbeta^* \rangle$. 
By an application of Theorem \ref{ld thm 2 bahadur representation} and Lemma \ref{ld lem:huber} with $\gamma = \sigma_\varepsilon \sqrt{n/(d+t)}$ and $t = \log n$, we obtain
\begin{align}
\label{ld thm 3a}
&\big| n^{1/2} \langle \bu , \hat{\bbeta} - \bbeta^* \rangle - S_n^0 \big| \nn \\
&\leq
n^{1/2} \Big| \Big\langle \bSigma^{1/2} {\mathbf{J}}^{-1} \bu , \bSigma^{-1/2}{\mathbf{J}}(\hat{\bbeta} - \bbeta^*)  - \frac{1}{n} \sum_{i=1}^n w_\tau(\varepsilon_i) \ell_\gamma'(\varepsilon_i) \bSigma^{-1/2}\bx_i   \Big\rangle \Big | 
+ \big| \EE S_n \big|   \nn \\
&\leq
n^{1/2} \big|\big| {\mathbf{J}}^{-1} \bu \big|\big|_{\bSigma} \cdot \big|\big| \bSigma^{-1/2}{\mathbf{J}} (\hat{\bbeta} - \bbeta^*) -  \frac{1}{n} \sum_{i=1}^n  w_\tau(\varepsilon_i) \ell_\gamma'(\varepsilon_i) \bSigma^{-1/2}\bx_i \big|\big|_2   + n^{1/2} \Big | \EE \Big[ \langle {\mathbf{J}}^{-1} \bu , \bx \rangle \EE \Big\{ w_\tau(\varepsilon)\ell_\gamma'(\varepsilon)|\bx \Big\}  \Big] \Big|  \nn \\
&\leq
n^{1/2} \big|\big| {\mathbf{J}}^{-1} \bu \big|\big|_{\bSigma} C \cdot \frac{d + \log n}{n}
+ n^{1/2} \frac{\bar{\tau} v_3}{\gamma^2} \Big( \EE \langle {\mathbf{J}}^{-1} \bu , \bx \rangle^2 \Big)^{1/2}
\nn \\
&\leq
C_1 \big|\big| {\mathbf{J}}^{-1} \bu \big|\big|_{\bSigma} \frac{d + \log n}{\sqrt{n}},
\end{align}
with probability at least $1 - 3n^{-1}$, where $C_1 = C + {\bar{\tau} v_3}/{\sigma_\varepsilon^2}$.

Next, we show that the centered partial sum $S_n^0 = n^{-1/2} \sum_{i=1}^n (1 - \EE)a_i b_i$ is approximately normally distributed. It follows from Berry-Esseen inequality (e.g., see \cite{Tyurin2011}) that
\begin{align}
\label{ld thm 3b}
\sup_{x \in \RR} ~\big| \PP(S_n^0 \leq \var(S_n^0)^{1/2} x) - \Phi(x) \big| \leq \frac{\EE |a_i b_i - \EE a_i b_i |^3}{2 \var(S_n^0)^{3/2} \sqrt{n}}.
\end{align}
Thus, it suffices to obtain a lower bound for $\var(S_n^0)$ and an upper bound for $\EE ( |a_i b_i - \EE a_i b_i |^3)$.
By an application of Lemma~\ref{ld lem:huber}, we have 
$\EE (a_i b_i) \leq \bar{\tau}v_3 || {\mathbf{J}}^{-1} \bu ||_{\bSigma}/\gamma^2$ and $\EE(a_i b_i)^2 \geq \underline{\tau}^2 (\sigma_\varepsilon^2 - 2v_3/\gamma) \| {\mathbf{J}}^{-1} \bu \|_{\bSigma}^2 $.
Thus, $\var(S_n^0) = \EE(a_i b_i)^2 - (\EE a_i b_i)^2 \geq \| {\mathbf{J}}^{-1} \bu \|_{\bSigma}^2 ( \underline{\tau}^2 \sigma_\varepsilon^2 - 2\underline{\tau}^2 v_3/\gamma - \bar{\tau}^2v_3^2/\gamma^4 )$. 
For sufficiently large $\gamma$ (i.e.,  $n \gtrsim d$), we obtain the lower bound $\var(S_n^0)^{3/2} \geq || {\mathbf{J}}^{-1} \bu ||_{\bSigma}^3(\underline{\tau}^3 \sigma_\varepsilon^3/2)$. 
Next, we proceed to obtain an upper bound for the centered third moment $\EE |a_i b_i - \EE (a_i b_i) |^3$. Recall that $m_3 = \sup_{\bu \in \SSS^{d-1}}\EE |\langle \bu , \bSigma^{-1/2}\bx \rangle|^3$, we have $\EE  |a_i b_i|^3 \leq \EE \big [ |\langle  {\mathbf{J}}^{-1} \bu, \bx_i  \rangle |^3 \EE \big \{ |w_\tau(\varepsilon_i) \ell_\gamma'(\varepsilon_i)|^3 | \bx_i \big \} \big ] 
\leq \bar{\tau}^3 v_3 m_3 || {\mathbf{J}}^{-1} \bu ||_{\bSigma}^3$. Along with Minkowski's inequality $|a+b|^p \leq 2^{p-1} (|a|^p + |b|^p)$, we obtain $\EE |a_i b_i - \EE a_i b_i |^3 \leq 4 \bar{\tau}^3 v_3 m_3  (1 + {v_3^2}/{m_3 \gamma^6} )|| {\mathbf{J}}^{-1} \bu ||_{\bSigma}^3$. Therefore $\EE |a_i b_i - \EE a_i b_i |^3 \leq 8 \bar{\tau}^3 v_3 m_3 || {\mathbf{J}}^{-1} \bu ||_{\bSigma}^3$ provided that $n \gtrsim d$. 
Substituting the above inequalities into \eqref{ld thm 3b}, we have
\begin{align}
\label{ld thm 3c}
\sup_{x \in \RR}~ \big| \PP(S_n^0 \leq \var(S_n^0)^{1/2} x) - \Phi(x) \big| \leq C_2 n^{-1/2},
\end{align}
where $C_2 = 8v_3 m_3 \bar{\tau}^3/(\underline{\tau}\sigma_\varepsilon)^3 $.

Let  $\sigma^2 = \EE (a_i b_i)^2 = \bu^{\T} {\mathbf{J}}^{-1} \EE \big[\big\{w_\tau(\varepsilon)\ell_\gamma'(\varepsilon) \big\}^2 \bx \bx^{\T} \big] {\mathbf{J}}^{-1} \bu$. 
By an application of Lemma~\ref{ld lem:huber}, we have $ \underline{\tau}^2 (\sigma_\varepsilon^2 - 2v_3/\gamma) \| {\mathbf{J}}^{-1} \bu \|_{\bSigma}^2 \leq \sigma^2  \leq \bar{\tau}^2 \sigma_\varepsilon^2 || {\mathbf{J}}^{-1} \bu ||_{\bSigma}^2$. 
Moreover,  $| \var(S_n^0) - \sigma^2| = |\EE a_i b_i|^2 \leq \bar{\tau}^2 v_3^2 || {\mathbf{J}}^{-1} \bu ||_{\bSigma}^2/\gamma^4$.
Provided that  $n \gtrsim d$, we obtain
\begin{align*}
\bigg| \frac{\var(S_n^0)}{\sigma^2} - 1 \bigg| \leq 
\bigg(1 - \frac{2v_3}{\sigma_\varepsilon^2 \gamma} \bigg)^{-1} \cdot
\frac{\bar{\tau}^2 v_3^2 }{\underline{\tau}^2 \sigma_\varepsilon^2} \cdot \frac{1}{\gamma^4}
\leq \frac{2\bar{\tau}^2 v_3^2 }{\underline{\tau}^2 \sigma_\varepsilon^2} \cdot \frac{1}{\gamma^4}.
\end{align*}
An application of Lemma A.7 in the supplement of \cite{SZ2015} indicates that
\begin{align}
\label{ld thm 3d}
\sup_{x \in \RR} \big| \Phi(x/\var(S_n^0)^{1/2}) - \Phi(x/\sigma) \big| \leq C_3 \gamma^{-4},
\end{align}
where $C_3 = (\bar{\tau} v_3 / \underline{\tau} \sigma_\varepsilon)^2$.

Let $G \sim \cN(0,1)$. Applying the inequalities in \eqref{ld thm 3a}, \eqref{ld thm 3c}, and \eqref{ld thm 3d}, along with the fact that for all $a<b$ and $\sigma > 0$, $\Phi(b/\sigma) - \Phi(a/\sigma) \leq (2\pi)^{-1/2}(b-a)/\sigma$, we obtain that for any $x \in \RR$ and $\bu \in \RR^d$,
\begin{align*}
\PP(n^{1/2} \langle \bu , \hat{\bbeta} - \bbeta^* \rangle \leq x) 
&\leq
\PP \bigg( S_n^0 \leq x + C_1\big|\big| {\mathbf{J}}^{-1} \bu \big|\big|_{\bSigma} \frac{d + \log n}{\sqrt{n}} \bigg) + \frac{3}{n} \\
&\leq
\PP \bigg( \var(S_n^0)^{1/2} G \leq x + C_1\big|\big| {\mathbf{J}}^{-1} \bu \big|\big|_{\bSigma} \frac{d + \log n}{\sqrt{n}} \bigg) + \frac{3}{n} + \frac{C_2}{\sqrt{n}} \\
&\leq
\PP \bigg( \sigma G \leq x + C_1\big|\big| {\mathbf{J}}^{-1} \bu \big|\big|_{\bSigma} \frac{d + \log n}{\sqrt{n}} \bigg) + \frac{3}{n} + \frac{C_2}{\sqrt{n}} + \frac{C_3}{\gamma^4} \\
&\leq
\PP \bigg( \sigma G \leq x  \bigg) +\frac{C_1 ||{\mathbf{J}}^{-1} \bu ||_{\bSigma}}{\sqrt{2\pi} \sigma}	 \frac{d + \log n}{\sqrt{n}} +\frac{3}{n} + \frac{C_2}{\sqrt{n}} + \frac{C_3}{\gamma^4} \\
&\lesssim
\PP \bigg( \sigma G \leq x  \bigg) + \frac{d + \log n}{\sqrt{n}} + \frac{1}{n} + \frac{1}{\sqrt{n}} + \frac{(d + \log n)^2}{n^2},
\end{align*}
where the last inequality follows from the fact that $\sigma \asymp || {\mathbf{J}}^{-1} \bu ||_{\bSigma}$ and taking $\gamma = \sigma_\varepsilon \sqrt{n/(d+\log n)}$. A similar argument leads to a series of reverse inequalities. Since the above bounds are independent of $x$ and $\bu$, they hold uniformly over $x \in \RR$ and $\bu \in \RR^d$.

Putting together all the pieces, we conclude that by taking $\gamma = \sigma_\varepsilon \sqrt{n/(d+\log n)}$, we have
\begin{align*}
\sup_{\bu \in \RR^d, x \in \RR} \big| \PP(n^{1/2} \langle \bu , \hat{\bbeta} - \bbeta^* \rangle \leq \sigma x) - \Phi(x) \big| \lesssim \frac{d + \log n}{\sqrt{n}},
\end{align*}
 as long as $n \gtrsim d$.

\end{proof}

\subsubsection{Proof of Theorem \ref{thm:l1.retire}}
\begin{proof}
Let $\hat{\bbeta} :=\hat{\bbeta}^{(1)} $ be a minimizer of~\eqref{retire.est.convex} with $p'_\lambda(0) = \lambda$, i.e., optimization problem~\eqref{retire.est.convex} reduces to the $\ell_1$-penalized robustified expectile regression, i.e., 
 \#
\hat{\bbeta} \in \underset{\bbeta\in\RR^d}{\mathrm{minimize}}~\left\{\frac{1}{n} \sum_{i=1}^n L_{\tau,\gamma}(y_i-\bx_i^{\T}\bbeta)+ \lambda \sum_{j=2}^d  |\beta_j| \right\}.
\label{retire.est.convex.l1}
 \# 
    Let $\cS =\{1,\ldots,d\}$ be the active set of $\bbeta^*$, i.e., the index set $\cS$ contains indices for which $\bbeta^*_j\ne 0$.
Let $s=|\cS|$ be the cardinality of $\cS$. Recall the definition of the symmetric Bregman divergence in~\eqref{eq:symmetricdivergence}.  The main crux of the proof of Theorem~\ref{thm:l1.retire} involves establishing upper and lower bounds for 	$\cB(\hat{\bbeta}, \bbeta^* )$. We start with deriving an upper bound for $\cB(\hat{\bbeta}, \bbeta^* )$.  Throughout the proof, we write $\hat{\bdelta} = \hat{\bbeta}-\bbeta^*$.

Since $\hat{\bbeta}$ is a minimizer of~\eqref{retire.est.convex.l1}, we have 
\begin{align}
\cR_n(\hat{\bbeta})-\cR_n(\bbeta^*) 
&\leq 
\lambda(\|\bbeta^* \|_1-\|\hat{\bbeta} \|_1) \label{thm_claim_1}  \\
&\leq 
\lambda(\|\bbeta_\cS^* \|_1-\|\bbeta_\cS^* + \hat{\bdelta}_{\cS}\|_1 - \|\hat{\bdelta}_{\cS^{\cc}} \|_1) \nn \\
&\leq 
\lambda( \|\hat{\bdelta}_\cS \|_1 -  \|\hat{\bdelta}_{\cS^{\cc}} \|_1). \label{thm_claim_2}
\end{align}
By the optimality condition of $\hat{\bbeta}$, we have $\langle \nabla \cR_n(\hat{\bbeta})+\lambda \hat{\bz}, \hat{\bbeta}-\bbeta^*\rangle \leq 0$, where $\hat{\bz} \in \partial \|\hat{\bbeta} \|_1$ and $\langle \hat{\bz} ,\hat{\bbeta} \rangle = \| \hat{\bbeta} \|_1$.
Thus, conditioned on the event $\cE_{\rm{score}} := \{\lambda \geq 2\|\nabla \cR_n(\bbeta^*) \|_\infty \}$, $\cB(\hat{\bbeta}, \bbeta^* )$ can be upper bounded by 
\begin{align}
\cB(\hat{\bbeta}, \bbeta^* ) &= \langle \nabla  \cR_n(\hat{\bbeta}) -  \nabla \cR_n (\bbeta^*) , \hat{\bbeta} - \bbeta^* \rangle \\
&= 
\langle \nabla  \cR_n(\hat{\bbeta}) + \lambda \hat{\bz} , \hat{\bbeta} - \bbeta^* \rangle +\langle -\lambda \hat{\bz} -  \nabla \cR_n (\bbeta^*) , \hat{\bbeta} - \bbeta^* \rangle \nn \\
&\leq 
0 + \lambda(\|\bbeta^* \|_1-\|\hat{\bbeta} \|_1) + \|\nabla \cR_n(\bbeta^*) \|_\infty \cdot \|\hat{\bdelta} \|_1 \nn \\
&\leq 
\lambda(\|\hat{\bdelta}_\cS \|_1 -  \|\hat{\bdelta}_{\cS^{\cc}} \|_1) + \frac{\lambda}{2}(\|\hat{\bdelta}_\cS \|_1 +  \|\hat{\bdelta}_{\cS^{\cc}} \|_1)  \nn \\
&\leq
\frac{3}{2} \lambda s^{1/2} \| \hat{\bdelta} \|_2.
 \label{thm_part_a_3}
\end{align}

We now obtain a lower bound for $\cB(\hat{\bbeta}, \bbeta^* )$. To this end, we apply the restricted strong convexity result in Lemma~\ref{lem:RSC}. First, from the proof of Lemma~\ref{lem:RSC}, we know that the result in Lemma~\ref{lem:RSC} is applicable for any $\bbeta \in \bbeta^*+\BB(r)\cap \CC_1$, for which $\hat{\bbeta}$ does not necessarily satisfies.
To this end, we define an intermediate quantity to help facilitate the proof.  
Let $A_1$ be a constant that satisfies $\EE  (\bu^{\T} \bSigma^{-1/2}\bx) ^4 \leq A_1^4 \|\bu\|^4_2$ for all $\bu \in \RR^d$ and let $r_{\rm{loc}} = \gamma /(8\sqrt{2} \lambda_u A_1^2 ) $.
Consider $\hat{\bbeta}_\eta = \eta \hat{\bbeta} +(1-\eta)\bbeta^*$, where $\eta = \sup \big\{ u \in [0,1] : (1-u)\bbeta^* + u \hat{\bbeta} \in \bbeta^* +  \BB(r_{\rm{loc}}) \big\}$. 
Then, $\hat{\bbeta}_\eta \in \bbeta^* + \partial \BB(r_{\rm{loc}})$ whenever $\hat{\bbeta} \notin \bbeta^* + \BB(r_{\rm{loc}})$ where $\partial \BB(r_{\rm{loc}})$ is the boundary of $\BB(r_{\rm{loc}})$, and $\hat{\bbeta}_\eta = \hat{\bbeta}$ whenever $\hat{\bbeta} \in \bbeta^* + \BB(r_{\rm{loc}})$. 
Let $\hat{\bdelta}_{\eta}=\hat{\bbeta}_{\eta}-\bbeta^*$.  It remains to show that $\hat{\bbeta}_{\eta} \in \bbeta^* + \CC_1$, i.e., $\|(\hat{\bdelta}_{\eta})_{S^c}\|_1 \le 3 \|(\hat{\bdelta}_{\eta})_{S}\|_1$.
By convexity of $\cR_n(\cdot)$, we have
\begin{align}
\cR_n(\hat{\bbeta})-\cR_n(\bbeta^*) \geq \langle \nabla \cR_n(\bbeta^*), \hat{\bdelta} \rangle \geq -\|\nabla \cR_n(\bbeta^*) \|_\infty \cdot \|\hat{\bdelta} \|_1 \geq -\frac{\lambda}{2}(\|\hat{\bdelta}_\cS \|_1 +  \|\hat{\bdelta}_{\cS^{\cc}} \|_1) \label{thm_claim_3}
\end{align}
Combining \eqref{thm_claim_2} and \eqref{thm_claim_3}, we have $ \|\hat{\bdelta}_{\cS^{\cc}} \|_1 \leq 3\|\hat{\bdelta}_\cS \|_1$, conditioned on the event $\cE_{\rm{score}}$.  
Since $\eta \hat{\bdelta} = \hat{\bdelta}_\eta$, we have verified that $\hat{\bbeta}_\eta \in \bbeta^* + \BB(r_{\rm{loc}}) \cap \CC_1$, conditioned on $\cE_{\rm{score}}$.
 Applying Lemma \ref{lem:RSC} with $\bbeta = \hat{\bbeta}_\eta$, the following bound holds with probability at least $1-e^{-t}$
\begin{align} \label{thm_part_a_1}
\cB(\hat{\bbeta}_{\eta}, \bbeta^* ) = \langle \nabla  \cR_n(\hat{\bbeta}_\eta) -  \nabla \cR_n (\bbeta^*) , \hat{\bbeta}_\eta - \bbeta^* \rangle  \geq \frac{1}{2} \kappa_1 \underbar{$\tau$}   \|\hat{\bdelta}_\eta \|^2_2,
\end{align}
as long as $(\gamma, n, d)$ satisfies $\gamma \geq 4\sqrt{2}\lambda_u  \sigma_{\varepsilon}$ and $n \gtrsim s\log d + t$. 
For notational convenience, we denote the event at \eqref{thm_part_a_1} as $\cE_{\mathrm{rsc}}$ with $\PP(\cE_{\mathrm{rsc}}) \ge 1-e^{-t}$.

We now combine the lower and upper bounds in \eqref{thm_part_a_3} and \eqref{thm_part_a_1}.
Since $\cR_n(\cdot)$ is convex, by Lemma C.1 in \cite{SZF2020} we have
$\cB(\hat{\bbeta}_{\eta}, \bbeta^* ) \le \eta \cB(\hat{\bbeta}, \bbeta^* ),$
and thus implies that 
\[
\begin{cases}
\| \hat{\bdelta}_\eta \|_2 \leq 3(\kappa_1 \underbar{$\tau$} )^{-1} s^{1/2} \lambda;\\
\| \hat{\bdelta}_\eta \|_1  
\leq  4s^{1/2} \| \hat{\bdelta}_\eta \|_2
\leq 12(\kappa_1 \underbar{$\tau$} )^{-1} s \lambda.
\end{cases}
\]
We now show that with proper choice of $\lambda$ and $\gamma$, $\hat{\bbeta} \in \bbeta^* + \BB(r_{\rm{loc}})$, implying $\hat{\bbeta}_\eta = \hat{\bbeta}$. 
Let $\gamma = \sigma_{\varepsilon}\sqrt{{n}/{(\log d + t)}}$.  By Lemma \ref{lem:grad},  
\[
\|  \nabla \cR_n (\bbeta ^*) \|_\infty \leq \bar{\tau} \sigma_{\bx} \sigma_{\varepsilon }(3 \nu_0 + 1) \sqrt{{(\log d + t)}/{n}}
\]
 with probability at least $ 1-2e^{-t}$, suggesting that  $\lambda = 2c \bar{\tau} \sqrt{{(\log d + t)}/{n}}$ where $c=\sigma_{\bx} \sigma_{\varepsilon }(3 \nu_0 + 1)$.
Moreover, it can be verified that $\gamma \geq 4\sqrt{2}\lambda_u \sigma_{\varepsilon}$ under the scaling condition $n \gtrsim s \log d + t$.
Finally, we have 
 $\| \hat{\bbeta}_\eta - \bbeta^*\|_2 \leq 3(\kappa_1 \underbar{$\tau$} )^{-1} s^{1/2} \lambda \lesssim \sqrt{s(\log d + t)/n} < \sqrt{n/ (\log d + t )} \asymp r_{\rm{loc}}  $, i.e., $\hat{\bbeta}_\eta \in \bbeta^* +  \BB(r_{\rm{loc}})$. 
 This further implies that $\hat{\bbeta}=\hat{\bbeta}_\eta$ by construction. 
Thus, we obtain the desired results
$$
\begin{cases}
\| \hat{\bbeta} - \bbeta^* \|_2 \leq 3(\kappa_1 \underbar{$\tau$} )^{-1} s^{1/2} \lambda;\\
\| \hat{\bbeta} - \bbeta^* \|_1\leq 12(\kappa_1 \underbar{$\tau$} )^{-1} s \lambda,
\end{cases}
$$
 with probability $\PP \big( \cE_{\rm{score}} \cap \cE_{\rm{rsc}} \big) \geq 1-3e^{-t} $.

\end{proof}


\subsubsection{Proof of Theorem \ref{thm:random}}
Recall that  $\cR (\bbeta^*) = \EE \{\cR_n(\bbeta^*)\}$  be the population loss evaluated at $\bbeta^*$ and let $\bw^* = \nabla \cR_n(\bbeta^*) - \nabla \cR(\bbeta^*)$
be the centered score function.
We first show that given an estimator at the $(T-1)th$ iteration, $\hat{\bbeta}^{(T-1)}$, the estimation error of the subsequent estimator $\hat{\bbeta}^{(T)}$ can be improved sequentially by a $\delta$-fraction for some constant $\delta\in (0,1)$, under a beta-min condition on $\|\bbeta_{\cS}^*\|_{\min}$.
We establish a deterministic claim in the following proposition, where we conditioned on events that are related to the local restricted strong convexity property and the gradient of the loss function, $\cE_{\rm{rsc}}(r,L,\kappa)$ and 
$\{p_0'(a_0) \lambda \geq2 \bw^* \}$, respectively.

\begin{proposition}
 \label{prop:further steps}
 Let $p_0(\cdot)$ be a penalty function that satisfies Condition~\ref{def:concavepenalty}.
 Given $\kappa>0$, assume that there exists some constant $a_0 > 0$ such that $p_0'(a_0) >0$ and $\kappa > {\sqrt{5}}/{(2 a_0)}$ . Let $c>0$ be a constant that is the solution to the equation 
\begin{align} \label{choice of c}
0.5 p'_0(a_0)(c^2 + 1)^{1/2} + 2 = c \kappa a_0.
\end{align} 
Assume the beta-min condition $\| \bbeta^*_{\cS} \|_{\min} \geq a_0 \lambda $ and let  $ r^{\rm{crude}}=ca_0 s^{1/2} \lambda $. Conditioned on the event $\cE_{\rm{rsc}}(r,L,\kappa) \cap \{p_0'(a_0) \lambda \geq 2\|\bw^* \|_\infty \}$ with 
\[
L =  \{ 2 +2/p_0'(a_0) \}  (c^2 + 1)^{1/2} s^{1/2} + 2 s^{1/2}/p_{{0}}'(a_0),~r > r^{\rm{crude}}, ~\mathrm{and}~ \lambda \geq s^{-1/2}  \| \nabla \cR (\bbeta^* )\|_2,
\]
the sequence of solutions $\hat{ \bbeta}^{(1)},\ldots,\hat{ \bbeta}^{(T)}$ obtained from solving~\eqref{retire.est.convex} satisfies
\begin{align} \label{contraction.inequality}
  \| \hat{ \bbeta}^{(T)} - \bbeta^* \|_2 & \leq \delta  \| \hat{ \bbeta}^{(T-1)} - \bbeta^* \|_2  +  \kappa^{-1} \left\{ \|  p_\lambda'\{(|\bbeta^*_j|-a_0 \lambda)_+\} \|_2 +  \| \bw^* _{ \cS } \|_2 + \| \nabla \cR (\bbeta^* )\|_2 \right\},
\end{align}
where $\delta = \sqrt{5}/(2a_0 \kappa) \in (0,1)$ and $z_+ = \max(z,0)$. Furthermore, we have
\begin{align}\label{contraction.inequality2}
  \| \hat{ \bbeta}^{(T)} - \bbeta^* \|_2 \leq     {  \delta ^{T-1}   }  r^{\rm{crude}} +  \{(1- \delta  ) \kappa\}^{-1} \left\{\|  p_\lambda'\{(|\bbeta^*_j|-a_0 \lambda)_+\}  \|_2 +  \|\bw^*_{\cS } \|_2 +  \| \nabla \cR (\bbeta^* )\|_2  \right\} .
\end{align}

\end{proposition}

Proposition~\ref{prop:further steps} establishes the fact that every additional iteration of the proposed iteratively reweighted method shrinks the estimation error of the solution obtained from the previous iteration by a factor of $\delta \in (0,1)$, at the cost of inducing some extra terms  $\| p_\lambda'\{(|\bbeta^*_j|-a_0 \lambda)_+ \|_2$,   $\| \bw^* _{ \cS } \|_2$, and $\| \nabla \cR (\bbeta^* )\|_2 $, which can be shown to be smaller than $r^{\mathrm{crude}}$.
Such a phenomenon is also known as the contraction property and has been studied in different contexts \citep{FLSZ2018,PSZ2021}.  We refer the reader to \citet{PSZ2021} for a detailed discussion on the various terms that appear in~\eqref{contraction.inequality2}.
For completeness, we also provide the proof of Proposition~\ref{prop:further steps} in Section~\ref{proof:prop:further steps}.  

The results in Proposition~\ref{prop:further steps} are deterministic, conditioned on some events. 
In the following proof of Theorem~\ref{thm:random}, we provide an appropriate choice of the set of tuning parameters $(\lambda,\gamma)$ such that the event $\cE_{\rm{rsc}}(r,L,\kappa)  \cap \{p_0'(a_0) \lambda \geq 2\|\bw^* \|_\infty \}$ holds with high probability.
Moreover, we will control the shrinkage bias $\|  p_\lambda'\{(|\bbeta^*_j|-a_0 \lambda)_+ \|_2$ in~\eqref{contraction.inequality2} by proposing slightly stronger conditions on the minimum signal strength $\| \bbeta^*_{\cS} \|_{\min }$ as well as the first derivative of the penalty function $p_{\lambda}(\cdot)$. 


\begin{proof}
The proof is based on Proposition~\ref{prop:further steps}.
We will show that under the stated conditions in Theorem \ref{thm:random}, the events $\cE_{\rm{rsc}}(r,L,\kappa)$ and $ \{p_0'(a_0) \lambda \geq 2\|\bw^* \|_\infty \}$ in Proposition~\ref{prop:further steps} hold with high probabilities. 
We then show that the terms $p_\lambda'\{(|\bbeta^*_j|-a_0 \lambda)_+\} \|_2$, $\| \bw^* _{ \cS } \|_2$, and $\| \nabla \cR (\bbeta^* )\|_2$ can be upper bounded with high probabilities.

Picking  $\gamma = \sigma_{\varepsilon} \sqrt{{n}/(s + \log d + t )}$ and applying Lemma~\ref{lem:grad} indicates that 
\[
 \| \nabla \cR (\bbeta ^*) \|_2 \leq \bar{\tau} \sigma_{\varepsilon} \lambda_u^{1/2} \sqrt{(s + \log d +t)/n}.
\]
and 
\[
\|\nabla \cR_n(\bbeta^*)-\nabla \cR(\bbeta^*) \|_\infty\le  3 \nu_0 \bar{\tau} \sigma_{\bx} \sigma_{\varepsilon}  \sqrt{{(\log d + t) }/{n}},
\]
with probability at least $1 - 2e^{-t}$. 
Picking $\lambda \asymp \sigma_{\varepsilon}\sqrt{(\log d + t)/n}$, we have $\lambda \geq s^{-1/2}  \| \nabla \cR (\bbeta ^*) \|_2$ and the event $ \{p_0'(a_0) \lambda \geq 2\|\bw^* \|_\infty \}$ holds with probability at least $1 - 2e^{-t}$.

Next, we set $\kappa = 0.5 \kappa_1 \underbar{$\tau$}$ and set the constant $c$ to be the solution to~\eqref{choice of c}. 
Picking $r = \gamma / (8\sqrt{2} \lambda_u A_1^2)$, it can be shown that  $r \asymp \sigma_{\varepsilon} \sqrt{{n}/(s + \log d + t )} > \sigma_{\varepsilon} \sqrt{s(\log d + t)/n} \asymp r^{\rm{crude}}  $ and $\delta = \sqrt{5}/( a_0 \kappa_1 \underbar{$\tau$}) < 1$. Thus, setting $L =  \{ 2 +\frac{2}{p_{{0}}'(a_0) } \}  (c^2 + 1)^{1/2} s^{1/2} + \frac{2}{p_{{0}}'(a_0) } s^{1/2} $, Lemma \ref{lem:RSC} indicates that the event $\cE_{\rm{rsc}}(r,L,0.5 \kappa_1 \underbar{$\tau$}  )$ holds with probability at least $1 - e^{-t}$.

Moreover, by Lemma \ref{lem:grad_l2} and the choice of  $\gamma = \sigma_{\varepsilon} \sqrt{{n}/(s + \log d + t )}$, we obtain
\begin{align}\label{thm:random part3}
\| \bw^*_{\cS} \|_2  
\lesssim
\sigma_{\varepsilon}  \sqrt{\frac{s+ t}{n}},
\end{align}
with probability at least $1-e^{-t}$.

Finally, we obtain an upper bound for the term $\|  p_\lambda'(|\bbeta^*_\cS| - a_0 \lambda )_+ \|_2$. Since $|\beta^*_j| \geq (a_0 + a_1)\lambda$ for any $j\in \cS$, we have $p'_{\lambda}(|\beta^*_j| - a_0\lambda) = 0$.
Combining the aforementioned inequalities to~\eqref{contraction.inequality2}, we obtain
\begin{align*}
\| \hat{ \bbeta}^{(T)} - \bbeta^* \|_2
\lesssim
\delta^{T-1} \sigma_{\varepsilon} \sqrt{\frac{s (\log d + t)}{n}} + \frac{\sigma_{\varepsilon}}{1-\delta}    \sqrt{\frac{s+\log d +t}{n}}, 
\end{align*}
with probability at least $1-4e^{-t}$.
Setting $T \gtrsim \frac{\log\{ \log (d+  t)\}}{\log (1/\delta)}$ leads to the desired results in \eqref{test} and \eqref{oracle1}.

\end{proof}

\subsection{Proof of Lemmas}

\subsubsection{Proof of Lemma~\ref{ld lem:RSC}}
\begin{proof}
The proof is a simplified version of the proof of Lemma~\ref{lem:RSC}.  
In the following, we outline the slight difference of the proof.  
Let $\bdelta = \bSigma^{1/2}(\bbeta - \bbeta^*)$ and $\bz_i =  \bSigma^{-1/2}\bx_i$.  
Using the arguments from the beginning of the proof of Lemma~\ref{lem:RSC} to~\eqref{lem RSC EB}, it can be shown that $\EE \{\cB(\balpha)\} \geq 3/4$ provided that $\gamma \geq 4\sqrt{2}  \max \{\sigma_{\varepsilon}, 2A_1^2r \}$, where $\cB(\balpha)$ is as defined in \eqref{eq:RSC2.51}.  
Moreover, since $d<n$, by Cauchy-Schwarz inequality, we have 
\begin{align*}
\EE (\Delta )
&\leq
\frac{\gamma}{r} \EE \bigg\{ \sup_{\balpha \in \BB(r)} \frac{1}{n} \sn \langle e_i   \bz_i ,\balpha \rangle \bigg\}  \\
&\leq 
\frac{\gamma}{rn}  \EE  \sup_{\balpha \in \BB(r)} \Big|\Big| \sn e_i   \bz_i \Big|\Big|_2 \cdot || \balpha ||_2 \\
&\leq
\frac{\gamma}{r} \sqrt{\frac{d}{n}}.
\end{align*}
Consequently, we have $\Delta \leq 1/4$ with high probability provided that $n \gtrsim  ({\gamma}/{r})^2(d+t)$. Combining the above pieces finishes the proof.

\end{proof}

\subsubsection{Proof of Lemma~\ref{lem:RSC}}
\begin{proof}
For notational convenience, throughout the proof we let $\bdelta= \bbeta-\bbeta^*$.  
Recall from Definition~\ref{def:rsc} that $\BB (r) = \{ \bdelta \in \RR^d: \| \bdelta   \|_2 \leq r \}$ is a ball and  $\CC(L)=\{ \bdelta: \| \bdelta  \|_1 \leq L\| \bdelta   \|_2 \}$ is an $\ell_1$-cone.
In the following proof, we will provide a lower bound for the symmetrized Bregman divergence $\cB(\bbeta,  \bbeta^* ) $ under the constraint $\bbeta \in \bbeta^* + \BB(r) \cap \CC (L)$.

We start by defining the events
\begin{equation}
\label{rsc:event}
E_i(\bdelta, r,\gamma) = \{ | \varepsilon_i | \leq \gamma/2\} \cap \left\{ | \bx_i^{{\T}} \bdelta | \leq \frac{\gamma \| \bdelta \|_2}{2r}  \right\}
\end{equation}
for $i=1,\ldots,n$.
 The symmetrized Bregman divergence can then be low bounded by 
 \begin{equation}
 \label{RSC1}
 \begin{split}
 \cB(\bbeta,  \bbeta^* ) & = 
\frac{1}{n} \sn   \bigl\{   L'(\varepsilon_i) -   L' (\varepsilon_i -   \bx_i^{{\T}} \bdelta)  \bigr\} \cdot \bx_i^{{\T}} \bdelta\\
& \ge \frac{1}{n} \sn \bigl\{   L'(\varepsilon_i) -   L' (\varepsilon_i - \bx_i^{{\T}} \bdelta)  \bigr\} \cdot \bx_i^{{\T}} \bdelta  \cdot   \mathbbm{1}_{E_i(\bdelta,r,\gamma)},   
 \end{split}
 \end{equation}
 where $\mathbbm{1}_{E_i(\bdelta,r,\gamma)}$ is an indicator function that takes value one when the event in \eqref{rsc:event} holds and zero otherwise. 
 Thus, it suffices to obtain a lower bound on~\eqref{RSC1} for any $\bbeta \in \bbeta^* + \BB(r) \cap \CC (L)$.

Recall  from Section~\ref{Sec:Prelim}  that $L'(u) = \gamma w_\tau(u) \ell'(u/\gamma)$ with $w_\tau(u)= |\tau -  I(u<0)|$.
Conditioned on the event $E_i(\bdelta, r,\gamma)$, for any $\bdelta \in  \BB(r)$, we have 
 $|\varepsilon_i| \leq \gamma$ and $|\varepsilon_i -  \bx_i^{{\T}} \bdelta| \leq \gamma/2 + \gamma/2 = \gamma$.   
 For notational convenience, let $u_i = \varepsilon_i$ and let $v_i = \varepsilon_i-\bx_i^{\T} \bdelta$.  Then, the term $\bigl\{   L'(\varepsilon_i) -   L' (\varepsilon_i - \bx_i^{{\T}} \bdelta)  \bigr\} \cdot \bx_i^{{\T}} \bdelta$  can be rewritten as $\{ L'(u_i)  - L'(v_i)  \} (u_i-v_i)$.
 In the following, we obtain a lower bound for the term $\{ L'(u_i)  - L'(v_i)  \} (u_i-v_i)$  for any $u_i,v_i \in [-\gamma,\gamma]$.
 Let $\kappa_1 = \min_{|t|\leq 1} \ell''(t)$.  To this end, we consider three possible cases:
\begin{enumerate}
\item[(i)] ($u_iv_i=0$).   If $v_i=0$,  we have  $\{ L'(u_i)  - L'(v_i)  \} (u_i-v_i) \ge \gamma w_\tau(u_i) \{ \ell'(u_i/\gamma) - \ell'(0) \}u_i \geq \kappa_1 \underbar{$\tau$} u_i^2$, where the last inequality hold by the mean value theorem.
Similarly if $u_i=0$, $\{ L'(u_i)  - L'(v_i)  \} (u_i-v_i) \geq \kappa_1 \underbar{$\tau$} v_i^2$.

\item[(ii)] ($u_iv_i>0$). In this case, $w_\tau(u_i)=w_\tau(v_i) $ and hence $\{ L'(u_i)  - L'(v_i)  \} (u_i-v_i)  =  \gamma  w_\tau(u_i) \{ \ell'(u_i/\gamma) - \ell'(v_i/\gamma) \} (u_i-v_i)   \geq \kappa_1 \underbar{$\tau$} (u_i-v_i)^2$.

\item[(iii)] ($u_iv_i <0$).  In this case, we have either $u>0, v<0$ or $u<0, v>0$. For the former, $\{ L'(u_i)  - L'(v_i) \} (u_i-v_i) =  \gamma  \{ \tau   \ell'(u_i/\gamma) - (1-\tau) \ell'(v_i/\gamma) \}(u_i-v_i) \geq \kappa_1 \underbar{$\tau$} (u_i-v_i)^2$, where the last inequality holds by the mean value theorem. The latter can be shown in a similar fashion.
\end{enumerate} 
Combining all three cases, we conclude that $\{L'(u_i) - L'(v_i) \}(u_i-v_i)\geq  \kappa_1 \underbar{$\tau$} (u_i-v_i)^2$ for all $u_i, v_i\in[-\gamma, \gamma]$. Substituting this into \eqref{RSC1} yields
\begin{align}
	\cB(\bbeta,  \bbeta^* ) \geq   \frac{ \kappa_1 \underbar{$\tau$} }{n} \sn  (\bx_i^{{\T}} \bdelta )^2  \mathbbm{1}_{E_i(\bdelta, r,\gamma)}
\label{RSC2}
\end{align}
for any $\bdelta \in  \BB(r)$. 

Next, we will derive a lower bound for $(1/n)\sn  (\bx_i^{{\T}} \bdelta )^2  \mathbbm{1}_{E_i(\bdelta,r, \gamma)}$, uniformly over $\bdelta\in \BB(r)$.
To this end, we smooth the discontinuous indicator function $\mathbbm{1}_{E_i(\bdelta,r, \gamma)} = \mathbbm{1}_{\{ |\bx_i^{{\T}} \bdelta | \leq \gamma \| \bdelta \|_2 /(2r) \}} \cdot \mathbbm{1}_{\{ |\varepsilon_i| \leq \gamma/2 \}} $ by a Lipschitz continuous function. Using similar ideas from the proof of Proposition~2 in \cite{L2017}, for any $R\geq 0$, we define the truncated squared function  as
$$
 \varphi_R(u) = u^2 \mathbbm{1}(|u|\leq R/2) + (|u|-R)^2\mathbbm{1}(R/2 < |u| \leq R) , \ \ u \in \RR.
$$
It can be verified that the function $\varphi_R(\cdot)$ is $R$-Lipschitz continuous and satisfies the following:
\begin{align} \label{RSC2.5}
	u^2\mathbbm{1}(|u|\leq R/2) \leq \varphi_R(u) \leq  \min\bigl\{ u^2 \mathbbm{1}(|u| \leq R), (R/2)^2 \bigr\} ~~\mbox{ and }~~ \varphi_{cR}(cu) = c^2 \varphi_R(u)~\mbox{ for any } c\geq 0.
\end{align}
 It then follows from \eqref{RSC2} and \eqref{RSC2.5} that
\begin{equation}
\label{eq:RSC2.51}
\cB(\bbeta,  \bbeta^* ) \geq  \kappa_1 \underbar{$\tau$} \,\| \bdelta \|_2^2 \cdot  \underbrace{     \frac{1}{n} \sn \mathbbm{1}(|\varepsilon_i| \leq \gamma/2) \cdot \varphi_{\gamma/(2r)} (\bx_i^{{{\T}}} \balpha  )    }_{ =: \cB(\balpha ) },  ~~\mbox{ where }~ \balpha := \bdelta / \| \bdelta \|_2 \in \mathbb{S}^{d-1}. 
\end{equation}

Next, we bound the random quantity $\cB(\balpha )$ from below.  Let  $\Delta = \sup_{\balpha \in \mathbb{S}^{d-1}}-\cB(\balpha)+\EE \{\cB(\balpha)\}$.
Then, we have $\cB(\balpha )\ge \EE \{\cB(\balpha)\}-\Delta$.  It suffices to obtain a lower bound for $\EE \{\cB(\balpha)\}$ and an upper bound for the random fluctuation $\Delta$.
We start with obtaining a lower bound for  $\EE \{\cB(\balpha)\}$.

Recall that $A_1$ is a constant that satisfies $\EE  \{(\bu^{\T} \bx )^4\} \leq A_1^4 \|\bu\|^4_{\bSigma} \leq \lambda_u^2 A_1^4 \|\bu\|^4_2$ for all $\bu \in \RR^d$. 
Applying the inequality in~\eqref{RSC2.5}, for any $\balpha \in \mathbb{S}^{d-1}$, we have
\begin{align} \label{lem RSC EB}
\EE\{ \cB(\balpha) \}
&\geq 
\EE \big\{(\bx_i^{\T} \balpha)^2 \mathbbm{1}(|\varepsilon_i| \leq \gamma /2) \mathbbm{1}(|\bx_i^{\T} \balpha| \leq \gamma/{4r})\big\} \nn \\
&\geq 
\EE \Big[(\bx_i^{\T}\balpha)^2 \big\{ 1-\mathbbm{1}(|\bx_i^{\T}\balpha|> \gamma / 4r) - \mathbbm{1}(|\varepsilon_i|>\gamma/2) \big\} \Big] \nn \\
&\geq 
1-({4r}/{\gamma})^2 \EE(\bx_i^{\T}\balpha)^4 - \EE \Big[(\bx_i^{\T}\balpha)^2 \EE \big\{({2|\varepsilon_i|}/{\gamma})^2|\bx_i \big \} \Big ] \nn \\
&\geq 
1 - ({4r}/{\gamma})^2 \lambda_u^2 A_1^4 - ({2}/{\gamma})^2 \sigma^2_\varepsilon \lambda_u
\end{align}
Provided $\gamma \geq 4 \sqrt{2} \lambda_u \max \{\sigma_\varepsilon, 2A_1^2r \}$, we obtain $\EE \{\cB(\balpha)\} \geq 3/4$.

Next, we obtain an upper bound for $\Delta = \sup_{\balpha \in \mathbb{S}^{d-1}}-\cB(\balpha)+\EE\{ \cB(\balpha)\}$. 
Applying the inequality in \eqref{RSC2.5} on $\varphi_{\gamma / (2r)}(\cdot)$, we have $\cB(\balpha) \leq (\gamma / 4r)^2$. Applying Theorem 7.3 in \cite{B2003} and the inequality $ab \leq a^2/4 + b^2 $, for any $t \geq 0$, we obtain
\begin{align} \label{lem RSC partial 2}
\Delta 
&\leq 
\EE (\Delta) + \sqrt{\frac{\gamma^2 t}{4r^2 n}\EE( \Delta)}  + \lambda_u A_1^2 \sqrt{\frac{2t}{n}} + \frac{\gamma^2}{48r^2} \cdot \frac{t}{n} \nn \\
&\leq 
1.25 \EE (\Delta) + \lambda_u A_1^2 \sqrt{\frac{2t}{n}} + \frac{\gamma^2}{3r^2} \cdot \frac{t}{n},
\end{align} 
with probability at least $1-e^{-t}$. 

It remains to bound $\EE (\Delta)$. Let $\cB_i(\balpha)= \mathbbm{1}(|\varepsilon_i| \leq \gamma/2)  \cdot \varphi_{\gamma/(2r)} (\bx_i^{{{\T}}} \balpha  )$ and note that $\EE (\Delta) = \EE \big[ \sup_{\balpha \in \mathbb{S}^{d-1}} \big\{ -({1}/{n}) \sn \cB_i(\balpha) +({1}/{n}) \sn \EE \cB_i(\balpha) \big\}\big]$. By the symmetrization inequality for empirical process,  $\EE( \Delta) \leq 2\EE \big\{ \sup_{\balpha \in \mathbb{S}^{d-1}} ({1}/{n}) \sn e_i  \cB_i(\balpha) \big \}$,  where $ e_1,\dots,e_n$ are independent Rademacher random variables. 
Recall that $\CC_1= \{\bdelta : \|\bdelta_{\cS^{\cc}} \|_1 \leq  3\| \bdelta_{\cS} \|_1\}$ where $\cS=\supp (\bbeta^*)$. For all $\bbeta \in \bbeta^* + \BB(r) \cap \CC_1$, we have $\|\bbeta-\bbeta^* \|_1 \leq 4 \| (\bbeta - \bbeta^*)_{\cS}  \|_1 \leq 4s^{1/2}\|\bbeta-\bbeta^* \|_2$.
Since $\cB_i(\balpha)$ is ${\gamma}/{(2r)}$-Lipschitz, applying the Talagrand's contraction principle \citep{LT1991} and Holder's inequality, we have 
\begin{align}
\EE (\Delta )
&\leq
\frac{\gamma}{r} \EE \Bigg\{ \sup_{\bbeta \in \bbeta^* + \BB(r) \cap \CC_1} \frac{1}{n} \sn \Big\langle e_i  \bx_i ,\frac{\bbeta - \bbeta^*}{\|\bbeta-\bbeta^*\|_2} \Big\rangle \Bigg\} \nn \\
&\leq 
\frac{\gamma}{rn} 4s^{1/2} \EE \left\|\sn e_i \bx_i \right\|_\infty. \label{lem RSC partial 3b1}
\end{align}

Let $S_j=\sn e_i x_{ij}$ for $j=1,\dots,d$. It remains to bound $\EE \|\sn e_i \bx_i \|_\infty = \EE (\max_j |S_j|)$.
Since $\bx$ is sub-exponential, by Condition~\ref{cond:covariates}, we have $\PP(|x_{ij}|\geq \nu_0 \sigma_{jj}^{1/2} t)\leq e^{-t}$. Consequently, we obtain
\begin{align*}
    \EE (|e_i x_{ij}|^k) \leq \int_{0}^{\infty} \PP(|x_{ij}|^k \geq t)dt \leq k! \nu_0^k \sigma_{jj}^{k/2} \mbox{~ for all ~} k \geq 2.
\end{align*}
Along with the fact that $\EE (e_i x_{ij})=0$, for any $0 \le \lambda \le (\nu_0 \sigma_{\bx})^{-1}  $, the moment generating function of $e_i x_{ij}$ can be upper bounded by
\begin{align*}
\EE \left(e^{\lambda e_i x_{ij}} \right)
&\leq 
1 + \sum_{k \geq 2} \frac{\lambda^{k}}{k!} \EE |e_i x_{ij}|^{k} \\
&\leq 
1+ \sum_{k \geq 2} (\nu_0 \sigma_{jj}^{1/2} \lambda)^k \\
&\leq 
1+ \frac{\nu_0^2 \sigma_{\bx}^2 \lambda^2}{1-\nu_0 \sigma_{\bx}\lambda}
\end{align*}
Using the inequality $\log(1+x)\le x$ for all $x>0$, we have $\log \{\EE (e^{\lambda S_j})\} \leq \sn \log \{ 1+ \nu_0^2 \sigma_{\bx}^2 \lambda^2/(1-\nu_0 \sigma_{\bx}\lambda)\} \leq  (2n \nu_0^2 \sigma^2_{\bx} \lambda^2)/\{2(1-\nu_0\sigma_{\bx} \lambda)\}$ for any $1 \leq j \leq d$ and $0 \leq \lambda \leq (\nu_0 \sigma_{\bx})^{-1}$. 
Consequently $S_1,...,S_d$ are  sub-gamma $\Gamma_{+} (v,c)$ with $v=2n\nu_0^2 \sigma^2_{\bx}$ and $c=\nu_0 \sigma_{\bx}$. 
Applying Corollary 2.6 in \cite{BLM2013}, we obtain
\begin{align} \label{lem RSC partial 3b2}
\EE \left\| \sn e_i \bx_i \right\|_\infty = \EE \left(\max_j |S_j|\right) \leq \sqrt{2v\log 2d} + c\log 2d = \nu_0 \sigma_{\bx} \Big(2\sqrt{n\log 2d}+\log 2d\Big)
\end{align}

Combining \eqref{lem RSC partial 2}, \eqref{lem RSC partial 3b1}, and \eqref{lem RSC partial 3b2}, we obtain
\begin{align*}
\Delta \leq 5s^{1/2} \frac{\gamma \nu_0 \sigma_{\bx}}{r}  \Bigg(2\sqrt{\frac{\log 2d}{n}}+\frac{\log 2d}{n}\Bigg) + \lambda_u A_1^2 \sqrt{\frac{2t}{n}} + \frac{\gamma^2}{3r^2}\frac{t}{n},
\end{align*}
with probability at least $1-e^{-t}$.  
Provided that $n \gtrsim ({\sigma_{\bx} \nu_0 \gamma}/{r})^2 s (\log d + t)$, we have $\Delta \leq 1/8$ with probability at least $ 1-e^{-t}$. Putting all pieces together, as long as $\gamma \geq 4 \sqrt{2} \lambda_u \max \{\sigma_{\varepsilon}, 2A_1^2r \}$ and $n \gtrsim ({\sigma_{\bx} \nu_0 \gamma}/{r})^2 (s\log d + t)$, the  following bound holds uniformly over $\bbeta \in \bbeta^* + \BB(r) \cap \CC_1$:
\begin{align*}
\cB(\bbeta,\bbeta^*) \geq \frac{1}{2} \kappa_1 \underbar{$\tau$}\|\bbeta-\bbeta^* \|_2^2,
\end{align*}
 with probability at least $1-e^{-t}$.
The final result is obtained by replacing $4s^{1/2}$ by $L$.
\end{proof}

\subsection{Proof of Propositions}

\subsubsection{Proof of Proposition \ref{prop:further steps}}
\label{proof:prop:further steps}
\begin{proof}

We start by obtaining an upper bound for $\hat{\bbeta}^{(1)}$ obtained by solving~\eqref{retire.est.convex}, or equivalently, solving~\eqref{general.lasso2}, with an initial estimator $\hat{\bbeta}^{(0)}=\textbf{0}$ and $\blambda^{(0)}=p'_{\lambda}(\textbf{0})=(\lambda,\ldots,\lambda)^{\T}$.
Conditioned on the event $ \cE_{\rm{rsc}}(r,L_0,\kappa) \cap \cE_{\rm{score}}(\lambda)$ with $L_0 = 6  s^{1/2} $, and from the proof of Lemma~\ref{lem:deterministic.error.bound} with parameters $r, \kappa, \lambda >0$ such that $r >   2.5 \kappa^{-1}  s^{1/2} \lambda   $, any solution $\hat{ \bbeta}^{(1)}$ to~\eqref{general.lasso2} satisfies
\begin{align} \label{prop.ineq}
 \| \hat{ \bbeta}^{(1)} - \bbeta^* \|_2 \leq   2.5 \kappa^{-1}  s^{1/2} \lambda    .
\end{align}

We now continue to establish an upper bound on the estimation error for the subsequent estimators $\hat{\bbeta}^{(t)}$ for $t\ge2$.
For $t=1,2,\ldots$, we first construct a series of augmented sets 
\begin{align*}
\cA_{t} = \cS \cup \big\{ 1 \leq j \leq d : \lambda_j^{(t -1)} < p_0'(a_0)\lambda \big\}.
\end{align*}
Let $c>0$ be a constant such that $0.5 p_0'(a_0)(c^2 + 1)^{1/2} + 2 = c \kappa a_0$.
In the following, using mathematical induction, we will show that the cardinality of $\cA_{t}$ can be upper bounded as 
\begin{equation}
\label{eq:Acard}
|\cA_{t}| \le (c^2+1)s.
\end{equation}  
 For $t=1$, the inequality holds trivially, i.e., $|\cA_1| = |\cS| = s \leq (c^2 + 1)s$.
Now, assume that~\eqref{eq:Acard} holds for some integer $t \ge 2$.
We aim to show that $|\cA_{t+1}| \le (c^2+1)s$.
To this end, we first obtain an upper bound of the cardinality of the set $\cA_{t+1}\setminus \cS$.
Since $p'_{\lambda}(\cdot)$ is monotonically decreasing on $\RR^+$, by the definition of $\cA_{t+1}$, for each $j\in \cA_{t+1}\setminus \cS$, we have $p'_{\lambda}(|\hat{\bbeta}_j^{(t)}|) = \lambda _j^{(t)} \leq p_0'(a_0)\lambda = p'_{\lambda}(a_0 \lambda)$, which implies $|\hat{\bbeta}_j^{(t)}| \geq a_0 \lambda$.
Moreover, the monotonicity of $p'_{\lambda}(\cdot)$ on $\RR^+$ and the definition of $\cA_t$ imply that $\| \blambda^{(t-1)}  \|_{\infty}  =\| p'_{\lambda}(|\hat{\bbeta}^{(t-1)}|)  \|_{\infty} \leq \| p'_{\lambda}(\textbf{0})  \|_{\infty} = \lambda  $ and $\big\|  \blambda_{\cA_{t}^{\cc}}^{(t -1)} \big\|_{\min} \geq p_0'(a_0)\lambda $, respectively. 

Conditioned on the event $\cE_{\rm{rsc}}(r,L,\kappa) \cap  \{p_0'(a_0) \lambda \geq 2\|\nabla \cR_n(\bbeta^*)-\nabla \cR(\bbeta^*) \|_\infty \} $ with $L =  \{ 2 +2/p_0'(a_0)  \}  (c^2 + 1)^{1/2} s^{1/2} + 2  s^{1/2}/p_0'(a_0) $, it follows from the proof of Lemma \ref{lem:deterministic.error.bound} that 
\begin{align}
\| \hat{\bbeta}^{(t)}-\bbeta^* \|_2
&\leq 
\frac{ \| \blambda^{(t -1)}_{\cS} \|_2 + \| \bw^*_{\cA_{t}} \|_2 +  \| \nabla \cR (\bbeta^* )\|_2  }{\kappa} \label{furstep.ineq.basic}  \\
&\leq
\frac{\big\{ 0.5{p_{{0}}'(a_0)} (c^2 + 1)^{1/2} +2 \big\} s^{1/2}\lambda }{\kappa} \nn \\
&=
c a_0 s^{1/2}\lambda = r^{\rm{crude}}< r. \label{furstep.ineq.1}
\end{align}
Along with the fact that $\bbeta^*_j=0$ for all $j \in {\cA_{t+1}} \setminus \cS$, we obtain
\begin{align}
|{\cA_{t+1}} \setminus \cS|^{1/2} = \| \mathbf{1}_{{\cA_{t+1}} \setminus \cS} \|_2 
&\leq 
\bigg\| \Big(\frac{\hat{\bbeta}^{(t)}}{a_0	\lambda} \Big)_{{\cA_{t+1}} \setminus \cS} \bigg\|_2  \nn \\
&\leq
\frac{1}{a_0 \lambda} \big\| (\hat{\bbeta}^{(t)} - \bbeta^*)_{\cA_{t+1} \setminus \cS} \big\|_2 \label{furstep.ineq.2.5} \\
&\leq
c s^{1/2}, \nn
\end{align}
where the last inequality holds by applying~\eqref{furstep.ineq.1}. 
Therefore $|\cA_{\ell +1}|=|{\cA_{\ell+1}} \setminus \cS| + |\cS| \leq (c^2 +1)s$. By induction,  $|\cA_{t}| \leq (c^2 + 1)s$ holds for all $t \geq 1$. Consequently,~\eqref{furstep.ineq.basic} holds for all $t \geq 1$. 

We note that the upper bound~\eqref{furstep.ineq.1} is not sharp and is mainly derived for proving~\eqref{furstep.ineq.2.5}.
 We now derive a sharper upper bound for $\hat{\bbeta}^{(t)}$ by controlling the terms $\| \blambda^{(t -1)}_{\cS} \|_2$ and  $\| \bw^*_{\cA_{t}} \|_2$ more carefully.
 We start with providing a tighter upper bound for $\| \blambda^{(t -1)}_{\cS} \|_2$.  
 For each $j \in \cS$, we consider the following two cases:~(i) if $|\hat{\bbeta}^{(t -1)}_j - \bbeta^*_j| \geq a_0 \lambda$, then the inequality $\lambda_j^{(\ell -1)} \leq \lambda \leq a_0^{-1}|\hat{\bbeta}^{(t -1)}_j - \bbeta^*_j|$ holds trivially; (ii) if $|\hat{\bbeta}^{(t -1)}_j - \bbeta^*_j| < a_0 \lambda$, then along with minimal signal strength condition $\| \bbeta^*_{\cS}  \|_{\min} \geq a_0 \lambda$ and the monotonicity of $p'_{\lambda}(\cdot)$ on $\RR^{+}$, we have $0 \leq |\bbeta^*_j| - a_0 \lambda \leq |\hat{\bbeta}^{(t -1)}_j|$, thus $\lambda_j^{(t -1)} = p'_{\lambda}(|\hat{\bbeta}^{(t -1)}_j|) \leq p'_{\lambda}\{(|\bbeta^*_j|-a_0 \lambda)_+\}$.
 Combining the two cases above, we obtain
\begin{align} \label{careful.control.1}
\| \blambda^{(t -1)}_{\cS} \|_2 \leq 
\| p'_{\lambda}\{(|\bbeta^*_j|-a_0 \lambda)_+\} \|_2 + 
a_0^{-1}\| (\hat{\bbeta}^{(t -1)}-\bbeta^*)_{\cS} \|_2
\end{align}

We now obtain an upper bound for $\| \bw^*_{\cA_{t}} \|_2$.
Since ${\cA_{t}} = \cS \cup (\cA_{t} \setminus \cS)$, we have
\begin{align}
\| \bw^*_{\cA_{\ell}} \|_2 
&= 
\| \bw^*_{\cS} \|_2 + \| \bw^*_{\cA_{t} \setminus \cS} \|_2 \nn \\
&\leq
\| \bw^*_{\cS} \|_2 + |\cA_{t} \setminus \cS|^{1/2}\| \bw^* \|_{\infty} \nn \\
&\leq
\| \bw^*_{\cS} \|_2 +\frac{p_0'(a_0)}{2a_0}\| (\hat{\bbeta}^{(t -1)}-\bbeta^*)_{\cA_{t} \setminus \cS} \|_2 \label{careful.control.2}\\
&\leq
\| \bw^*_{\cS} \|_2 +\frac{1}{2 a_0}\| (\hat{\bbeta}^{(t -1)}-\bbeta^*)_{\cA_{t} \setminus \cS} \|_2, \label{careful.control.3}
\end{align}
where \eqref{careful.control.2} holds from applying~\eqref{furstep.ineq.2.5}, and \eqref{careful.control.3} holds from the fact that ${p'_\lambda({a_0})} \leq 1$. 

Putting \eqref{furstep.ineq.basic}, \eqref{careful.control.1}, and \eqref{careful.control.3} together, and applying the inequality $\sqrt{a} + \sqrt{b/4} \leq \sqrt{5(a+b)/4}$ for $a,b \geq 0$, we obtain
\begin{align}
\| \hat{\bbeta}^{(t)}-\bbeta^* \|_2
&\leq 
\frac{\| \blambda^{(t -1)}_{\cS} \|_2 + \| \bw^*_{\cA_{\ell}} \|_2 +  \| \nabla \cR (\bbeta^* )\|_2  }{\kappa}\nn \\
&\leq
\frac{\| p'_{\lambda}\{(|\bbeta^*_j|-a_0 \lambda)_+\} \|_2 + \| \bw^*_{\cS} \|_2 +  \| \nabla \cR (\bbeta^* )\|_2  }{\kappa} + \frac{\sqrt{5}}{2 a_0 \kappa}\| (\hat{\bbeta}^{(t -1)}-\bbeta^*)_{{\cA_{\ell}}} \|_{2}  \nn \\
&\leq
\frac{\| p'_{\lambda}\{(|\bbeta^*_j|-a_0 \lambda)_+\}+ \| \bw^*_{\cS} \|_2 + \| \nabla \cR (\bbeta^* )\|_2  }{\kappa} + \delta \| \hat{\bbeta}^{(t -1)}-\bbeta^* \|_2, \label{final.ineq.1}
\end{align}
for all $t\ge 2$. 
The result in \eqref{contraction.inequality2} can then be obtained by applying~\eqref{final.ineq.1} iteratively.
\end{proof}

\subsubsection{Proof of Proposition \ref{ld prop approx err}}
\begin{proof}
Let $\bdelta = \bbeta^* - \bbeta_\gamma^*$. The optimality of $\bbeta_\gamma^*$ and the mean value theorem indicate respectively that $\nabla \cR(\bbeta_\gamma^*) = \bm{0}$, and
\begin{align}
\label{ld prop 1a}
\bdelta^{\T} \nabla^2\cR(\tilde{\bbeta}_\gamma^*) \bdelta 
&=
\langle \nabla \cR(\bbeta^*) -\nabla \cR(\bbeta_\gamma^*), \bdelta \rangle  \nn \\ 
&=
\langle \nabla \cR(\bbeta^*) , \bdelta \rangle 
=
-\frac{1}{n} \sum_{i=1}^n \EE \{w_\tau (\varepsilon_i) \ell_\gamma'(\varepsilon_i)\bx_i^{\T}\bdelta\},
\end{align}
where $\tilde{\bbeta}_\gamma^* = \lambda \bbeta^* + (1-\lambda) \bbeta_\gamma^*$ for some $0\leq \lambda \leq 1$.

We start with an upper bound on the right-hand side of \eqref{ld prop 1a}. 
By the fact that $\EE \{ w_\tau(\varepsilon)\varepsilon|\bx \} = 0$ and $|\ell'(u) - u| \leq u^2$, we have  $ \EE \{ w_\tau(\varepsilon)\ell_\gamma'(\varepsilon)|\bx \}  \leq \EE[ \gamma w_\tau(\varepsilon)\{ \ell'(\varepsilon/{\gamma}) - {\varepsilon}/{\gamma} \}|\bx ] \leq \bar{\tau} \sigma_\varepsilon^2 / \gamma$. Consequently
\begin{align}
\label{ld prop 1b}
\EE \{w_\tau (\varepsilon_i) \ell_\gamma'(\varepsilon_i)\bx_i^{\T}\bdelta \}
\leq
\EE |\bx^{\T} \bdelta|\cdot \bar{\tau} \sigma_\varepsilon^2 / \gamma \leq || \bSigma^{1/2} \bdelta ||_2 \cdot \bar{\tau} \sigma_\varepsilon^2 / \gamma.
\end{align}

Next, we obtain a lower bound for $\bdelta^{\T} \nabla^2\cR(\tilde{\bbeta}_\gamma^*) \bdelta$.
Let $L_{\tau,\infty} (\cdot)$ be the resulting asymmetric $\ell_2$ loss when taking $\gamma = \infty$ in $L_{ \tau, \gamma}(\cdot)$.
Moreover, let $\cR_\infty(\bbeta) = \EE\{ L_{\tau,\infty}(y - \bx^{\T} \bbeta)\}$. 
Since $\cR(\cdot)$ is convex and minimized at $\bbeta_\gamma^*$, we have 
\[
\cR(\tilde{\bbeta}_\gamma^*) \leq \lambda \cR(\bbeta^*) + (1-\lambda)\cR(\bbeta_\gamma^*) \leq \cR(\bbeta^*) \leq \cR_\infty(\bbeta^*) \leq \bar{\tau}\sigma_\varepsilon^2 / 2.
\] 
On the other hand, by the definition of Huber loss, for all $\bbeta \in \RR^d$,  we have 
\[
\cR(\bbeta) \geq n^{-1} \sum_{i=1}^n \EE w_\tau(y_i - \bx_i^{\T} \bbeta)(\gamma|y_i - \bx_i^{\T} \bbeta| - \gamma^2/2)\mathbbm{1}(|y_i - \bx_i^{\T} \bbeta| > \gamma).
\] 
Let $\tilde{\varepsilon_i} = y_i - \bx_i^{\T} \tilde{\bbeta}_\gamma^*$.
Combining the above inequalities, we have 
\begin{align*}
\frac{\gamma}{n} \sum_{i=1}^n \EE \{w_\tau(\tilde{\varepsilon_i}) |\tilde{\varepsilon_i}| \mathbbm{1}(|\tilde{\varepsilon_i}|>\gamma)\}
&\leq
\frac{\gamma^2}{2n} \sum_{i=1}^n \EE \{w_\tau(\tilde{\varepsilon_i})  \mathbbm{1}(|\tilde{\varepsilon_i}|>\gamma)\} + \frac{\bar{\tau} \sigma_\varepsilon^2}{2} \\
&\leq
\frac{\gamma}{2n} \sum_{i=1}^n \EE \{w_\tau(\tilde{\varepsilon_i}) |\tilde{\varepsilon_i}| \mathbbm{1}(|\tilde{\varepsilon_i}|>\gamma)\} + \frac{\bar{\tau} \sigma_\varepsilon^2}{2},
\end{align*}
which further implies that
\begin{align}
\label{ld prop 1c}
\frac{1}{n} \sum_{i=1}^n \EE \{w_\tau(\tilde{\varepsilon_i})\mathbbm{1}(|\tilde{\varepsilon_i}|>\gamma)\} \leq \frac{1}{n\gamma} \sum_{i=1}^n \EE \{w_\tau(\tilde{\varepsilon_i}) |\tilde{\varepsilon_i}| \mathbbm{1}(|\tilde{\varepsilon_i}|>\gamma) \}\leq \frac{\bar{\tau}\sigma_\varepsilon^2}{\gamma^2}.
\end{align}

Moreover, note that $\nabla^2 \cR(\tilde{\bbeta}_\gamma^*) = n^{-1} \sum_{i=1}^n \EE \{w_\tau(\tilde{\varepsilon_i})\bx_i \bx_i^{\T}\} - n^{-1} \sum_{i=1}^n \EE \{w_\tau(\tilde{\varepsilon_i})\mathbbm{1}(|\tilde{\varepsilon_i}| > \gamma) \bx_i \bx_i^{\T}\}$. It then follows from the Cauchy–Schwarz inequality and \eqref{ld prop 1c} that
\begin{align*}
\bdelta^{\T} \nabla^2 \cR(\tilde{\bbeta}_\gamma^*) \bdelta 
&=
\frac{1}{n} \sum_{i=1}^n \EE \{w_\tau(\tilde{\varepsilon_i}) (\bdelta^{\T} \bx_i)^2 \}- 
\frac{1}{n} \sum_{i=1}^n \EE \{w_\tau(\tilde{\varepsilon_i})\mathbbm{1}(|\tilde{\varepsilon_i}| > \gamma)  (\bx_i^{\T} \bdelta)^2\} \\
&\geq
\underline{\tau} || \bSigma^{1/2}\bdelta ||_2^2 - \Bigg\{ \frac{1}{n} \sum_{i=1}^n \EE w_\tau^2(\tilde{\varepsilon_i})\mathbbm{1}^2(|\tilde{\varepsilon_i}| > \gamma) \Bigg\}^{1/2} \Bigg\{ \frac{1}{n} \sum_{i=1}^n \EE (\bx_i^{\T} \bdelta)^4 \Bigg\}^{1/2} \\
&\geq
\underline{\tau} || \bSigma^{1/2}\bdelta ||_2^2 - \Bigg\{ \frac{1}{n} \sum_{i=1}^n \bar{\tau} \EE  w_\tau(\tilde{\varepsilon_i})\mathbbm{1}(|\tilde{\varepsilon_i}| > \gamma) \Bigg\}^{1/2} \Bigg\{ \frac{1}{n} \sum_{i=1}^n \EE \langle  \bSigma^{1/2}\bdelta, \bSigma^{-1/2}\bx_i \rangle^4 \Bigg\}^{1/2} \\
&\geq
\underline{\tau} || \bSigma^{1/2}\bdelta ||_2^2 - \frac{\bar{\tau} \sigma_\varepsilon}{\gamma}A_1^2|| \bSigma^{1/2}\bdelta ||_2^2.
\end{align*}
Picking $\gamma \geq 2\sigma_\varepsilon A_1^2 \bar{\tau}/\underline{\tau}$, we have
\begin{align}
\label{ld prop 1d}
\bdelta^{\T} \nabla^2 \cR(\tilde{\bbeta}_\gamma^*) \bdelta \geq \underline{\tau} || \bSigma^{1/2}\bdelta ||_2^2 /2.
\end{align}

Putting together \eqref{ld prop 1a}, \eqref{ld prop 1b}, and \eqref{ld prop 1d} completes the proof.
\end{proof}

\section{Proof of Technical Lemmas~\ref{ld lem:B1B2}--\ref{lem:deterministic.error.bound}}
\label{appendix:technicallemmaproof}

\subsection{Proof of Lemma~\ref{ld lem:B1B2}}
\begin{proof}
The proof is a simplified version of the proof of Lemmas \ref{lem:grad} and \ref{lem:grad_l2}, and is thus omitted.

\end{proof}

\subsection{Proof of Lemma~\ref{ld lem:huber}}
\begin{proof}

We start with obtaining an upper bound for $\EE \{w_\tau(\varepsilon)\ell_\gamma'(\varepsilon)\}$. Denote $\ell(\cdot)$ as the Huber loss with $\gamma=1$.  By the fact that $|\ell'(u) - u|\leq |u|^3$ for all $u \in \RR$, we have
\begin{align*}
\big| \EE \{w_\tau(\varepsilon)\ell_\gamma'(\varepsilon) \}\big|
=
\big| \EE \gamma w_\tau(\varepsilon) \big\{ \ell'(\varepsilon/\gamma) - (\varepsilon/\gamma) \big\} \big|
\leq   \bar{\tau} \gamma^{-2}   \EE |\varepsilon|^3 =   \bar{\tau} \gamma^{-2} v_3   .
\end{align*}

Turning to $\EE \big\{ w_\tau(\varepsilon)\ell_\gamma'(\varepsilon) \big\}^2$, note that $\EE \ell_\gamma'(\varepsilon)^2 = \sigma_\varepsilon^2 - \EE \varepsilon^2 \mathbbm{1}(|\varepsilon| > \gamma ) + \gamma^2 \PP(|\varepsilon| > \gamma)$.  By Markov's inequality,  $
 \EE ( \varepsilon^2 - \gamma^2 ) \mathbbm{1}(|\varepsilon| > \gamma )  \leq \gamma^{-1}  \EE | \varepsilon |^3  = \gamma^{-1} \nu_3$. Combining this with the fact that $\underline{\tau}\leq w_\tau(\varepsilon) \leq \bar{\tau}$ and $|\ell_\gamma'(\varepsilon)| \leq |\varepsilon|$ completes the proof.

\end{proof}

\subsection{Proof of Lemma~\ref{lem:grad}}
\begin{proof}
We start with an upper bound for the term $\|   \nabla \cR (\bbeta ^*) \|_2 =\sup_{\bu \in \mathbb{S}^{d-1}} \EE \{L'(\varepsilon) \bu^{\T} \bx\}$. 
Under Condition~\ref{def:general.loss} on $\ell(\cdot)$ and  Condition~\ref{cond:randomnoise} on the random noise $\varepsilon$,  we have $\EE (   \varepsilon^2 | \bx ) \le  \sigma_{\varepsilon}^2$ and  $|\ell'(u)-u |\leq u^2$. 
Since $\EE[w_\tau (\varepsilon)  \varepsilon |\bx ]= 0$ and $L'(\varepsilon) = \gamma w_\tau(\varepsilon) \ell'(\varepsilon/\gamma)$, we have 
\begin{align*}
\big|\EE \{L'(\varepsilon)|\bx\} \big| 
\leq 
\big| \gamma \EE \big[ w_\tau (\varepsilon) \{ \ell'(\varepsilon/\gamma)-\varepsilon / \gamma \} |\bx \big] \big|
\leq 
\big| \gamma^{-1} \EE \big\{ w_\tau (\varepsilon) \varepsilon^2|\bx \big\} \big| 
\leq 
\gamma^{-1} \bar{\tau} \sigma_{\varepsilon}^2.
\end{align*}
Therefore,  
\[
\EE \big\{L'(\varepsilon) \bu^{\T} \bx \big\} 
= 
\EE \big[ \EE \{L'(\varepsilon)|\bx\}  \bu^{\T} \bx   \big] 
\leq 
\gamma^{-1} \bar{\tau} \sigma_{\varepsilon}^2 \EE ( {|} \bu^{\T} \bx {|}  ) 
\leq 
\gamma^{-1} \bar{\tau} \sigma_{\varepsilon}^2 \| \bu \|_{\bSigma}.
\] 
Taking the supremum over all $\bu \in \mathbb{S}^{d-1}$, we have $ \|  \nabla \cR (\bbeta ^*) \|_2 \leq \gamma^{-1} \bar{\tau} \sigma_{\varepsilon}^2 \lambda_{u}^{1/2}$, as desired.

Next, we obtain an upper bound for the centered score $\bw^* = \nabla \cR_n (\bbeta ^*) -  \nabla \cR (\bbeta ^*)=-({1}/{n}) \sn  \big[ L'( \varepsilon_i ) \bx_i-\EE \{L'( \varepsilon_i ) \bx_i \}\big]$ using the Bernstein's inequality.  We start with establishing an upper bound on the $k$th moment of $L'( \varepsilon_i ) \bx_i $.
Let 
 $\eb_j \in \RR^d $ be the canonical basis vector, i.e., the $j$th entry equals one and all other entries equal zero.  
 Setting $\bu=\eb_j$ in Condition~\ref{cond:covariates} yields $\PP \big(|x_{ij}| \geq \nu_0 \sigma^{1/2}_{jj} t \big) \leq e^{-t}$. 
Therefore, 
\begin{equation*}
\begin{split}
\EE|x_{ij}|^k 
&= 
\int_{0}^{\infty} k u^{k-1} \PP(|x_{ij}| \geq u)\mathrm{d} u \\
&=  
\int_{0}^{\infty} k \nu_0^{k} \sigma_{jj}^{k/2} \PP \Big( |x_{ij}| \geq \nu_0 \sigma_{jj}^{1/2} t \Big) t^{k-1}\mathrm{d} t\\  
&\leq 
\nu_0^k \sigma_{jj}^{k/2} k \int_{0}^{\infty} t^{k-1} e^{-t} \mathrm{d}t \\
&= 
k! \nu_0^k \sigma_{jj}^{k/2} . 
\end{split}
\end{equation*}
In addition, $|\ell'(u)|\leq \min (1,|u|) $ for all $u \in \RR$, thus $|L'(\varepsilon_i)| = |\gamma w_\tau (\varepsilon_i) \ell'(\varepsilon_i/\gamma) | \leq \min \big\{ \bar \tau \gamma, \bar{\tau} |\varepsilon_i| \big\}$.  Combining the above inequalities, for all $k \geq 2$ and $1 \leq j \leq d$, we have
\begin{align*}
\EE |L'(\varepsilon_i)x_{ij}|^k 
&\leq 
\EE \Big\{ (\bar \tau \gamma)^{k-2} |x_{ij}|^k\cdot \EE (\bar{\tau}^2\varepsilon_i^2|\bx_i)  \Big\} \\
&\leq 
\bar{\tau}^k \gamma^{k-2} \sigma_{\varepsilon}^2 \EE |x_{ij}|^k \\
&\leq
\bar{\tau}^k \gamma^{k-2} \sigma_{\varepsilon}^2 \nu_0^k \sigma_{jj}^{k/2}k! \\
&\leq 
\frac{k!}{2} ( 2\bar{\tau}^2 \sigma_{\varepsilon}^2 \nu_0^2 \sigma^2_{\bx})(\nu_0 \bar{\tau} \sigma_{\bx} \gamma)^{k-2}.
\end{align*}

By Bernstein's inequality, for every $u>0$ and $j \in \{1, \dots ,d  \}$, we obtain
\[
\bigg|\frac{1}{n} \sn \Big[ L'(\varepsilon_i)x_{ij}-\EE \{L'(\varepsilon_i)x_{ij}\} \Big]  \bigg| \le
\nu_0 \sigma_{\bx} \bar{\tau} \Bigg( 2\sigma_{\varepsilon}\sqrt{\frac{u}{n}}+\gamma \frac{u}{n} \Bigg) 
\]
with probability at least $1-2e^{-u}$.
Applying the union bound yields
\begin{align*}
\|  \nabla \cR_n (\bbeta ^*) -  \nabla \cR (\bbeta ^*)\|_\infty \leq \nu_0 \sigma_{\bx} \bar{\tau} \Bigg( 2\sigma_{\varepsilon}\sqrt{\frac{u}{n}}+\gamma \frac{u}{n} \Bigg)
\end{align*}
with probability at least $1-2de^{-u}$. We then set $u = \log d + t$ to reach 
\begin{equation}
\label{lemmaB1.1}
\|  \nabla \cR_n (\bbeta ^*) - \nabla \cR(\bbeta ^*)\|_\infty \le \nu_0 \sigma_{\bx} \bar{\tau} \Bigg(2\sigma_{\varepsilon} \sqrt{\frac{\log d + t }{n}}+ \gamma \frac{\log d + t }{n}\Bigg)
\end{equation}
with probability at least $1-2e^{-t}$.

Finally, we now obtain an upper bound for $\|  \nabla \cR_n (\bbeta ^*) \|_\infty$.  By the triangle inequality, we have $\|  \nabla \cR_n (\bbeta ^*) \|_\infty \leq \|  \nabla \cR_n (\bbeta ^*) - \nabla \cR (\bbeta ^*)\|_\infty + \|   \nabla \cR (\bbeta ^*) \|_\infty$.  It suffices to obtain an upper bound for $\|   \nabla \cR (\bbeta ^*) \|_\infty$. We have 
$$
\|   \nabla \cR (\bbeta ^*) \|_\infty = \max_j \EE \{L'(\varepsilon_i)x_{ij} \}
\leq
\max_j \EE \big[ x_{ij} \EE \{ L'(\varepsilon_i)|\bx_i \} \big]
\leq
\max_j \EE ({|} x_{ij} {|} \gamma^{-1} \bar{\tau} \sigma_{\varepsilon}^2)
\leq
\sigma_{\bx} \gamma^{-1} \bar{\tau} \sigma_{\varepsilon}^2.
$$
Combining the above and~\eqref{lemmaB1.1}, we have 
$$
\|  \nabla \cR_n (\bbeta ^*) \|_\infty 
\leq
\sigma_{\bx} \bar{\tau}  \Bigg(2  \nu_0 \sigma_{\varepsilon} \sqrt{\frac{\log d + t }{n}} + \nu_0 \gamma \frac{\log d + t }{n} +\gamma^{-1} \sigma^2_{\varepsilon} \Bigg) 
$$
with probability at least  $1-2e^{-t}$, as desired.
\end{proof}

\subsection{Proof of Lemma~\ref{lem:grad_l2}}
\begin{proof}
Recall that $\bw^* = \nabla \cR_n (\bbeta ^*) -  \nabla \cR (\bbeta ^*)$.  The goal is to obtain an upper bound for the oracle centered loss function $\bw_{\cS}^*$ under the $\ell_2$ norm.
To this end, we employ a covering argument. 
Specifically, for any $\epsilon  \in (0,1)$, there exists an $\epsilon$-net $\cN_{\epsilon}$ of the unit sphere in $\RR^s$  with cardinality $| \cN_{\epsilon}| \leq (1+2/\epsilon)^s$ such that
\begin{align}\label{lem:gram_l2 part 1}
\| \bw^*_{\cS} \|_2 
\leq 
\frac{1}{1-\epsilon} \max_{\bu \in \cN_{\epsilon}} \big\langle -\bw^*_{\cS},\bu \big\rangle
=
\frac{1}{1-\epsilon} \max_{\bu \in \cN_{\epsilon}} \frac{1}{n} \sn \Big [ L'(\varepsilon_i)  \bx_{i\cS}^{\T} \bu  - \EE \left\{ L'(\varepsilon_i)  \bx_{i\cS}^{\T} \bu \right\}  \Big ]
\end{align}
From Condition \ref{def:general.loss} on the loss function $\ell(\cdot)$, we have $|\ell'(u)|\leq \min (1,|u|) $ for all $u \in \RR$.
Thus, we have $|L'(\varepsilon_i)| = |\gamma w_\tau (\varepsilon_i) \ell'(\varepsilon_i/\gamma) | \leq \min \big( \bar \tau \gamma, \bar{\tau} |\varepsilon_i| \big)$. 
Since $\bx$ is sub-exponential, by Condition~\ref{cond:covariates},
we have $\PP \big(|  \bu^{\T} \bx |\geq \nu_0\|\bu\|_{\bSigma}\cdot t \big)\leq e^{-t}$ for all $t \in \RR$ and $\bu \in \RR^d$. 
Thus, for all $k \geq 2$, and by a change of variable, we obtain
\begin{align*}
\EE \left(\big| L'(\varepsilon_i)  \bx_{i\cS}^{\T} \bu  \big|^k\right)
&\leq 
\EE \Big\{ (\bar{\tau}\gamma)^{k-2} | \bx_{i\cS}^{\T} \bu |^k  \EE \big(\bar{\tau}^2 \varepsilon_i^2|\bx_{i\cS} \big) \Big\} \\
&\leq
\bar{\tau}^k \gamma^{k-2} \sigma_{\varepsilon}^2 \EE \big| \bx_{i\cS}^{\T} \bu \big|^k \\
&\leq 
\bar{\tau}^k \gamma^{k-2} \sigma_{\varepsilon}^2 \int_{0}^{\infty} kt^{k-1} \PP \big(| \bx_{i\cS}^{\T} \bu | \geq t \big) \mathrm{d} t \\
&\leq
\frac{k!}{2} \Big(2\bar{\tau}^2 \sigma_{\varepsilon}^2 \nu_0^2 \| \bu \|_{\Sb}^2  \Big) \cdot \Big( \bar{\tau} \gamma \nu_0  \| \bu \|_{\Sb} \Big)^{k-2}.
\end{align*}

Applying the Bernstein's inequality with $a = 2\bar{\tau}^2 \sigma_{\varepsilon}^2 \nu_0^2 \| \bu \|_{\Sb}^2$ and $b=\bar{\tau} \gamma \nu_0  \| \bu \|_{\Sb} $, along with the inequality $\| \bu \|_{\Sb} \leq \lambda^{1/2}_{\max}(\Sb)\| \bu \|_2 = \lambda^{1/2}_{\max}(\Sb)$, we have for all $x>0$,
\begin{align}
\label{lem:gram_l2 part 2}
\frac{1}{n} \sn \Big [L'(\varepsilon_i) \bx_{i\cS}^{\T} \bu - \EE \{L'(\varepsilon_i) \bx_{i\cS}^{\T} \bu \}\Big ]
\leq 
\bar{\tau} \nu_0 \lambda^{1/2}_{\max}(\Sb)  \Bigg(2 \sigma_{\varepsilon} \sqrt{\frac{x}{n}} + \gamma \frac{x}{n} \Bigg ),
\end{align}
with probability at least $1-e^{-x}$. 
Combining \eqref{lem:gram_l2 part 1} and \eqref{lem:gram_l2 part 2}, and applying the union bound over all vectors $\bu \in \cN_{\epsilon}$, we have 
\begin{align*}
\| \bw^*_{\cS} \|_2 \leq \frac{\bar{\tau} \nu_0 \lambda^{1/2}_{\max}(\Sb)}{1-\epsilon} \Bigg(2 \sigma_{\varepsilon} \sqrt{\frac{x}{n}} + \gamma \frac{x}{n} \Bigg)
\end{align*}
with probability at least $1-(1+2/\epsilon)^se^{-x}$. Selecting $\epsilon = 1/3$ and $x=2s+t$,  we obtain
\begin{align*}
\| \bw^*_{\cS} \|_2 
\leq 3\bar{\tau} \nu_0 \lambda^{1/2}_{\max}(\Sb) \Bigg( \sigma_{\varepsilon} \sqrt{\frac{2s+t}{n}} + \gamma \frac{2s+t}{2n} \Bigg),
\end{align*}
 with probability at least $1 - e^{-t}$.
\end{proof}
\subsection{Proof of Lemma~\ref{lem:l1cone}}
\begin{proof} 
Let $\hat{\bbeta}$ be any solution to~\eqref{general.lasso2}. Since~\eqref{general.lasso2} is convex, there exists a subgradient $\bxi \in \partial \| \hat{\bbeta} \|_1$ such that $\nabla \cR_n(\hat{\bbeta})+\blambda \circ \bxi= \mathbf{0}$.  Thus, we have 
$$
\begin{aligned}
0 &= \langle \nabla \cR_n(\hat{\bbeta})+\blambda \circ \bxi,\hat{\bbeta} - \bbeta \rangle\\
 &= 
\langle \nabla \cR_n(\hat{\bbeta})-\nabla \cR_n(\bbeta),\hat{\bbeta} - \bbeta \rangle +
\langle \nabla \cR_n(\bbeta) -  \nabla \cR(\bbeta),\hat{\bbeta} - \bbeta \rangle +
\langle  \nabla \cR (\bbeta), \hat{\bbeta} - \bbeta\rangle +
\langle \blambda \circ \bxi,\hat{\bbeta} - \bbeta \rangle \\
&\geq
0+\langle \bw(\bbeta),\hat{\bbeta} - \bbeta \rangle +
\langle    \nabla \cR (\bbeta),\hat{\bbeta} - \bbeta \rangle + \langle \blambda \circ \bxi,\hat{\bbeta} - \bbeta \rangle \\
&\geq
-\| \bw(\bbeta) \|_{\infty} \| \hat{\bbeta} - \bbeta \|_1 -  \|  \nabla  \cR  (\bbeta )\|_2\| \hat{\bbeta} - \bbeta \|_2
+ \langle \blambda \circ \bxi,\hat{\bbeta} - \bbeta \rangle
\end{aligned}
$$
Since $\bbeta_{\cA^{\cc}}=\bf{0}, \| \bxi \|_{\infty}\leq 1$, and $\langle \bxi,\hat{\bbeta} \rangle = \| \hat{\bbeta} \|_1$, we can obtain a lower bound for  $\langle \blambda \circ \bxi,\hat{\bbeta} - \bbeta \rangle $ as
\begin{align*}
\langle \blambda \circ \bxi,\hat{\bbeta} - \bbeta \rangle 
&= \langle (\blambda \circ \bxi)_{\cA^{\cc}}, \hat{\bbeta}_{\cA^{\cc}}\rangle + \langle (\blambda \circ \bxi)_{\cA},(\hat{\bbeta} - \bbeta)_{\cA} \rangle \\
&\geq  \| \blambda_{\cA^{\cc}}  \|_{\min} \| \hat{\bbeta}_{\cA^{\cc}} \|_1 - \| \blambda_{\cA} \|_{\infty} \| (\hat{\bbeta} - \bbeta)_{\cA} \|_1 \\
&\geq \| \blambda_{\cA^{\cc}}  \|_{\min} \| (\hat{\bbeta} - \bbeta)_{\cA^{\cc}} \|_1 - \| \blambda \|_{\infty} \| (\hat{\bbeta} - \bbeta)_{\cA} \|_1.
\end{align*}
Combining the above inequalities yields
$$
\| \bw(\bbeta) \|_{\infty} \| \hat{\bbeta} - \bbeta \|_1 +  \|  \nabla  \cR  (\bbeta )\|_2\| \hat{\bbeta} - \bbeta \|_2 \geq
\| \blambda_{\cA^{\cc}}  \|_{\min} \| (\hat{\bbeta} - \bbeta)_{\cA^{\cc}} \|_1 -\| \blambda \|_{\infty} \| (\hat{\bbeta} - \bbeta)_{\cA} \|_1.
$$
The result~\eqref{eq:l1cone} can then be obtained by rearranging the terms. 
\end{proof}

\subsection{Proof of Lemma~\ref{lem:deterministic.error.bound}}
\begin{proof} 
The proof is similar to that of the proof of Theorem~\ref{thm:l1.retire}.  For some $r > 0$ to be specified, define an intermediate quantity $\hat{\bbeta}_\eta = \eta \hat{\bbeta} +(1-\eta)\bbeta^*$ where $\eta = \sup \{ u \in [0,1] : (1-u)\bbeta^* + u \hat{\bbeta} \in \bbeta^* + \BB(r) \}$.
When $\hat{\bbeta} \in \bbeta^* + \BB(r)$, we have $\hat{\bbeta}_\eta = \hat{\bbeta}$.  On the other hand, when $\hat{\bbeta} \notin \bbeta^* + \BB(r)$, $\hat{\bbeta}_\eta $ lies on  $\bbeta^* + \partial \BB(r)$ with $\eta <1$.   

We first show that $\hat{\bbeta}_\eta \in \bbeta^* + \BB(r) \cap \CC(L)$. 
Since $\cR_n(\cdot)$ is convex, by an application of Lemma C.1  in \cite{SZF2020}, we have
\begin{align}\label{basic.ineq}
0\le \langle \nabla \cR_n(\hat{\bbeta}_\eta) -\nabla \cR_n(\bbeta^*), \hat{\bbeta}_\eta-\bbeta^* \rangle \leq \eta \langle \nabla \cR_n(\hat{\bbeta})-\nabla \cR_n(\bbeta^*), \hat{\bbeta} - \bbeta^* \rangle.
\end{align}
Conditioned on the event 
$ \{a \lambda \geq 2\|\nabla \cR_n(\bbeta^*)-\nabla \cR(\bbeta^*) \|_\infty \}$ and the assumption that $\| \blambda \|_{\infty} \leq \lambda$ and $\| \blambda_{\cA^{\cc}}  \|_{\min} \geq a\lambda$, applying Lemma~\ref{lem:l1cone}, we have 

\begin{equation*}
\begin{split}
\| (\hat{\bbeta}-\bbeta^*)_{\cA^{\cc}} \|_1 &\leq
\frac{ \big\{ \| \blambda \|_{\infty} + \| \bw(\bbeta^*) \|_{\infty} \big\} \| (\hat{\bbeta}-\bbeta^*)_{\cA} \|_1 +  \|  \nabla  \cR  (\bbeta^* )\|_2 \|\hat{\bbeta}-\bbeta^*  \|_2}{\| \blambda_{\cA^{\cc}} \|_{\min}-\| \bw(\bbeta^*) \|_{\infty}}\\
&\le \left(1+\frac{2}{a}\right)\| (\hat{\bbeta} - \bbeta^*)_{\cA} \|_1 + \frac{2}{a\lambda} \|  \nabla  \cR  (\bbeta^* )\|_2 \| \hat{\bbeta} - \bbeta^* \|_2
\end{split}
\end{equation*}
By the assumption that $\lambda \geq s^{-1/2} \|  \nabla  \cR  (\bbeta^* )\|_2$, we have 
\begin{equation*}
\begin{split}
\|\hat{\bbeta} - \bbeta^* \|_1&\le \left(2+2/a\right)\| (\hat{\bbeta} - \bbeta^*)_{\cA} \|_1 + \frac{2}{a\lambda} \|  \nabla  \cR  (\bbeta^* )\|_2 \| \hat{\bbeta} - \bbeta^* \|_2\\
&\le \left(2+2/a\right) k^{1/2} \|(\hat{\bbeta}-\bbeta^*)_{\cA}\|_2+\frac{2}{a\lambda} \|  \nabla  \cR  (\bbeta^* )\|_2 \| \hat{\bbeta} - \bbeta^* \|_2\\
&\le \left\{\left(2+2/a\right) k^{1/2}+2s^{1/2}/a \right\} \| \hat{\bbeta} - \bbeta^* \|_2.
\end{split}
\end{equation*}
The above inequality implies that $\hat{\bbeta} \in \bbeta^* +  \CC(L)$ with $L=(2+2/a)k^{1/2} + 2 s^{1/2}/a$. 
Since $\hat{\bbeta}_\eta - \bbeta^*= \eta (\hat{\bbeta} -\bbeta^*)$ and $\hat{\bbeta}_\eta  \in \bbeta^* + \BB(r)$ by construction, we have $\hat{\bbeta}_\eta \in \bbeta^* + \BB(r) \cap \CC(L)$. Consequently, conditioned on the event $\cE_{\rm{rsc}}(r,L,\kappa)$, we have
\begin{align}\label{partial.ineq.1}
 \langle \nabla \cR_n(\hat{\bbeta}_\eta) -\nabla \cR_n(\bbeta^*), \hat{\bbeta}_\eta-\bbeta^* \rangle \ge \kappa \| \hat{\bbeta}_\eta - \bbeta^* \|_2 ^2. 
\end{align}
 
Next we upper bound the right-hand side of \eqref{basic.ineq}. Let  Since $\hat{\bbeta}$ is a solution to~\eqref{general.lasso2}, we have 
\begin{align*}
\langle \nabla \cR_n(\hat{\bbeta})-\nabla \cR_n(\bbeta^*), \hat{\bbeta} - \bbeta^* \rangle 
&= 
\langle \nabla \cR_n(\hat{\bbeta})+\blambda \circ \bxi,\hat{\bbeta} - \bbeta^* \rangle - \langle \blambda \circ \bxi, \hat{\bbeta} - \bbeta^* \rangle \\
&~~~~
- \langle \nabla \cR_n(\bbeta^*)-  \nabla \cR (\bbeta^*),\hat{\bbeta} - \bbeta^* \rangle - \langle  \nabla \cR (\bbeta^*),\hat{\bbeta} - \bbeta^* \rangle \\
&:=
\Pi_1 - \Pi_2 - \Pi_3 - \Pi_4
\end{align*}
We now obtain bounds for the terms $\Pi_1, \dots, \Pi_4$. 
For $\Pi_1$, since $\hat{\bbeta}$ is a solution to~\eqref{general.lasso2}, we have $\Pi_1 \leq 0$. 
We now obtain a lower bound for $\Pi_2$.  Since $[d]=\cS \cup (\cA \setminus\cS) \cup \cA^{\cc}$, $\bbeta^*_{\cS^{\cc}}=\textbf{0}, \| \bxi \|_{\infty}\leq 1$, and $\langle \bxi,\hat{\bbeta} \rangle = \| \hat{\bbeta} \|_1$, we have
\begin{equation*}
\begin{split}
\langle \blambda \circ \bxi,\hat{\bbeta} - \bbeta^* \rangle 
&=
\langle (\blambda \circ \bxi)_{\cS},(\hat{\bbeta} - \bbeta^*)_{\cS} \rangle + \langle (\blambda \circ \bxi)_{\cA \setminus\cS},\hat{\bbeta}_{\cA \setminus\cS} \rangle +\langle (\blambda \circ \bxi)_{\cA^{\cc}},\hat{\bbeta}_{\cA^{\cc}} \rangle \\
&\geq
-\| \blambda_{\cS} \|_2 \| (\hat{\bbeta} - \bbeta^*)_{\cS} \|_2 + \langle \blambda_{\cA \setminus\cS},| \hat{\bbeta}_{\cA \setminus\cS} | \rangle + \langle \blambda_{\cA^{\cc}},| \hat{\bbeta}_{\cA^{\cc}} | \rangle \\
&\geq
-\| \blambda_{\cS} \|_2 \| (\hat{\bbeta} - \bbeta^*)_{\cS} \|_2 + 0 + \|  \blambda_{\cA^{\cc}} \|_{\min} \| \hat{\bbeta}_{\cA^{\cc}} \|_1 \\
&\geq
-\| \blambda_{\cS} \|_2 \| (\hat{\bbeta} - \bbeta^*)_{\cS} \|_2 +\|  \blambda_{\cA^{\cc}} \|_{\min} \| (\hat{\bbeta}-\bbeta^*)_{\cA^{\cc}} \|_1.
\end{split}
\end{equation*}
For $\Pi_3$, it can be shown that 
\begin{equation*}
\begin{split}
\langle \nabla \cR_n(\bbeta^*)- \nabla \cR (\bbeta^*),\hat{\bbeta} - \bbeta^* \rangle  
& = 
\langle \bw^*_{\cA},\hat{\bbeta} - \bbeta^* \rangle  +  \langle \bw^*_{\cA^{\cc}},\hat{\bbeta} - \bbeta^* \rangle  \\
&\geq
-\| \bw^*_{\cA} \|_2 \| (\hat{\bbeta} - \bbeta^*)_{\cA} \|_2 - \| \bw^* \|_{\infty} \| (\hat{\bbeta} - \bbeta^*)_{\cA^{\cc}} \|_1.
\end{split}
\end{equation*}
Finally, for $\Pi_4$, we have 
$$
\big\langle   \nabla \cR (\bbeta^*),\hat{\bbeta} - \bbeta^* \big\rangle   
\geq
-\|  \nabla  \cR  (\bbeta^* )\|_2\| \hat{\bbeta} - \bbeta^* \|_2.
$$

Combining all of the above inequalities with $\| \blambda_{\cA^{\cc}}  \|_{\min} \geq a\lambda \geq  2\| \bw^* \|_{\infty}$, we obtain, 
\begin{align} 
\label{partial.ineq.2}
\langle \nabla \cR_n(\hat{\bbeta})-\nabla \cR_n(\bbeta^*), \hat{\bbeta} - \bbeta^* \rangle 
&\leq
\Big( -\| \blambda_{\cA^{\cc}}  \|_{\min} + \| \bw^* \|_{\infty} \Big) \| (\hat{\bbeta} - \bbeta^* )_{\cA^{\cc}} \|_1 \nn \\
&~~~~+\| \bw^*_{\cA} \|_2\| (\hat{\bbeta} - \bbeta^* )_{\cA} \|_2 + \| \blambda_{\cS} \|_2 \| (\hat{\bbeta} - \bbeta^* )_{\cS} \|_2 + \|  \nabla  \cR  (\bbeta^* )\|_2 \| \hat{\bbeta} - \bbeta^* \|_2 \nn \\
&\leq
\Big ( \| \blambda_{\cS} \|_2 + \| \bw^*_{\cA} \|_2 + \|  \nabla  \cR  (\bbeta^* )\|_2\Big ) \| \hat{\bbeta} - \bbeta^* \|_2.
\end{align}
Putting \eqref{basic.ineq}, \eqref{partial.ineq.1}, and  \eqref{partial.ineq.2} together, and using the fact that $\eta \| \hat{\bbeta} - \bbeta^* \|_2 = \| \hat{\bbeta}_\eta - \bbeta^* \|_2$, we obtain
\begin{align} \label{partial.ineq.3}
\kappa \| \hat{\bbeta}_\eta - \bbeta^* \|_2 \leq \| \blambda_{\cS} \|_2 + \| \bw^*_{\cA} \|_2 +\|  \nabla  \cR  (\bbeta^* )\|_2.
\end{align}
Furthermore, under the scaling conditions, we have $\| \blambda_{\cS} \|_2  \leq s^{1/2}\lambda$ and $ \| \bw^*_{\cA} \|_2 \leq k^{1/2}a\lambda/2$.
Putting these into~\eqref{partial.ineq.3}, we obtain $\| \hat{\bbeta}_\eta - \bbeta^* \|_2 \leq \kappa^{-1} \big\{ (2s^{1/2}+ {k^{1/2}a}/{2})\lambda \big\} < r$. Thus, $\hat{\bbeta}_\eta$ falls in the interior of $\bbeta^* + \BB(r)$, implying that $\eta = 1$ and that $\hat{\bbeta}_\eta=\hat{\bbeta}$. This completes the proof that $\| \hat{\bbeta} - \bbeta^* \|_2  \le \kappa^{-1} \big\{ (2s^{1/2}+ {k^{1/2}a}/{2})\lambda  \big\}$.
\end{proof}

\end{document}